\documentclass[twocolumn]{aastex63}

\newcommand{\gaia}{\it Gaia}
\newcommand{\tess}{\it TESS}
\newcommand{\Teff}{$T_{\text{eff}}$}

\newcommand{\exofast}{\tt EXOFASTv2}

\received{}
\revised{}
\accepted{}
\submitjournal{-}

\shorttitle{Testing the Limits of Precise Subgiant Characterization}
\shortauthors{Godoy-Rivera et al.}

\begin{document}

\title{Testing the Limits of Precise Subgiant Characterization with APOGEE and Gaia:\\Opening a Window to Unprecedented Astrophysical Studies}
\correspondingauthor{Diego Godoy-Rivera}
\email{godoyrivera.1@osu.edu}

\author[0000-0003-4556-1277]{Diego Godoy-Rivera}
\affiliation{Department of Astronomy, The Ohio State University, 140 West 18th Avenue, Columbus, OH 43210, USA}

\author[0000-0002-4818-7885]{Jamie Tayar}
\affiliation{Institute for Astronomy, University of Hawaii, 2680 Woodlawn Drive, Honolulu, Hawaii 96822, USA}
\affiliation{Hubble Fellow}

\author[0000-0002-7549-7766]{Marc H. Pinsonneault}
\affiliation{Department of Astronomy, The Ohio State University, 140 West 18th Avenue, Columbus, OH 43210, USA}

\author[0000-0003-1445-9923]{Romy Rodr\'iguez Mart\'inez}
\affiliation{Department of Astronomy, The Ohio State University, 140 West 18th Avenue, Columbus, OH 43210, USA}

\author[0000-0002-3481-9052]{Keivan G. Stassun}
\affiliation{Department of Physics \& Astronomy, Vanderbilt University, 6301 Stevenson Center Lane, Nashville, TN 37235, USA}

\author[0000-0002-4284-8638]{Jennifer L. van Saders}
\affiliation{Institute for Astronomy, University of Hawaii, 2680 Woodlawn Drive, Honolulu, Hawaii 96822, USA}

\author[0000-0002-1691-8217]{Rachael L. Beaton}
\affiliation{Hubble Fellow}
\affiliation{Carnegie-Princeton Fellow}
\affiliation{Department of Astrophysical Sciences, Princeton University, 4 Ivy Lane, Princeton, NJ 08544, USA}
\affiliation{The Observatories of the Carnegie Institution for Science, 813 Santa Barbara St., Pasadena, CA 91101, USA}

\author[0000-0002-1693-2721]{D. A. Garc\'ia-Hern\'andez}
\affiliation{Instituto de Astrof\'isica de Canarias (IAC), E-38205 La Laguna, Tenerife, Spain}
\affiliation{Universidad de La Laguna (ULL), Departamento de Astrof\'isica, E-38206 La Laguna, Tenerife, Spain}

\author{Johanna K. Teske}
\affiliation{Hubble Fellow}
\affiliation{The Observatories of the Carnegie Institution for Science, 813 Santa Barbara St., Pasadena, CA 91101, USA}
\affiliation{Earth and Planets Laboratory, Carnegie Institution of Washington, 5241 Broad Branch Road, N.W., Washington, DC 20015, USA}
\begin{abstract}
Given their location on the Hertzsprung-Russell (HR) diagram, thoroughly characterized subgiant stars can place stringent constraints on a wide range of astrophysical problems. Accordingly, they are prime asteroseismic targets for the Transiting Exoplanet Survey Satellite ({\tess}) mission. In this work, we infer stellar properties for a sample of 347 subgiants located in the {\tess} Continuous Viewing Zones (CVZs), which we select based on their likelihood of showing asteroseismic oscillations. We investigate how well they can be characterized using classical constraints (photometry, astrometry), and validate our results using spectroscopic values. We derive luminosities, effective temperatures, and radii with mean 1$\sigma$ random (systematic) uncertainties of 4.5\% (2\%), 33 K (60 K), and 2.2\% (2\%), as well as more model-dependent quantities such as surface gravities, masses, and ages. We use our sample to demonstrate that subgiants are ideal targets for mass and age determination based on HR diagram location alone, discuss the advantages of stellar parameters derived from a detailed characterization over widely available catalogs, show that the generally used 3D extinction maps tend to overestimate the extinction for nearby stars (distance $\lesssim$ 500 pc), and find a correlation that supports the rotation-activity connection in post main sequence stars. The complementary roles played by classical and asteroseismic data sets will open a window to unprecedented astrophysical studies using subgiant stars.
\end{abstract}

\section{Introduction}
\label{sec:intro}

The subgiant branch is an intermediate stage of evolution between the main sequence and the red giant branch (RGB). Subgiants are slightly more luminous than main sequence stars, and can therefore be studied at larger distances. In addition, their temperatures and gravities are not so different from the well-calibrated dwarf regime, and they have not yet undergone all the mixing and structural changes that happen on the RGB.

Given their intermediate evolutionary stage, subgiants can provide meaningful constraints on a number of astrophysical processes that depend on the stars' Hertzsprung-Russell (HR) diagram location. Some of these include: the transport of angular momentum between the core and envelope after the main sequence and as stars evolve towards the RGB \citep{deheuvels12,deheuvels14,deheuvels20}, the corresponding evolution of surface rotation rates \citep{simonian20}, the existence of a common dynamo mechanism across different evolutionary phases \citep{egeland18,lehtinen20,metcalfe20}, and mixing processes in stellar interiors such as atomic diffusion, gravitational settling, and the dredge up of nuclearly processed material due to the development of a deep convection zone \citep{boesgaard20,lind09,souto18,souto19}. Other areas that benefit from using subgiants include studies of: multiple populations in star clusters \citep{li16}, the occurrence rate of planets as a function of stellar mass \citep{johnson10,luhn19}, the origin of runaway stars \citep{hattori19}, and the distribution of stars around the Milky Way's central black hole \citep{schodel18}. Considering their many applications, a detailed and homogeneous characterization of large samples of subgiant stars to derive their stellar properties, including ages, is very relevant. 

Precise stellar ages for large samples of stars are crucial for answering several astrophysical questions, ranging from the evolution of extrasolar planetary systems \citep{silverstone06,meyer08,meng17,berger20b}, to the kinematic and chemical evolution history of our Galaxy \citep{bovy19,bensby13,bensby20,gallart19,rendle19,sharma19}. Measuring them {\it en masse}, however, has proved to be a challenge \citep{soderblom10,howes19,mints19,sahlholdt19}. Nevertheless, for subgiants, stellar evolution theory predicts precise ages as a function of position on the HR diagram, opening a window for a reliable age diagnostic.

Isochrone fitting, one of the oldest methods for estimating stellar ages, consists of comparing temperature and luminosity to the predictions of a stellar model. Although this technique is subject to significant confusion on many parts of the HR diagram, it is extremely well suited in the subgiant regime. Here, stars evolve rapidly at almost constant luminosity, and this method has the potential to provide extremely precise ages, so long as a precise luminosity can be determined. The release of parallaxes as part of the {\gaia} mission's Data Release 2 (DR2; \citealt{gaia18a,lindegren18}), in combination with precise photometric surveys such as the Two Micron All Sky Survey (2MASS; \citealt{cutri03,skrutskie06}) and ALLWISE \citep{wright10,cutri14}, provide that opportunity.

Ultimately, the stellar properties derived from isochrone fitting (luminosity, temperature, radius, mass, age) can be complemented and compared with those derived from spectroscopy (temperature, surface gravity, chemical abundances, rotational velocities; e.g., \citealt{holtzman15,holtzman18,jonsson20,meszaros13}), high precision light curves (surface rotation, activity indicators; e.g., \citealt{mcquillan14,santos19}), and asteroseismology (surface gravity, mass, age, core rotation; e.g., \citealt{chaplin13,gai17,li20b,li20a,serenelli17}). While some of these techniques are more suitable for dwarfs than giants (e.g., the measurement of rotational velocities from spectroscopy), and vice versa (e.g., the detection of seismic oscillations), subgiants inhabit an optimal intersection where often they all perform successfully. The simultaneous availability of all of these data will place subgiants as potentially the best characterized stars, and it will allow both unprecedented tests of stellar interiors theory, as well as extremely precise population studies for Galactic archaeology and planetary systems.

In this paper, we take a carefully selected sample of bright subgiant stars and examine how well they can be characterized using photometric and astrometric information. We compare these results to spectroscopic analyses, and have chosen the sample to be amenable to future asteroseismic comparisons as well. We demonstrate that the wide availability of precise all-sky data sets (e.g., {\gaia}, 2MASS, and ALLWISE) offers an opportunity to thoroughly characterize subgiant stars without the need for additional observations, and we show that accurate and precise temperatures and luminosities are sufficient to precisely estimate masses and ages for subgiants. This suggests a remarkable opportunity for a wide range of future astrophysical studies.

This paper is structured as follows. In \S\ref{sec:data} we present the sample of subgiants stars we study and gather the data we use throughout this work. In \S\ref{sec:method_and_results} we perform Spectral Energy Distribution (SED) fitting and derive accurate and precise stellar parameters. In \S\ref{sec:applications} we compare our results with those from large catalogs that are being extensively used by the community. In \S\ref{sec:rotation_and_activity} we study the rotation-activity connection in our subgiants. We conclude in \S\ref{sec:conclusions}.
\section{Data} 
\label{sec:data}
\subsection{Sample Selection}
\label{subsec:data_sample_selection}

For this investigation we select stars from both hemispheres with a high probability of having asteroseismic characterization from the Transiting Exoplanet Survey Satellite ({\tess}; \citealt{ricker15}), which in the future could be compared to the classical characterization we discuss here. Specifically, for the southern hemisphere we used the asteroseismic target list (ATL; \citealt{schofield19}) version 4, which used parallaxes from {\gaia} DR1 \citep{lindegren16}, and selected stars with a greater than 50\% probability of asteroseismic oscillation detection, located in the {\tess} southern continuous viewing zone (CVZ; ecliptic latitude $< -78{\degr}$), with estimated $\nu_{\text{max}} \geq 240$ $\mu$Hz to exclude more evolved giants, and with predicted temperatures cooler than 5600 K, where the asteroseismic mixed modes necessary for precise investigations of the stellar interior begin to develop. For the selection in the northern hemisphere, we used the updated ATL version 5, which incorporated parallaxes from {\gaia} DR2 \citep{lindegren18}, and applied the same cuts (oscillation probability $> 50\%$, ecliptic latitude $>78{\degr}$, $\nu_{\text{max}}\geq 240$ $\mu$Hz, predicted {\Teff} $<5600$ K). This resulted in 120 stars located in the southern hemisphere and 227 in the northern hemisphere, for a final sample of 347 stars in total, with some systematic differences in the distribution of stars between the two hemispheres (as we discuss in the remainder of \S\ref{sec:data} and \S\ref{sec:method_and_results}). Our target list is reported in Table \ref{tab:table_astrometry_photometry_spectroscopy}.

We note that the ATL temperatures (and thus the associated $\nu_{\text{max}}$ values and asteroseismic detection probabilities) were calculated using optical photometry from Tycho-2 \citep{hog00} and APASS \citep{henden09}, as well as 3D extinction maps \citep{bovy16}. As we discuss in \S\ref{subsec:data_photometry_and_astrometry} and \S\ref{subsec:methodresults_extinction}, we find all of these to be less precise than the level of precision that we can achieve with higher quality data in our sample. Therefore, many of the ATL temperatures are incorrect, particularly for stars with $G \gtrsim 10.5$ mag. This causes our sample to extend to {\Teff} $>$ 5600 K, at odds with the initially intended selection, and to cover the entirety of the subgiant branch. This effect is particularly pronounced for the northern sample, as it extends to fainter apparent magnitudes (see \S\ref{subsec:data_photometry_and_astrometry}).

\begin{table}
\tablenum{1}
\scriptsize
\centering
\caption{Target list and data compilation. \label{tab:table_astrometry_photometry_spectroscopy}}
\begin{tabular}{lll}
\hline
\hline
Column & Source & Description\\
\hline 
TIC & {\tess} Input Catalog & TIC ID\\
Hemisphere & This paper & Northern or Southern\\
{\gaia} DR2 & {\gaia} DR2 Catalog & {\gaia} DR2 Source ID\\
Parallax & {\gaia} DR2 Catalog & Parallax\\
$\sigma_{\text{Parallax}}$ & {\gaia} DR2 Catalog & Parallax error\\
$G$ & {\gaia} DR2 Catalog & $G$ band magnitude \\
$\sigma_{\text{G}}$ & {\gaia} DR2 Catalog & $G$ band magnitude error \\
$G_{\text{BP}}$ & {\gaia} DR2 Catalog & $G_{\text{BP}}$ band magnitude \\
$\sigma_{G_{\text{BP}}}$ & {\gaia} DR2 Catalog & $G_{\text{BP}}$ band magnitude error \\
$G_{\text{RP}}$ & {\gaia} DR2 Catalog & $G_{\text{RP}}$ band magnitude \\
$\sigma_{G_{\text{RP}}}$ & {\gaia} DR2 Catalog & $G_{\text{RP}}$ band magnitude error \\
2MASS & 2MASS Catalog & 2MASS ID\\
$J$ & 2MASS Catalog & $J$ band magnitude\\
$\sigma_{J}$ & 2MASS Catalog & $J$ band magnitude error\\
$H$ & 2MASS Catalog & $H$ band magnitude\\
$\sigma_{H}$ & 2MASS Catalog & $H$ band magnitude error\\
$K$ & 2MASS Catalog & $K$ band magnitude\\
$\sigma_{K}$ & 2MASS Catalog & $K$ band magnitude error\\
ALLWISE & ALLWISE Catalog & ALLWISE ID\\
W1 & ALLWISE Catalog & W1 magnitude\\
$\sigma_{\text{W1}}$ & ALLWISE Catalog & W1 magnitude error\\
W2 & ALLWISE Catalog & W2 magnitude\\
$\sigma_{\text{W2}}$ & ALLWISE Catalog & W2 magnitude error\\
W3 & ALLWISE Catalog & W3 magnitude\\
$\sigma_{\text{W3}}$ & ALLWISE Catalog & W3 magnitude error\\
W4 & ALLWISE Catalog & W4 magnitude\\
$\sigma_{\text{W4}}$ & ALLWISE Catalog & W4 magnitude error\\
GALEX & GALEX Catalog & GALEX ID\\
$FUV$ & GALEX Catalog & FUV magnitude\\
$\sigma_{FUV}$ & GALEX Catalog & FUV magnitude error\\
$NUV$ & GALEX Catalog & NUV magnitude\\
$\sigma_{NUV}$ & GALEX Catalog & NUV magnitude error\\
NUV Excess  & This paper & NUV excess\\
$\sigma_{\text{NUV Excess}}$ & This paper & NUV excess error\\
{\Teff} & APOGEE Catalog & Temperature\\
$\sigma_{T_{\text{eff}}}$ & APOGEE Catalog & Temperature error\\
$\log(g)$ & APOGEE Catalog & Surface gravity\\
$\sigma_{\log(g)}$ & APOGEE Catalog & Surface gravity error\\
$\log(g)$ cal & APOGEE Catalog & Surface gravity calibration\\
$[$M/H$]$ & APOGEE Catalog & Metallicity\\
$\sigma_{\text{[M/H]}}$ & APOGEE Catalog & Metallicity error\\
$v\sin i$ & APOGEE Catalog & Projected rotational velocity\\
\hline
\\
\end{tabular}
\tablecomments{(The full table is available online in machine-readable format.) Compilation of our subgiant target list, as well as the photometric, astrometric, and spectroscopic data used throughout this work from the {\gaia} DR2, 2MASS, ALLWISE, and APOGEE catalogs. Additionally, when available, we report photometry from GALEX and the NUV excess we calculate in \S\ref{sec:rotation_and_activity}.}
\end{table}
\subsubsection{{\tess} Objects of Interest in our Sample}
\label{subsubsec:data_sample_selection_TOI}

For completeness, we search for potential exoplanet candidates in our sample, and find four stars in the {\tess} Object of Interest (TOI) list (updated as of September 2020). These stars (and their respective TOIs) are: TIC 381976956 (TOI 866.01), TIC 349829627 (TOI 788.01), TIC 233735068 (TOI 1306.01), and TIC 229940491 (TOI 1642.01). We do not use this information throughout this paper, but note that we report revised stellar parameters for these stars as part of our analysis.
\subsection{Photometric and Astrometric Data}
\label{subsec:data_photometry_and_astrometry}

\begin{figure*}[ht!]
\gridline{
	    \fig{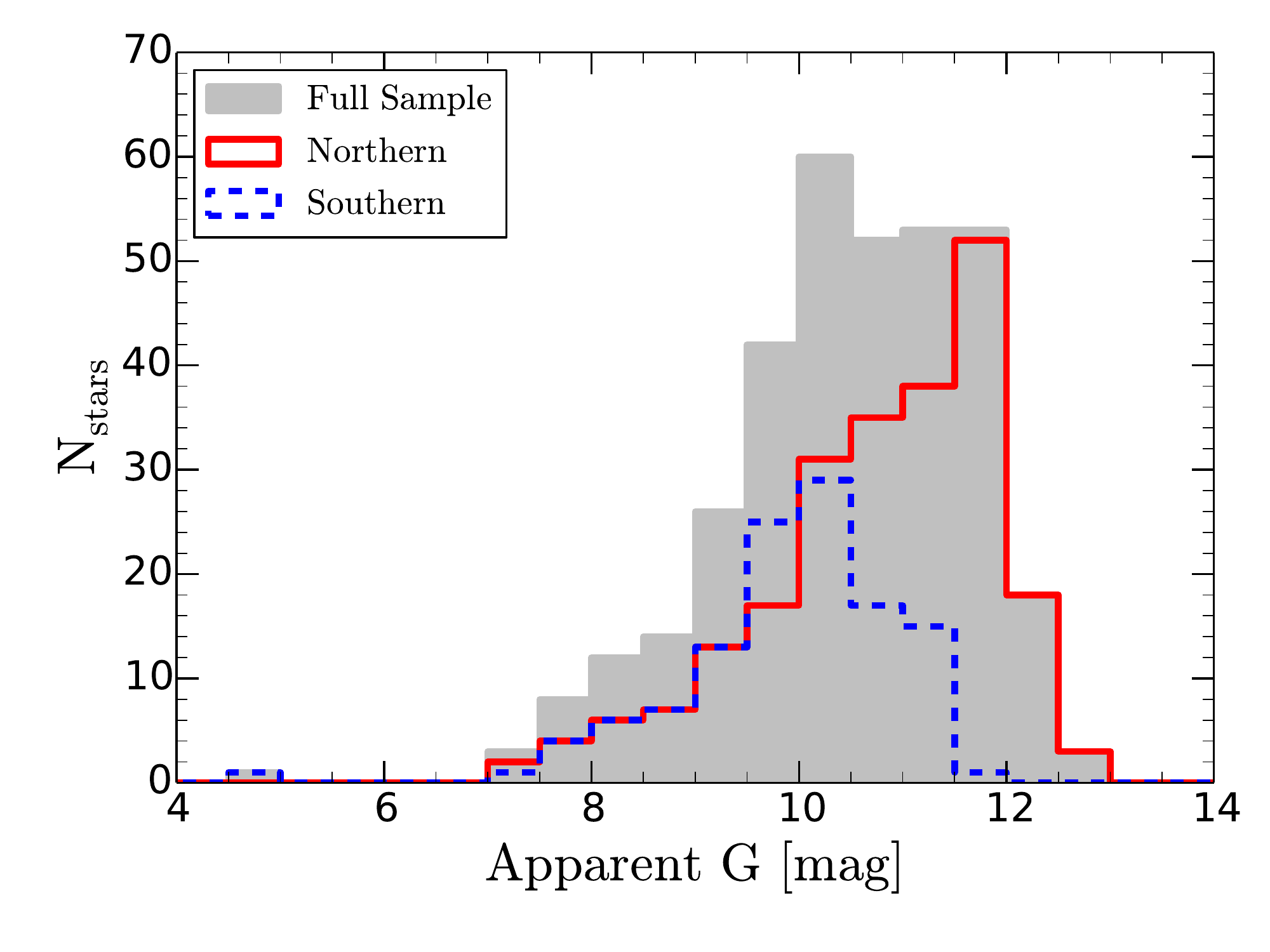}{0.45\textwidth}{(a)}
	    \fig{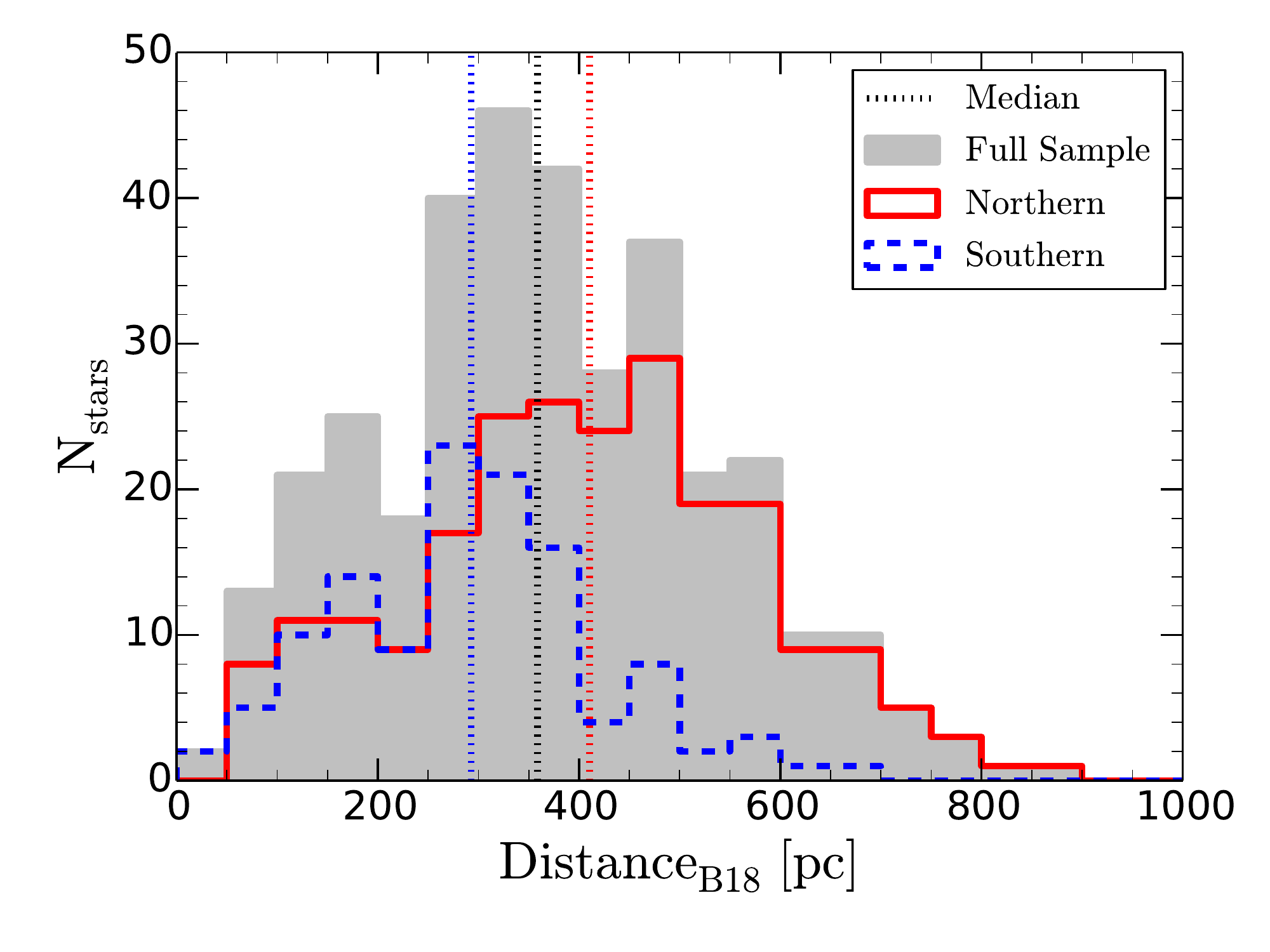}{0.45\textwidth}{(b)}
          }
\gridline{
	    \fig{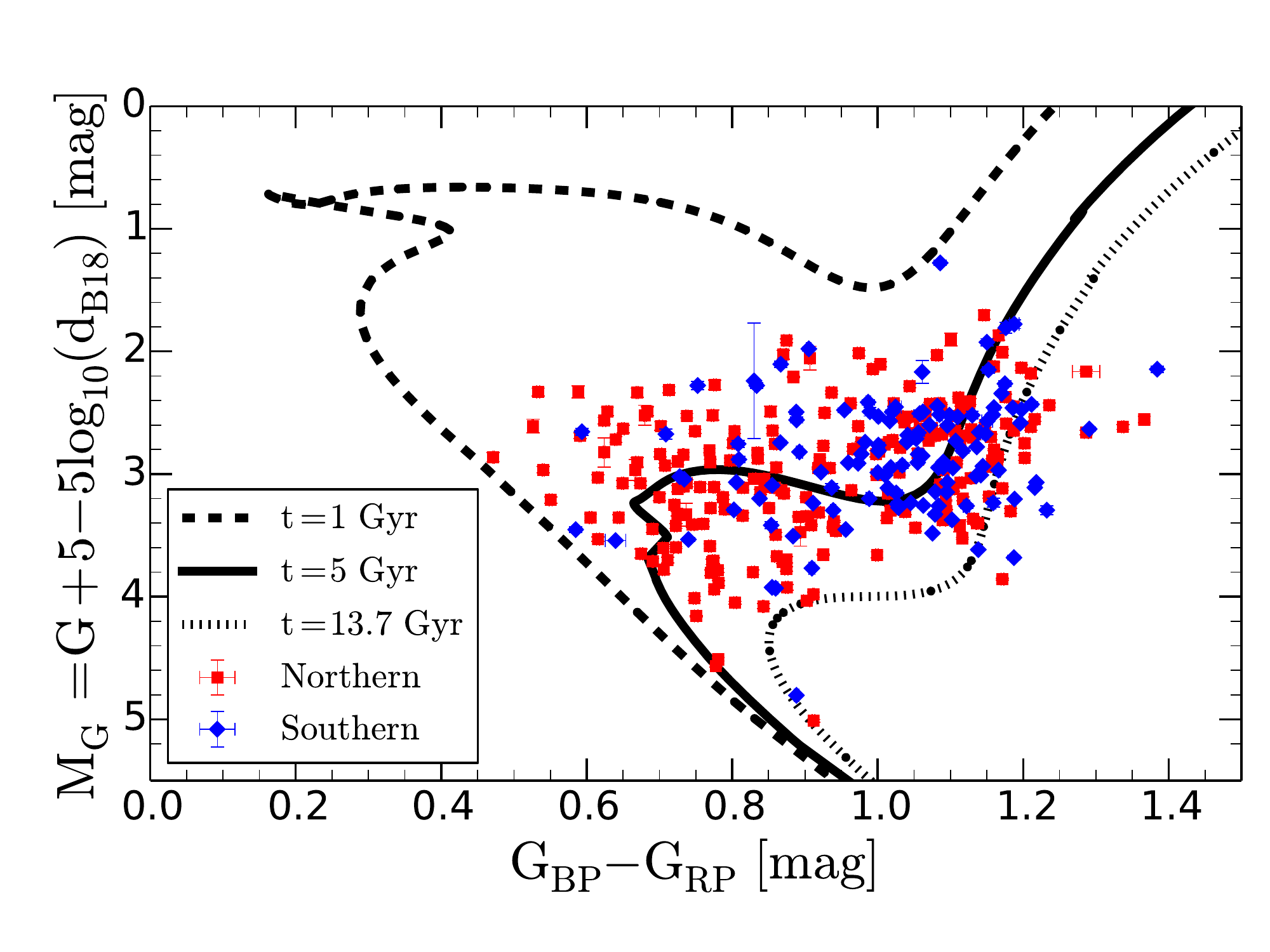}{0.45\textwidth}{(c)}
	    \fig{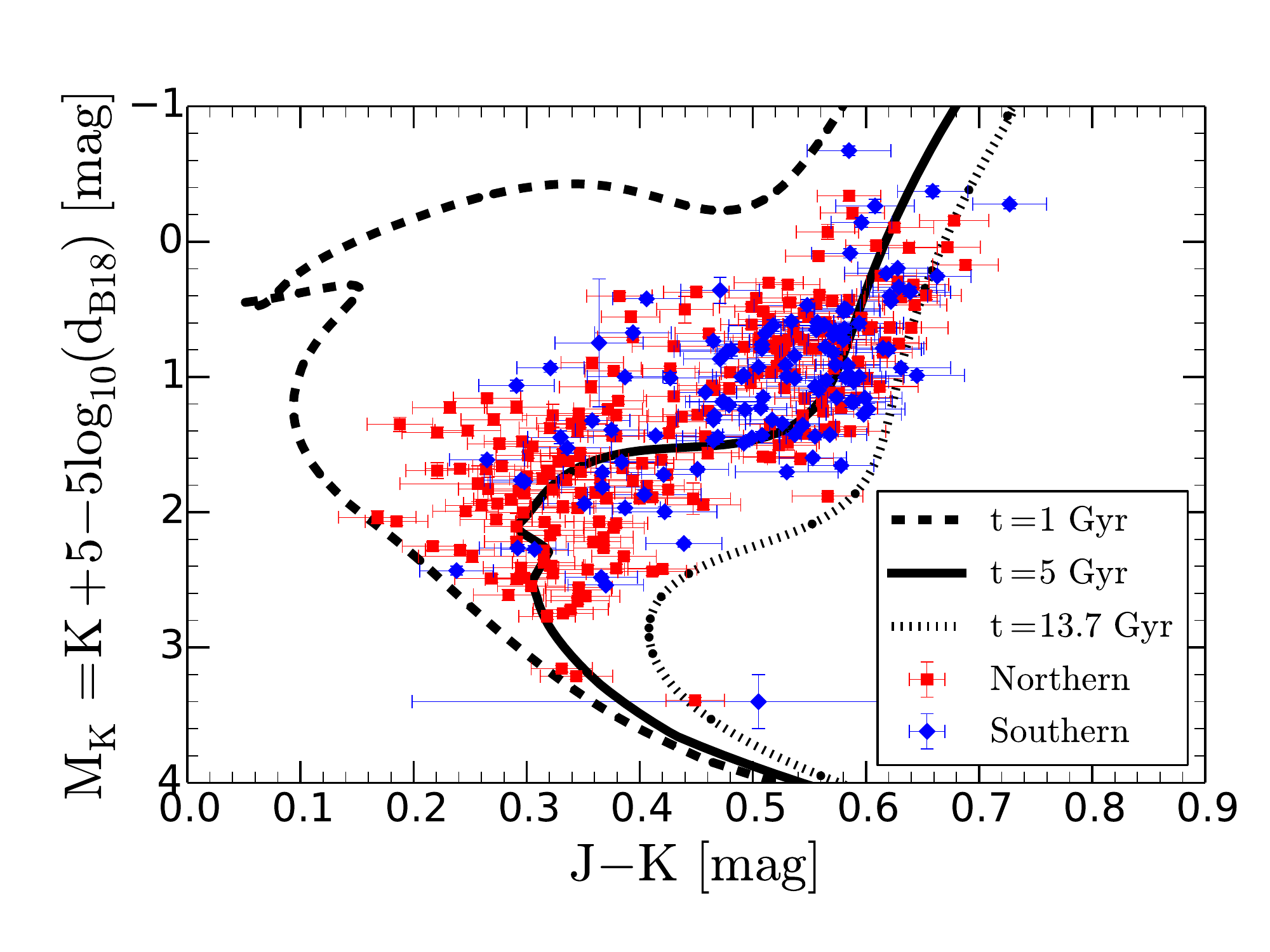}{0.45\textwidth}{(d)}
          }
\caption{Summary of photometric and astrometric data for our subgiant stars. The full sample is shown in grey in the background, while the northern sample is shown in red (solid line in the top panels, squares in the bottom panels) and the southern sample is shown in blue (dashed line in the top panels, diamonds in the bottom panels). \textbf{(a)} Distribution of apparent {\gaia} $G$ band magnitudes. The northern sample (which was selected using {\gaia} DR2) is larger and extends to fainter magnitudes than the southern sample (which was selected using {\gaia} DR1). \textbf{(b)} Distribution of {\gaia} DR2 distances from B18. The median values of the distributions are shown as the vertical dotted lines in the corresponding colors. The median distance of the southern sample is $\sim$ 100 pc closer than that of the northern sample. \textbf{(c)} {\gaia} absolute $M_G$ ($= G+5-5\log_{10}(d_{\text{B18}})$) versus $G_{\text{BP}}-G_{\text{RP}}$ CMD. \textbf{(d)} 2MASS absolute $M_K$ ($= K+5-5\log_{10}(d_{\text{B18}})$) versus $J-K$ CMD. Neither of the CMDs have been corrected for extinction. For reference, in both CMDs we show a sample of MIST models with solar metallicity and ages of 1 (dashed), 5 (solid), and 13.7 (dotted) Gyr. We truncate the models to include only hydrogen burning phases for clarity. Both CMDs show that the southern sample is more evolved on average than the northern sample.}
\label{fig:data_sample_astrometry_photometry}
\end{figure*}

We now proceed to gather photometric and astrometric data for our targets. We first query the 8th version of the {\tess} Input Catalog (TICv8; \citealt{stassun19}), which also contains information about the crossmatching between the TICv8 and a number of catalogs such as {\gaia} DR2, 2MASS, and ALLWISE.

We use the photometric data from these surveys ({\gaia} $G$, $G_{\text{BP}}$, and $G_{\text{RP}}$; 2MASS $J$, $H$, and $K$; and ALLWISE W1, W2, W3, and W4), as well as {\gaia} parallaxes (accounting for the parallax zero point following \citealt{schonrich19}) and distances from \citealt{bailerjones18} (hereafter B18). These data were available for virtually all of our targets. We find parallaxes and photometry in all the {\gaia}, 2MASS, and W1, W2, and W3 ALLWISE bands for 340/347 stars ($\approx$ 98\%). We also find photometry in the ALLWISE W4 band for 269/347 stars ($\approx$ 78\%), with the lower recovery rate likely due to the wavelength-dependent ALLWISE resolution. We report the photometric and astrometric data we compile in Table \ref{tab:table_astrometry_photometry_spectroscopy}.

While there are other sources of photometry available, particularly in the optical (such as those used in the ATL), they did not meet our requirements for homogeneous and reliable photometry in both hemispheres and across our magnitude range. We do not use the Tycho-2 \citep{hog00} photometry, as most of our stars are considered too faint for Tycho-2. Conversely, our stars are too bright for the Pan-STARRS photometry \citep{chambers16}, which saturates around 12\textendash14th mag. Additionally, we find the latest APASS photometry (DR10; \citealt{henden19}) to be heterogeneous, which makes it hard to control the systematics, especially for these relatively bright stars. Finally, we do not use the GALEX near-UV (NUV) photometry \citep{bianchi11,bianchi17} for homogeneous characterization purposes, as NUV detections are poor probes of the SED in this temperature range, because SED models are typically poorly calibrated in the UV \citep{eastman19}, and because NUV magnitudes are correlated with rotation and activity \citep{dixon20}. Nonetheless, we make use of the GALEX photometry later, to study the rotation-activity connection in our sample (see \S\ref{sec:rotation_and_activity}).

Figure \ref{fig:data_sample_astrometry_photometry} summarizes our photometric and astrometric data by showing the apparent {\gaia} $G$ band magnitude and distance distribution for our stars, as well as the absolute {\gaia} and 2MASS color magnitude diagrams (CMDs). The northern sample is larger and extends to fainter magnitudes, but all of our targets are brighter than $G=13$ mag. The median distances of the northern and southern sample are $\sim$ 400 pc and $\sim$ 300 pc, respectively. All of this is in agreement with the sample selection described in \S\ref{subsec:data_sample_selection}, as the southern stars were selected using the more restricted {\gaia} DR1 data, while the northern stars were selected using the more recent {\gaia} DR2. For reference, in the CMDs we also show a sample of representative non-rotating MESA Isochrones and Stellar Tracks (MIST) models \citep{choi16,dotter16,paxton11,paxton13,paxton15} with solar metallicity and ages of 1, 5, and 13.7 Gyr. Both the {\gaia} and 2MASS CMDs have similar topologies and show that our stars are distributed along the subgiant branch, with two concentrations of stars, one closer to the main sequence turn off (MSTO) and one closer to the lower RGB. This is consistent with expectations based on the lifetime of the different phases, and rapid evolution on the subgiant branch. Comparing between the hemispheres, the southern sample is more evolved on average than the northern sample. This is a consequence of the ATL temperatures being less accurate for the latter, as it extends to fainter magnitudes, causing it to unintendedly include hotter, less evolved stars (see \S\ref{subsec:data_sample_selection}).
\subsection{Spectroscopic Data}
\label{subsec:data_spectroscopy}

\begin{figure}[h]
\epsscale{1.2}  
\plotone{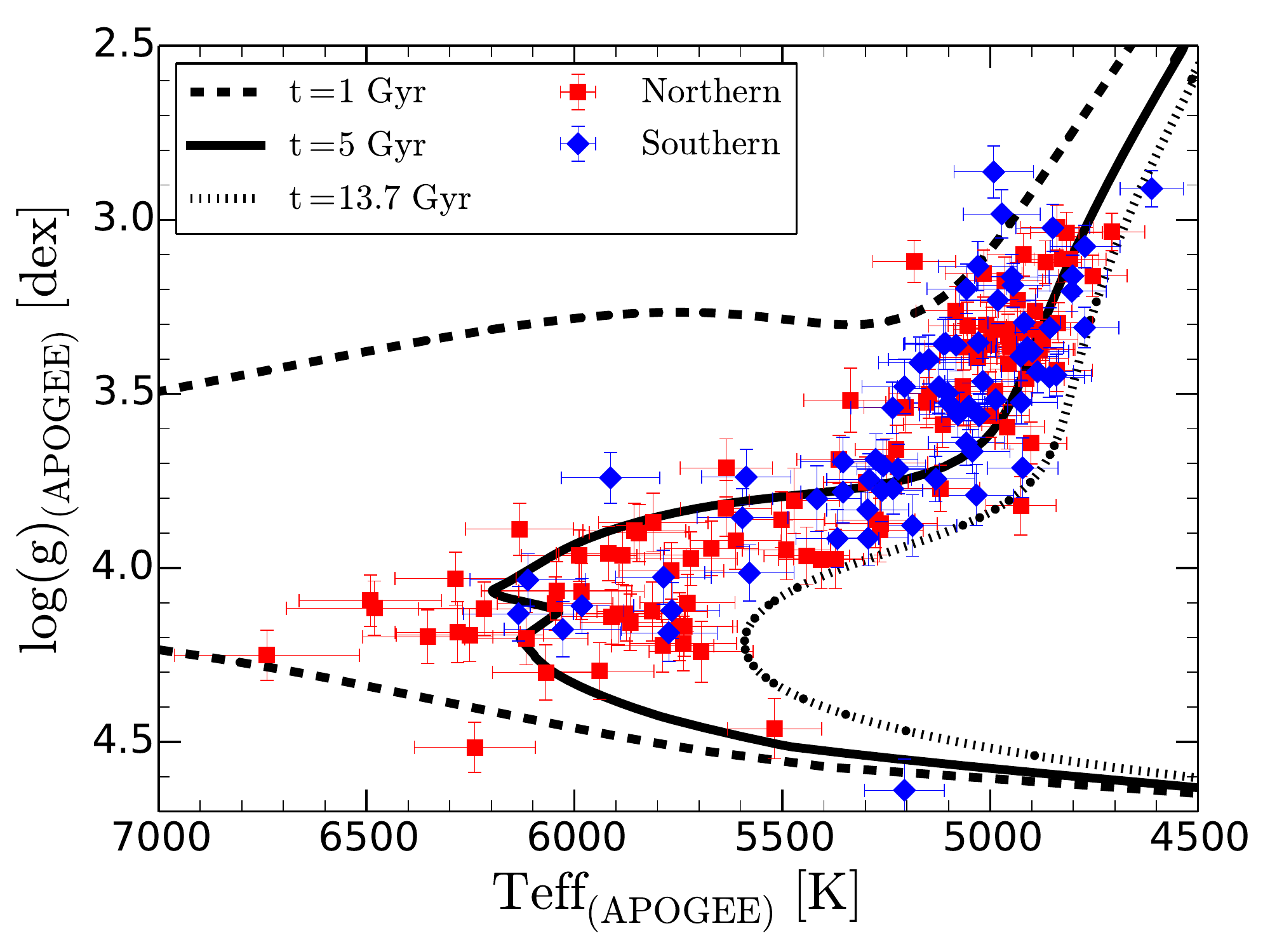}
\caption{$\log(g)$ versus {\Teff} (Kiel diagram) for the subgiants with APOGEE data (168/347 stars, or $\approx$ 48\% of our sample). The color and symbol scheme is the same as Figure \ref{fig:data_sample_astrometry_photometry}, and for reference we show the same set of MIST models. Similar to what is seen in the CMDs, our stars are distributed along the subgiant branch with two concentrations of points, one closer to the MSTO and one closer to the lower RGB, in agreement with expectations based on the lifetime of the different phases.}
\label{fig:data_sample_spectroscopy}
\end{figure}

The source of spectroscopic parameters for our sample of subgiants is the Apache Point Observatory Galactic Evolution Experiment (APOGEE; \citealt{majewski17,ahumada17,jonsson20}). APOGEE is one of the major spectroscopic programs of the Sloan Digital Sky Survey IV (SDSS-IV; \citealt{blanton17}), and its data are collected with the 2.5-m SDSS telescope \citep{gunn06} located at Apache Point Observatory and the 2.5-m du Pont telescope \citep{bowen73} at Las Campanas Observatory using twin $R \sim 22,500$ $H$ band spectrographs \citep{wilson19b}. The APOGEE data reduction, spectral parameters and chemical abundances pipeline (ASPCAP), and spectral line list, are presented in \citet{nidever15}, \citet{garciaperez16}, and \citet{shetrone15} and \citet{smith21}, respectively. The general APOGEE target selection is described in \citet{zasowski17}, and in this paper we also benefit from the APOGEE Bright Time Extension (BTX; Beaton et al., in prep) proposal as well as from a Contributed Program using access to APOGEE through the Carnegie Observatories Time Allocation Committee (Santana et al., in prep).

The enormous advantage of using data from APOGEE, in contrast to other spectroscopic surveys, is that we can treat our northern and southern hemisphere samples homogeneously. We use the internal APOGEE update that observations up to March 2020\footnote{This corresponds to the closure of the telescopes due to the COVID-19 pandemic.} (MJD=58932), and find data for 168/347 ($\approx$ 48\%) of our subgiant stars. The APOGEE stellar parameters include {\Teff}, $\log(g)$, and [M/H], with their mean uncertainties for our sample being $\approx$ 106 K, $\approx$ 0.073 dex, and $\approx$ 0.009 dex, respectively \citep{jonsson20}. We report the spectroscopic data we compile in Table \ref{tab:table_astrometry_photometry_spectroscopy}.

Figure \ref{fig:data_sample_spectroscopy} shows the $\log(g)$ versus {\Teff} (Kiel diagram) for the stars found in APOGEE, with the northern and southern samples shown in red and blue, together with the same set of representative solar metallicity MIST models for reference. The Kiel diagram shows a similar behaviour to the CMDs of Figure \ref{fig:data_sample_astrometry_photometry}. Additionally, although not explicitly shown, the metallicity distribution of our targets is centered at $\text{[M/H]}=0$ dex, with $\sigma_{\text{[M/H]}}=0.23$ dex, and covers the entire $-0.5 < \text{[M/H]} < +0.5$ dex range, which explains the stars located outside the isochrone bounds in Figures \ref{fig:data_sample_astrometry_photometry} and \ref{fig:data_sample_spectroscopy}.
\section{Method and Results}
\label{sec:method_and_results}

At this point we proceed to use the homogeneous and comprehensive data set we have assembled to derive stellar properties for our sample of subgiants. For this, we use the SED fitting technique.

The basis of this method is that stellar models predict the SED of stars as a function of temperature, composition, and surface gravity, and these model predictions can be compared with the observed photometry to solve for the intrinsic stellar properties (e.g., see \citealt{stassun16,stassun18b,stevens17}). Another component of this fit is interstellar reddening, which induces a wavelength-dependent SED distortion. Both extinction and {\Teff} can be inferred from SED fitting, although there are well known degeneracies in such solutions. In our case, because we have spectroscopic parameters from APOGEE, we can use spectroscopic priors for our solutions and test the accuracy of our SED fitting results.
\subsection{{\exofast}}
\label{subsec:methodresults_exofast}

{\exofast} \citep{eastman13,eastman17,eastman19} is an IDL exoplanet modeling software package that also provides stellar parameter estimates. We take advantage of such capabilities to perform SED fitting and derive physical properties for our subgiants.

To do this, {\exofast} requires as input broad band photometry, as well as priors provided by the user. Although {\exofast} has the capability of automatically querying a number of photometric catalogs when fitting an SED, many of these, however, are too heterogeneous or unreliable for our stars and purposes (see \S\ref{sec:data}). We override this capability and only include the trustworthy photometry from {\gaia}, 2MASS, and ALLWISE. 

Additionally, {\exofast} uses the MIST stellar evolutionary models in the parameter estimation. These models assume no rotation, solar-scaled abundances \citep{asplund09}, and their colors are based on the ATLAS/SYNTHE stellar atmosphere models (\citealt{kurucz70,kurucz79,kurucz93}; see \citealt{choi16} and \citealt{dotter16} for further details on the MIST models). We nevertheless note that, for most of the properties we derive, the dependence on models is minimal (see \S\ref{subsec:methodresults_results}). We discuss further details of our SED fitting with {\exofast} (such as the priors we use) in Appendix \ref{appendix:app_sed_fitting}.
\subsection{Interstellar Extinction}
\label{subsec:methodresults_extinction}

Interstellar extinction is a key component of the SED fitting procedure, and {\exofast} allows the user to either provide priors on the $V$ band extinction ($A_V$), or to derive it simultaneously with the stellar parameters, as long as informative inputs are provided. Given that our targets are nearby stars located in the {\tess} CVZs, we expect them to be on low-extinction sightlines. For reference, the foreground reddening of the Large Magellanic Cloud (LMC), which is located in the {\tess} southern CVZ, has been estimated by \citet{bessel91} to be $E(B-V)=0.05$ (or $A_V=0.155$ using $R_V=3.1$; \citealt{cardelli89}), and we do not expect the bulk of our southern targets to exceed this value.

In recent years several 3D extinction maps have become available, in principle superseding the classical 2D map by \citet{schlegel98}. Examples of these are {\tt Bayestar15/17/19} \citep{green15,green18,green19} and \citet{bovy16}. While homogeneous in nature, the {\tt Bayestar} maps only cover the sky north of Declination $\gtrsim -30{\degr}$. On the other hand, while all-sky, the \citet{bovy16} map is a combination of the maps by \citet{marshall06}, \citet{green15}, and \citet{drimmel03}. Additionally, although the TICv8 catalog also reports extinctions, these are based on a combination of the maps by \citet{green18} and \citet{schlegel98}. Ultimately, none of these maps cover the entire sky homogeneously, and given that our subgiant stars span both hemispheres, any 3D map we could adopt would inevitably use a combination of heterogeneous maps and techniques. We are therefore cautious of their accuracy, and proceed to assess their reliability.

\begin{figure}[h]
\epsscale{1.2}  
\plotone{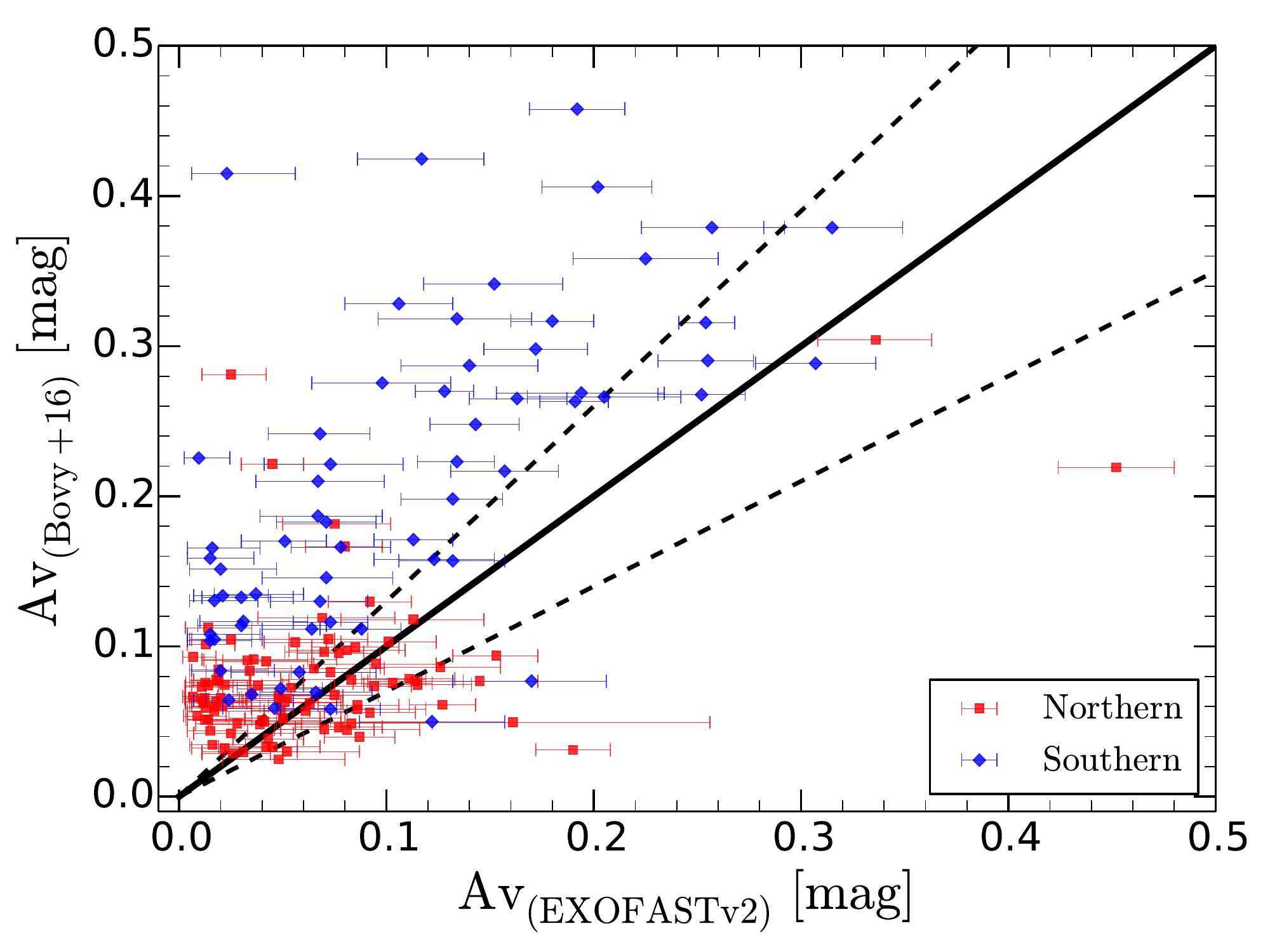}
\caption{Comparison of extinction values from the all-sky 3D \citet{bovy16} map with those inferred from {\exofast} (via SED fitting) for the subset of subgiants with spectroscopic parameters. The color and symbol scheme is the same as Figure \ref{fig:data_sample_astrometry_photometry}. The solid and dashed lines show the 1:1 and $\pm$ 30\% offsets for reference. Only the $x$-axis has error bars, as the \citet{bovy16} map does not report extinction uncertainties. For most of the sample the points are off the 1:1 relation and the \citet{bovy16} values overestimate the extinction.}
\label{fig:Av_map_comparison}
\end{figure}

In order to test these maps, we use the {\exofast} capabilities on the subset of stars with spectroscopic parameters, and calculate extinctions on a star-by-star basis. In other words, for the $\approx$ 48\% of the sample with APOGEE parameters, we {\it know} the values for {\Teff} and [M/H], and we fix them as answers when performing the SED fit, letting {\exofast} solve for the extinction as a free parameter. This sets the SED-derived temperatures to be, by construction, approximately on the APOGEE temperature scale. We provide further details of this calculation in Appendix \ref{appendix:app_sed_fitting}.

\begin{figure}[h]
\epsscale{1.2}  
\plotone{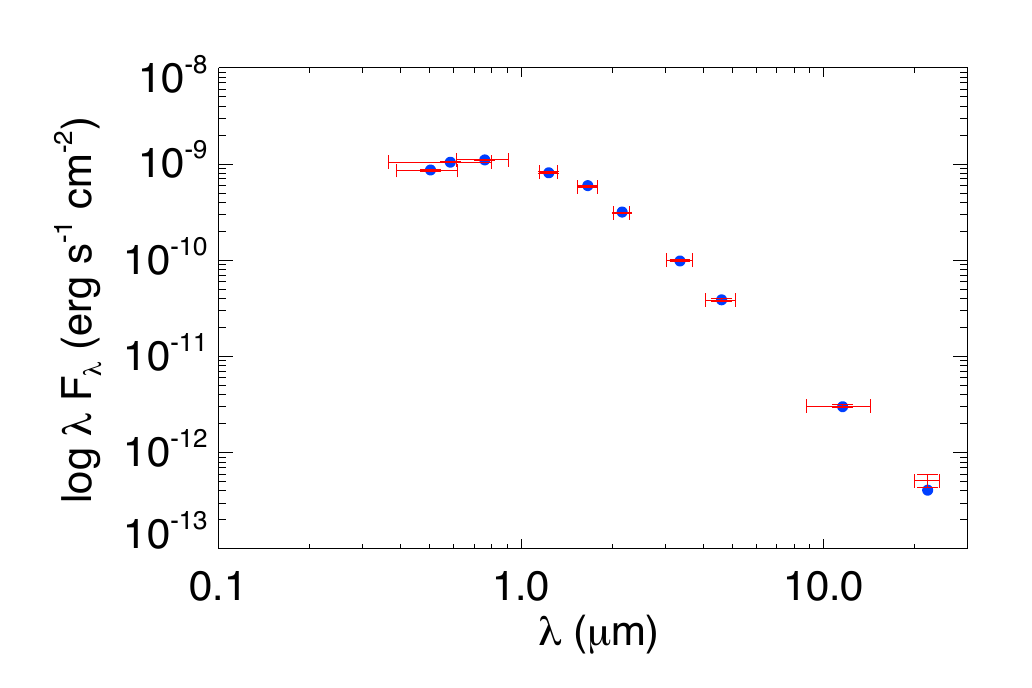}
\caption{Example of an {\exofast} SED fit for a representative star in our sample, TIC 441802076. The blue circles correspond to the passband integrated model fluxes, while the red crosses correspond to the observed broad band fluxes (with the error bars in wavelength denoting the width of the filters). In order of increasing wavelength the bands shown are {\gaia} $G_{\text{BP}}$, $G$, and $G_{\text{RP}}$, 2MASS $J$, $H$, and $K$, and ALLWISE $W1$, $W2$, $W3$, and $W4$. The {\gaia} photometry is imperative in our analysis, as it allows us to sample the peak of the SED.}
\label{fig:sed_example}
\end{figure}

In Figure \ref{fig:Av_map_comparison} we compare the resulting individual-star extinctions (available for $\approx$ 48\% of the sample), with those obtained by querying the \citet{bovy16} map (using the {\tt mwdust} package). For the northern sample, where the extinctions tend to be lower, some stars are close to the 1:1 line while for others \citet{bovy16} overestimates their extinctions. For the southern sample, where the extinctions tend to be higher, most of the \citet{bovy16} values also overestimate the extinctions by more than 30\%. Comparisons with the other aforementioned maps yielded similar disagreements. This suggests that adopting one of the extinction maps could potentially bias our results and introduce undesired systematic effects on the derived stellar parameters. Perhaps the inconsistent behaviours of the 3D maps can be explained by the spatial location of our stars, which are relatively nearby (see Figure \ref{fig:data_sample_astrometry_photometry}), and could therefore be too close for the maps to have enough foreground data to estimate reliable extinctions. Regardless, all of this raises serious concerns about the accuracy of the maps within $\sim$ 500 pc.

Given the above, we ultimately cease trying to obtain extinction values on a star-by-star basis from 3D maps. Instead, we use the fact that our targets are on low-extinction sightlines (Figure \ref{fig:Av_map_comparison}), and that we are observing tightly clustered stars that cover a small fraction of each hemisphere, to justify assuming that the extinction is uniform within a given hemisphere. We use the individual-star $A_V$ values derived by {\exofast} for stars with spectroscopic parameters to calculate global per-hemisphere extinctions, finding $A_V = 0.048 \pm 0.010$ and $A_V = 0.107 \pm 0.010$ for the northern and southern hemisphere, respectively. Further details on the calculation of these values can be found in Appendix \ref{appendix:app_sed_fitting}. Note that the global southern hemisphere extinction is, as expected, lower than the \citet{bessel91} value for the LMC.
\subsection{Results for the Full Sample}
\label{subsec:methodresults_results}

\begin{table*}
\tablenum{2}
\begin{minipage}{\textwidth}
\begin{center}
\caption{Table of stellar parameters derived from SED fitting using {\exofast}.\label{tab:table_exofast_derived_parameters}}
\begin{tabular}{c c c c c c c}
\hline
\hline
TIC & Luminosity & Temperature & Radius & $\log(g)$ & Mass & Age \\
-  & [$L_{\odot}$] & [K] & [$R_{\odot}$] & [dex] & [$M_{\odot}$] & [Gyr] \\
\hline
141757732 & $14.72^{+0.90}_{-0.78}$ & $4885^{+16}_{-16}$ & $5.36^{+0.16}_{-0.14}$ & $3.084^{+0.078}_{-0.085}$ & $1.28 ^{+0.23}_{-0.23}$ & $4.4^{+4.5}_{-2.0}$\\
142109390 & $5.64^{+0.15}_{-0.14}$ & $4892^{+22}_{-21}$ & $3.305^{+0.050}_{-0.045}$ & $3.469^{+0.064}_{-0.062}$ & $1.18^{+0.18}_{-0.16}$ & $6.5^{+4.2}_{-2.6}$\\
176875114 & $3.272^{+0.061}_{-0.056}$ & $5490^{+31}_{-27}$ & $1.998^{+0.022}_{-0.022}$ & $3.904^{+0.024}_{-0.027}$ & $1.166^{+0.060}_{-0.065}$ & $6.76^{+1.20}_{-0.91}$\\
\ldots & \ldots & \ldots & \ldots & \ldots & \ldots & \ldots \\
233658147 & $3.86^{+0.16}_{-0.14}$ & $5897^{+38}_{-34}$ & $1.881^{+0.036}_{-0.035}$ & $3.999^{+0.041}_{-0.047}$ & $1.29^{+0.11}_{-0.12}$ & $4.4^{+1.8}_{-1.3}$\\
441733094 & $4.8^{+0.31}_{-0.27}$ & $6734^{+54}_{-42}$ & $1.608^{+0.044}_{-0.043}$ & $4.194^{+0.027}_{-0.029}$ & $1.474^{+0.068}_{-0.066}$ & $1.15^{+0.55}_{-0.49}$\\
233155595 & $3.63^{+0.22}_{-0.20}$ & $5423^{+41}_{-39}$ & $2.157^{+0.054}_{-0.053}$ & $3.813^{+0.033}_{-0.032}$ & $1.105^{+0.066}_{-0.063}$ & $6.8^{+1.4}_{-1.0}$\\
\hline
\end{tabular}
\end{center}
\tablecomments{(The full table is available online in machine-readable format.) We report values for luminosity, temperature, radius, surface gravity, mass, and stellar age. The uncertainties listed in this table only include the contribution from random errors. While not explicitly included in Table \ref{tab:table_exofast_derived_parameters}, readers should be aware of systematic uncertainties, which we estimate to be 2\% for luminosity, 60 K for temperature, and 2\% for radius (see \S\ref{subsec:methodresults_errors}). We highlight that luminosity, temperature, and radius are mostly model-independent, while surface gravity, mass, and age rely more heavily on stellar models. We note that our mass and age values have not been compared with or calibrated to independent estimates from other methods (e.g., asteroseismology), and the reader should be mindful of potential inaccuracies.}
\end{minipage}
\end{table*}

\begin{figure*}
\gridline{
	    \fig{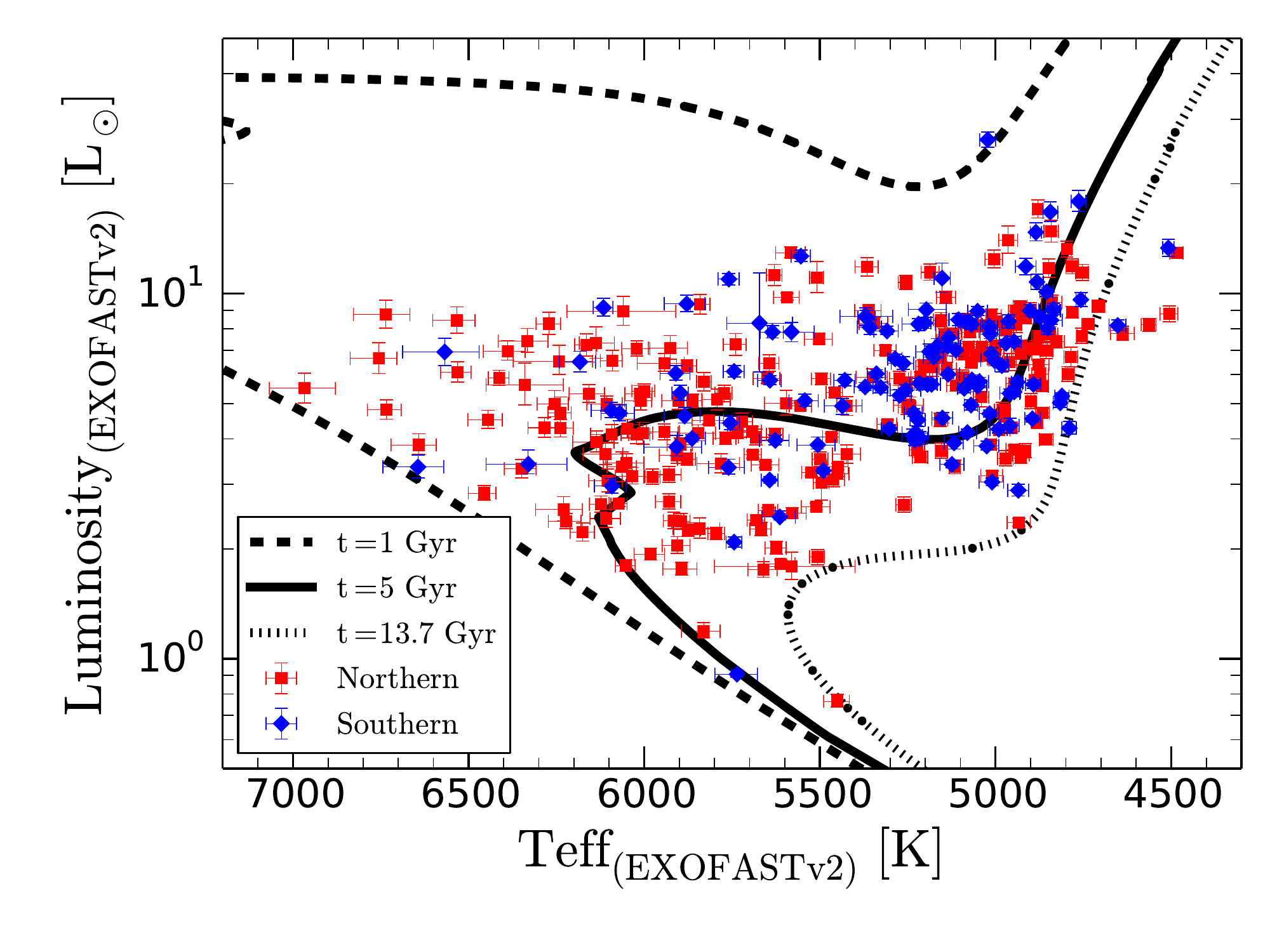}{0.50\textwidth}{(a)}
	    \fig{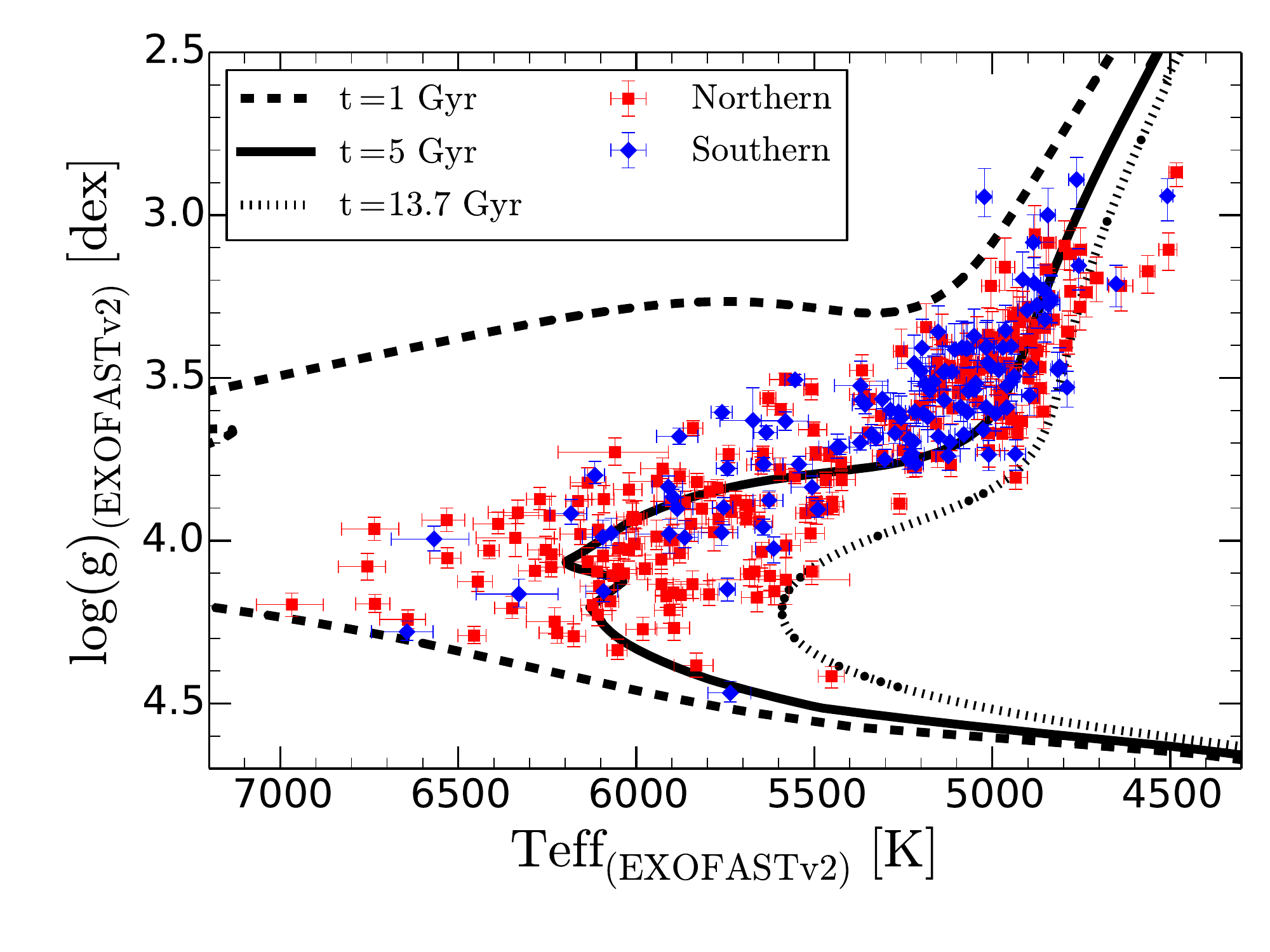}{0.50\textwidth}{(b)}
          }
\caption{Summary of luminosity, temperature, and surface gravity values derived using {\exofast} (via SED fitting) for the full sample. \textbf{(a)} Luminosity vs {\Teff} (HR diagram). \textbf{(b)} $\log(g)$ versus {\Teff} (Kiel diagram). The color and symbol scheme is the same as Figure \ref{fig:data_sample_astrometry_photometry}, and we show the same set of representative MIST models from \S\ref{sec:data}. The error bars correspond to the random errors reported by {\exofast}, and their mean values are $\sigma_{L}/L \approx 4.5\%$, $\sigma_{T_{\text{eff}}} \approx 33$ K, and $\sigma_{\log(g)} \approx 0.049$ dex. Our sample is distributed along the subgiant branch, spanning a range of ages, luminosities, temperatures, and surface gravities, and it therefore holds high scientific interest for a number of open astrophysical questions.}
\label{fig:exofast_results_LTlogg}
\end{figure*}

We now perform SED fitting on the full sample of subgiants using the global per-hemisphere extinctions derived in \S\ref{subsec:methodresults_extinction}. We highlight that for this we only feed {\exofast} the photometric and astrometric data (available for $\approx$ 98\% of the sample), but do not provide it the spectroscopic parameters (only available for $\approx$ 48\% of the sample). This allows us to perform a uniform, unbiased calculation that is valid for the entire sample. We make use of the spectroscopic parameters to assess the reliability of our results in \S\ref{subsec:methodresults_errors}. We provide further details of this calculation in Appendix \ref{appendix:app_sed_fitting}.

We run {\exofast} using these inputs and results are returned for 340/347 stars ($\approx$ 98\% of the sample; the remaining 7 stars lack the required data to perform the fit). We report our derived stellar parameters in Table \ref{tab:table_exofast_derived_parameters}. Figure \ref{fig:sed_example} shows the SED fit of a representative star as an example, comparing the input photometry (red crosses) with the resulting best-fit model (blue circles). Figure \ref{fig:sed_example} demonstrates how crucial the {\gaia} photometry is in our data set, as it allows us to sample the peak of the SED. 

For every star that we fit, {\exofast} produces a large number of output parameters. Of these, Table \ref{tab:table_exofast_derived_parameters} only reports the six that are more relevant in the context of our work, namely luminosity, temperature, radius, surface gravity, mass, and stellar age (but see Appendix \ref{appendix:app_sed_fitting} for further details). These parameters have different degrees of dependence on the adopted stellar models. Luminosity, temperature, and radius are only slightly model-dependent, as they are constrained by the shape of the SED, the flux normalization (set by the parallax), and bolometric corrections. Of these, only the bolometric corrections are model-dependent, and their dependence is small (e.g., \citealt{zinn19c}). Mass and age, on the other hand, have typically higher associated uncertainties as they rely more heavily on models and their underlying assumptions (see \S\ref{subsec:methodresults_exofast}). Surface gravity does have a strongly model-dependent component (mass), but given its logarithmic scale and the relatively narrow range of possible subgiant masses, it is not as model-dependent as stellar age.

In Figure \ref{fig:exofast_results_LTlogg} we show the resulting HR and Kiel diagrams for the full sample. While topologically similar to the previously shown CMDs, Figure \ref{fig:exofast_results_LTlogg} shows readily interpretable physical properties that can be directly compared with families of stellar models. Expanding on the spectroscopic diagram of Figure \ref{fig:data_sample_spectroscopy} (which only contained $\approx 48\%$ of the sample), Figure \ref{fig:exofast_results_LTlogg} populates the Kiel diagram for the full sample, virtually doubling the number of stars with $\log(g)$ values. Our subgiants span a wide range of ages, luminosities, temperatures, and surface gravities, making them ideal targets to provide useful constraints on a number of astrophysical problems (see \S\ref{sec:intro}).

\begin{figure}[h]
\epsscale{1.3}  
\plotone{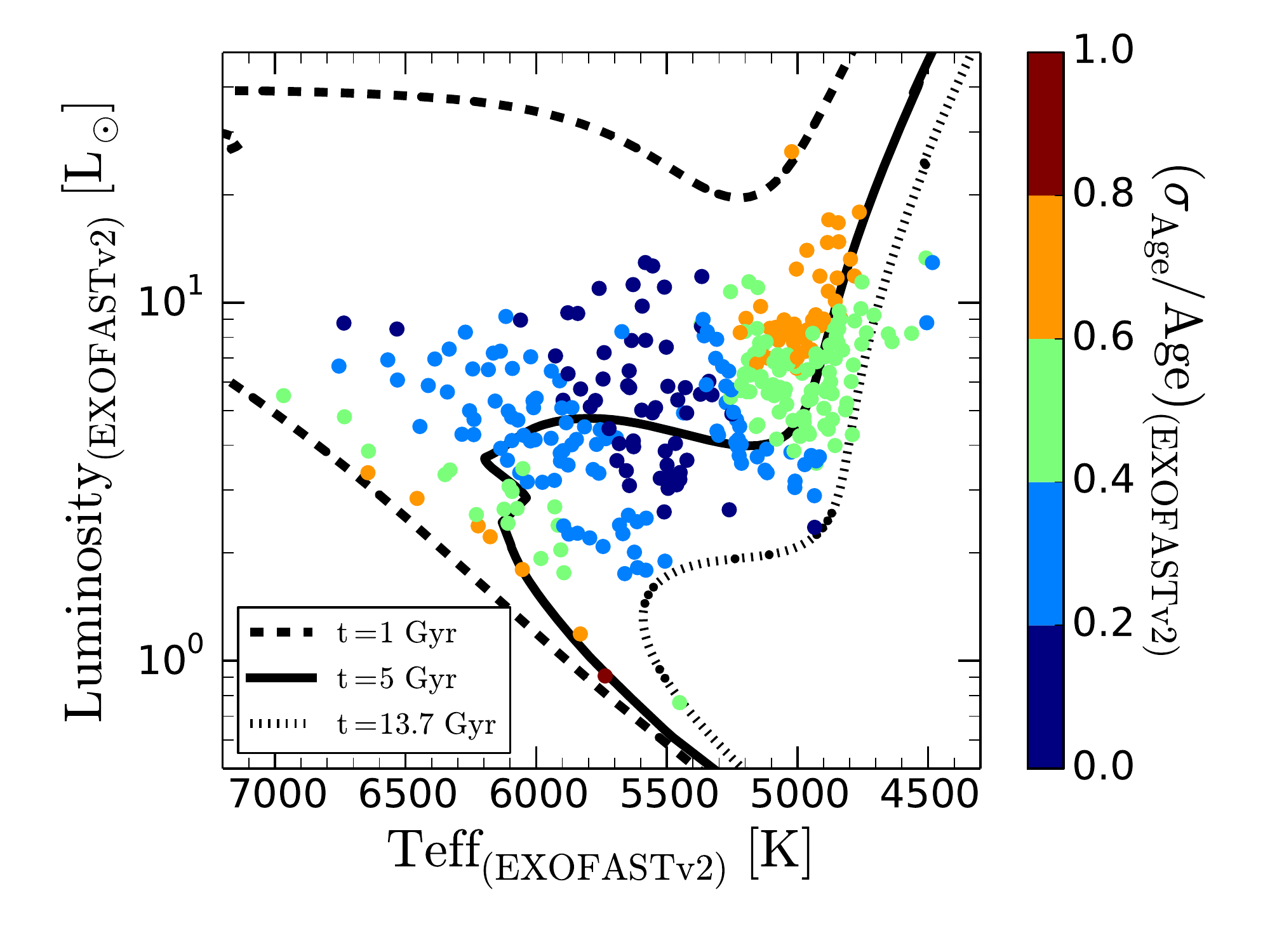}
\caption{HR diagram for the full subgiant sample, color-coded by fractional age uncertainty. The color bar values only account for the random errors. The precision on the derived ages is a strong function of HR diagram location, which illustrates that precise ages can be obtained for subgiant stars. The fractional age uncertainty reaches a minimum in the middle of the subgiant branch, where the models are the most separated from each other, and increases towards both the main sequence and the RGB.}
\label{fig:exofast_results_HRccfracage}
\end{figure}

Figure \ref{fig:exofast_results_HRccfracage} illustrates the prospects of our classical subgiant characterization by showing the HR diagram with the stars color-coded by their fractional age uncertainty. If fractional mass uncertainty is used for the color-coding, the figure is virtually identical, modulo a scaling factor of $\approx 1/3$ in the uncertainties. For subgiant stars, accurate and precise temperatures and luminosities are sufficient to estimate masses and ages. The precision of these is a strong function of position on the HR diagram, with the middle of the subgiant branch yielding the most precise results. The fractional age and mass uncertainties increase substantially for stars located closer to the main sequence or the RGB, where isochrones with considerably different ages are close together. We further discuss the role played by the knowledge of the targets' metallicity (or lack thereof) in the mass and age values in \S\ref{subsubsec:methodresults_errors_cautionmassage}.

\subsection{Error Analysis}
\label{subsec:methodresults_errors}

\begin{figure*}
\gridline{
	    \fig{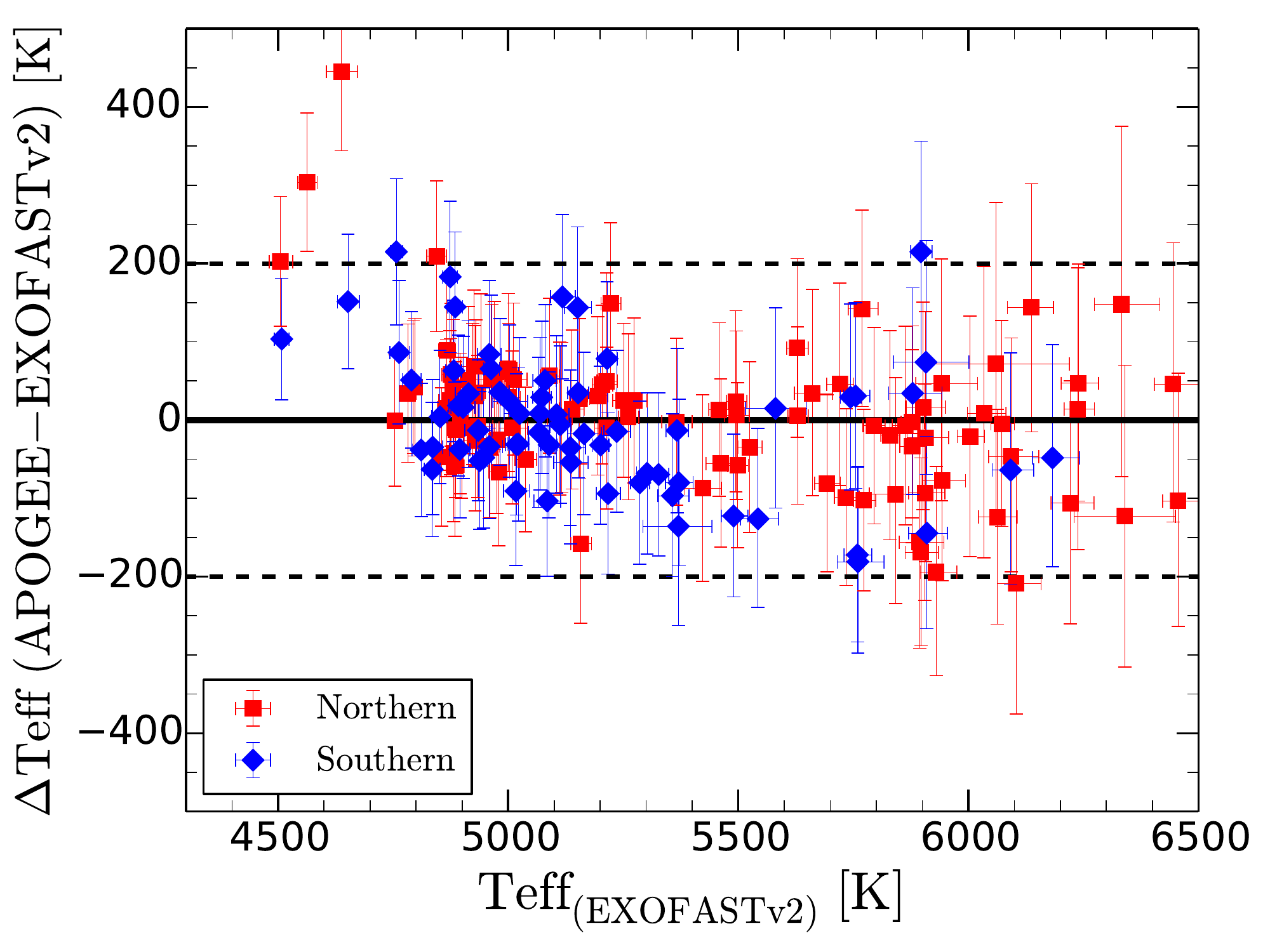}{0.45\textwidth}{(a)}
	    \fig{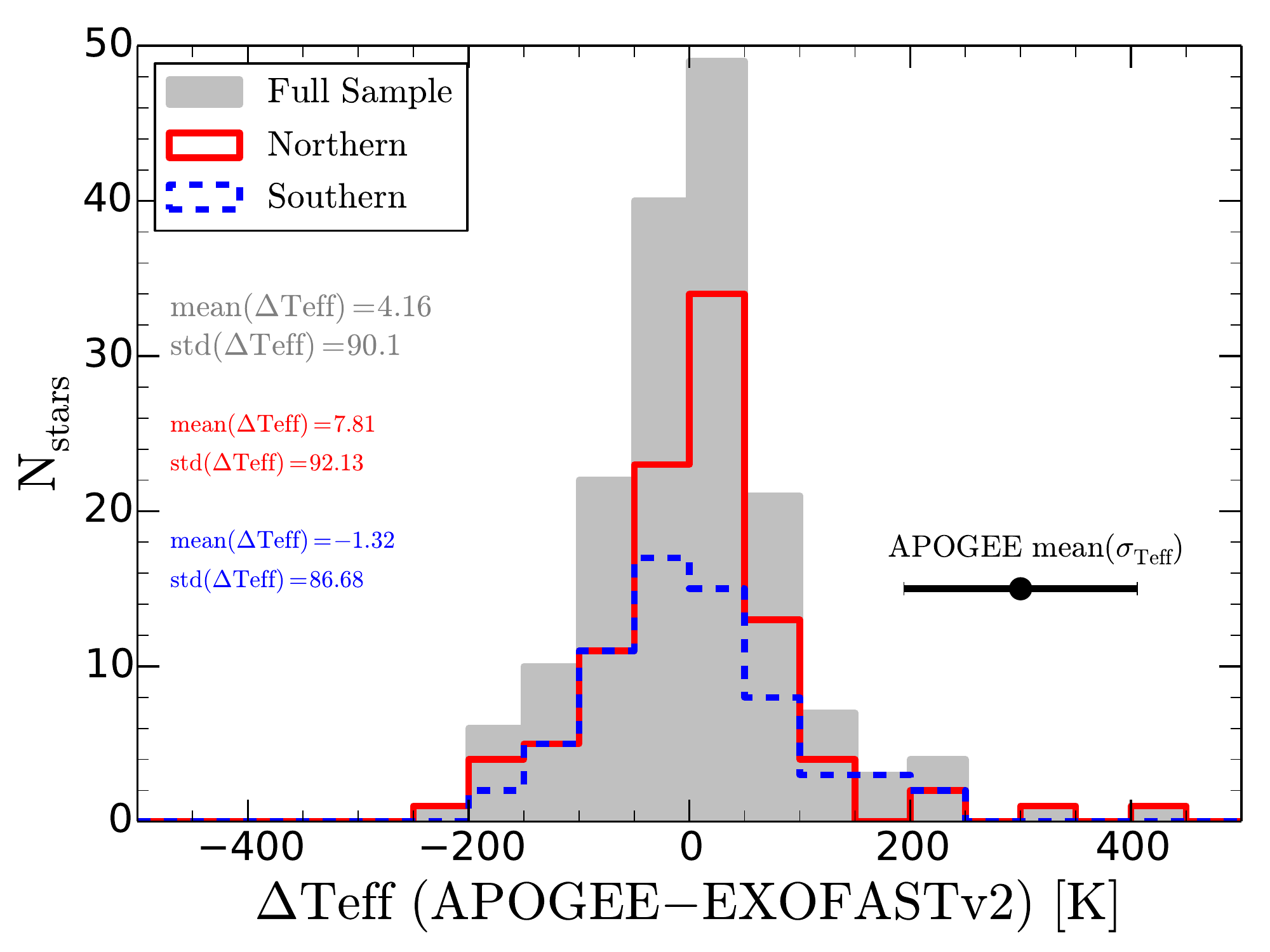}{0.45\textwidth}{(b)}
          }
\gridline{
	    \fig{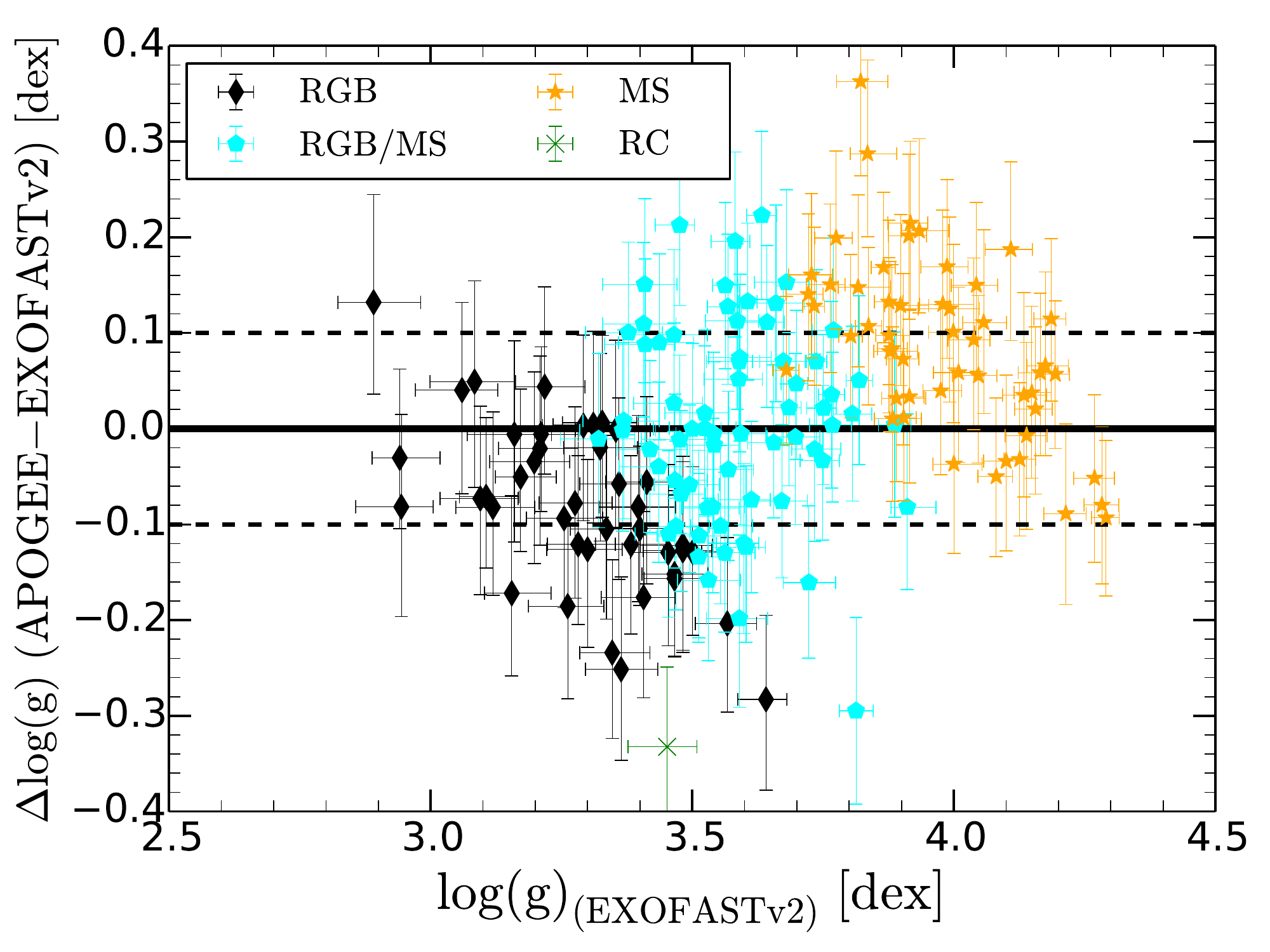}{0.45\textwidth}{(c)}
	    \fig{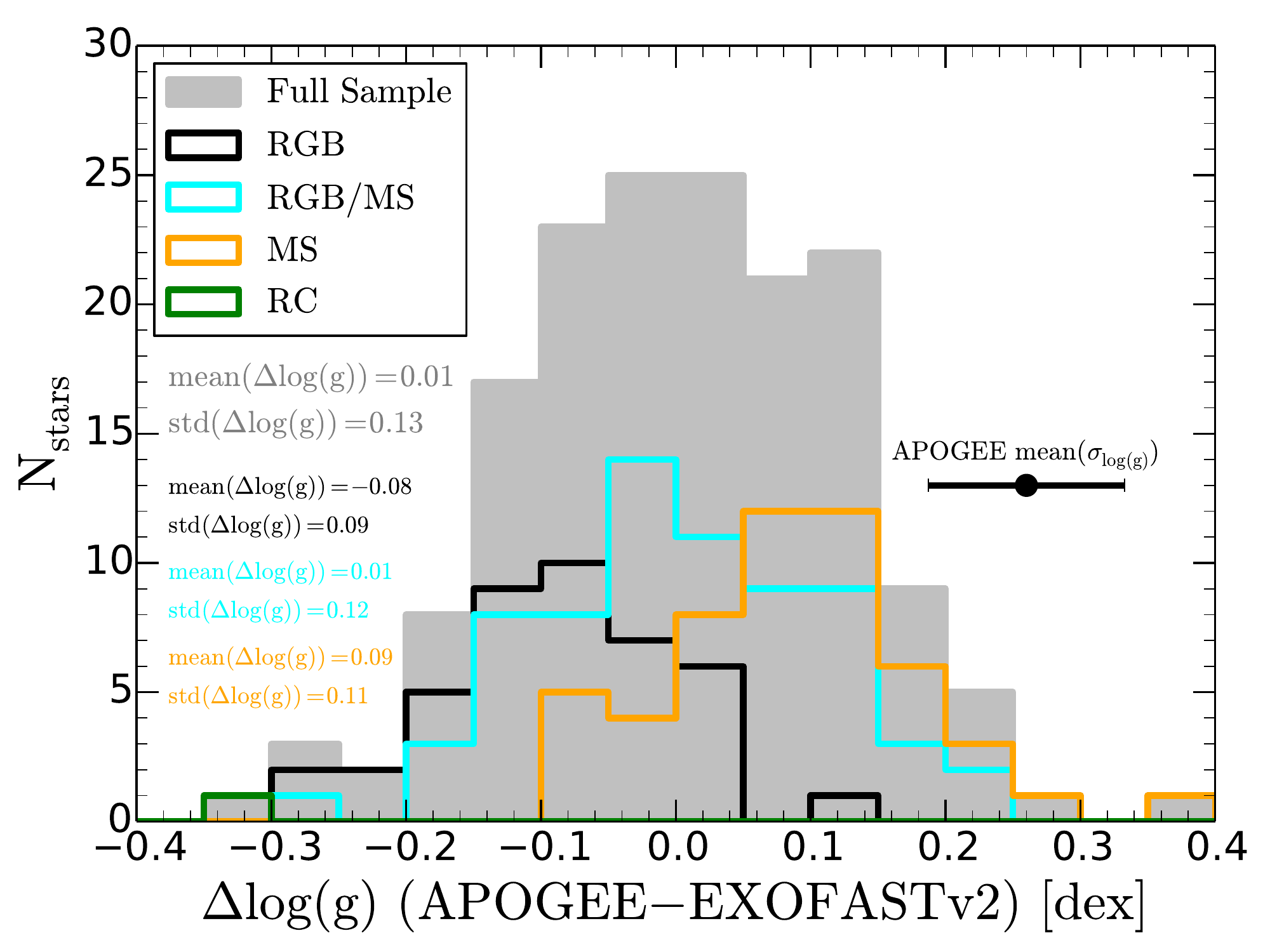}{0.45\textwidth}{(d)}
          }
\caption{Comparison of our derived temperatures (top) and surface gravities (bottom) with those from the APOGEE spectroscopic survey, for the subset of stars in common. \textbf{(a)} Temperature comparison, with the color and symbol scheme being the same as Figure \ref{fig:data_sample_astrometry_photometry}. The solid and dashed lines show the 1:1 and $\pm 200$ K relations for reference. \textbf{(b)} Distribution of $\Delta T_{\text{eff}}$. The mean APOGEE error is shown as the black circle for reference. There is good agreement between both temperatures and no strong {\Teff}-dependent offsets are seen. Panels \textbf{(c)} and \textbf{(d)} are analogs to panels \textbf{(a)} and \textbf{(b)} for the surface gravities (with the dashed line showing $\pm 0.1$ dex differences), but in this case the color and symbol scheme indicate the gravity calibration employed by APOGEE, where stars are classified as main sequence (orange stars), RGB (black diamonds), intermediate RGB/main sequence (cyan pentagons), or red clump (green cross). While there is a good global agreement, there are clear $\log(g)$-dependent differences within each of the different categories, as well as among them. We do not attempt to correct these differences in our gravities due to the heterogeneous APOGEE $\log(g)$ calibrations (see \S\ref{subsubsec:methodresults_errors_systematics}).}
\label{fig:exofast_vs_spectroscopy}
\end{figure*}

\begin{figure*}
\gridline{
	    \fig{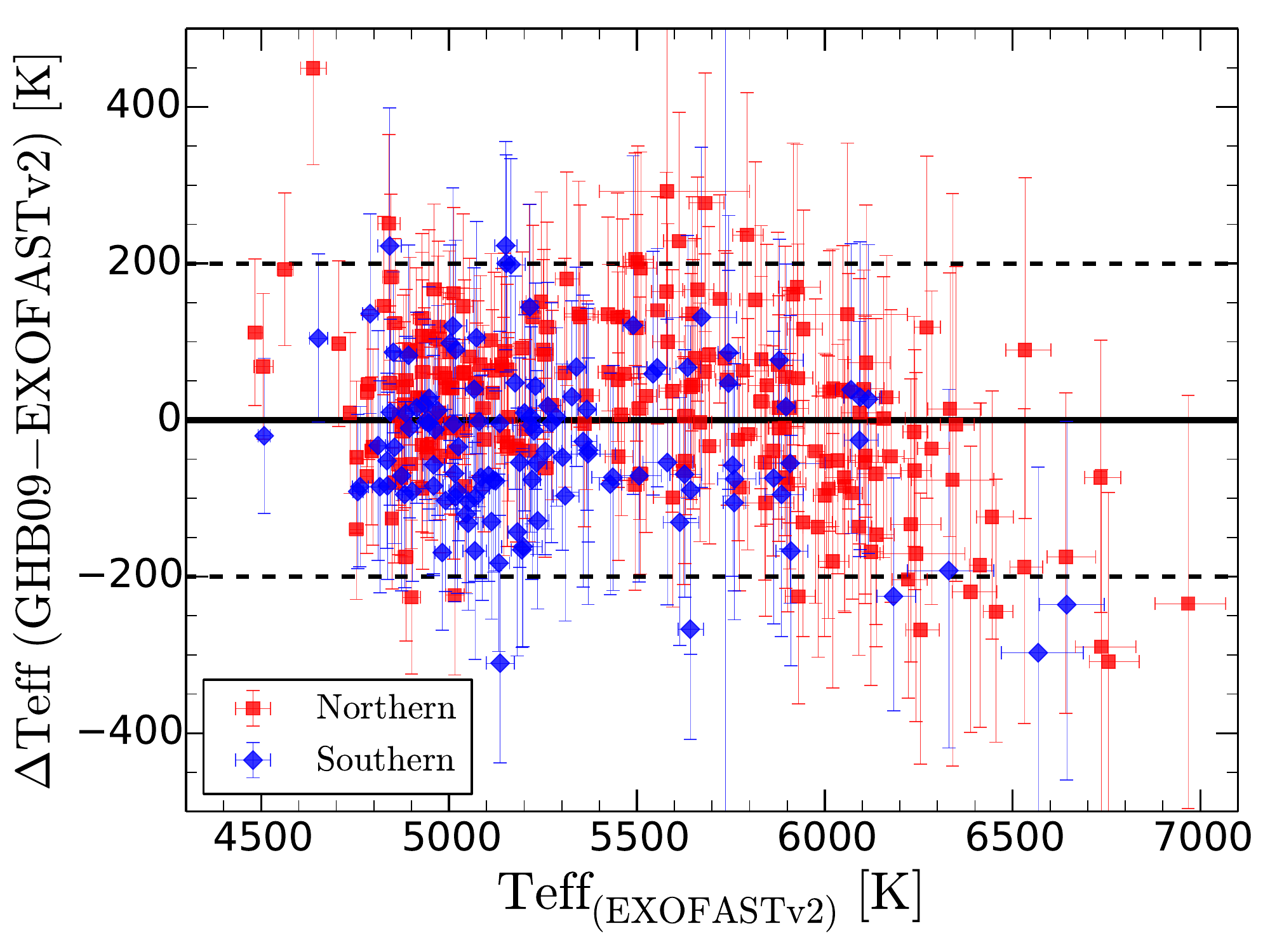}{0.45\textwidth}{(a)}
	    \fig{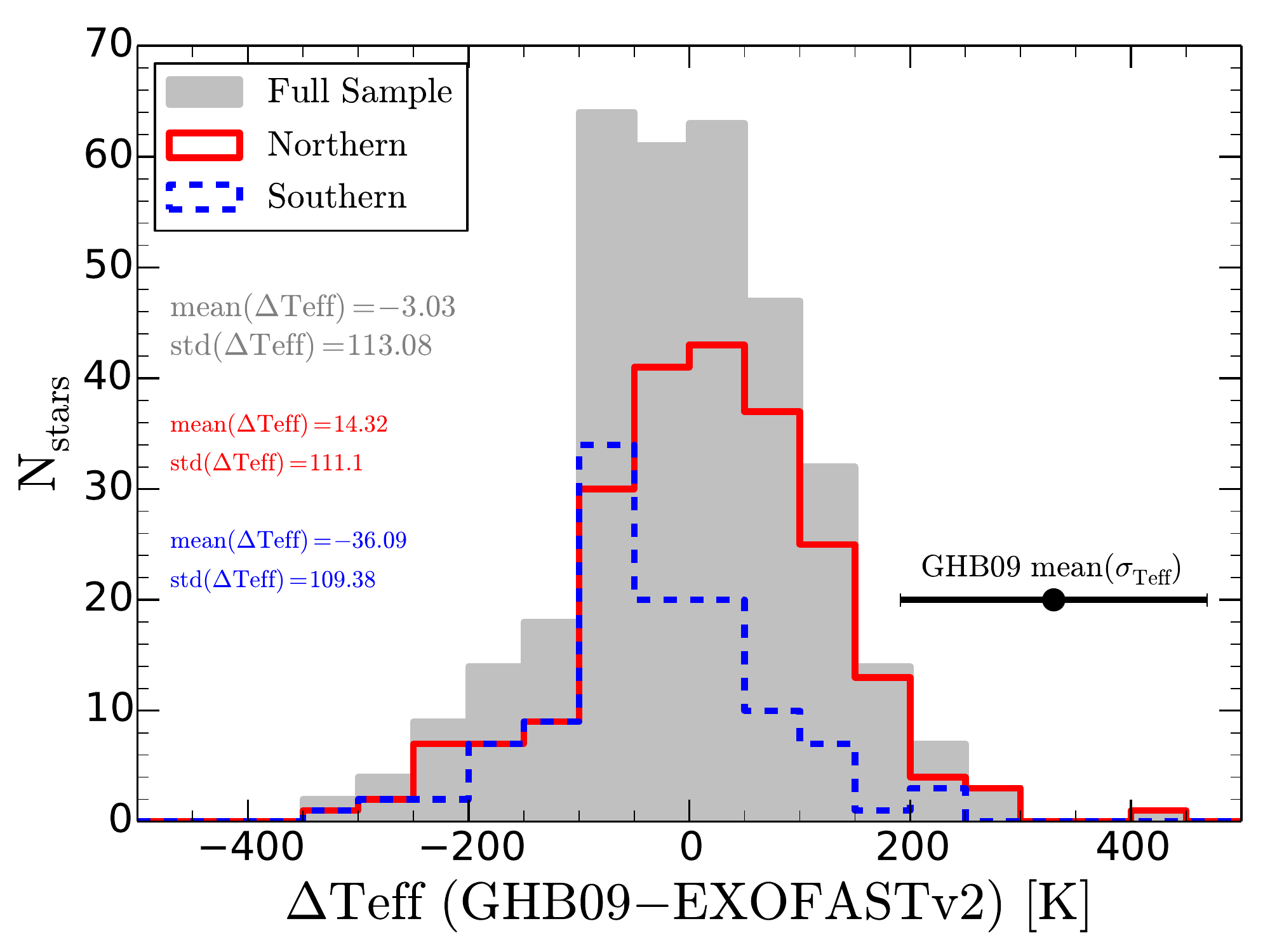}{0.45\textwidth}{(b)}
          }
\gridline{
	    \fig{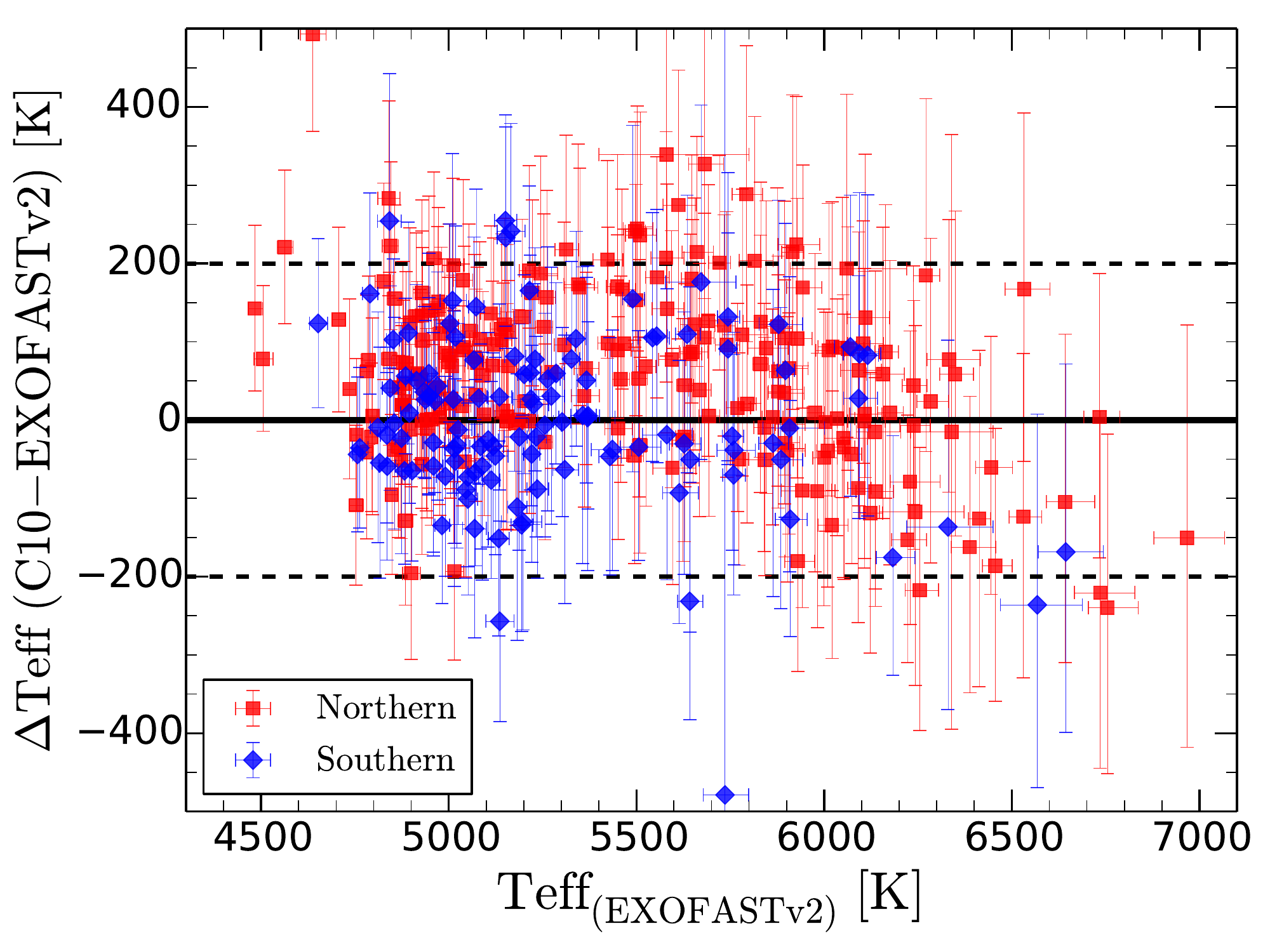}{0.45\textwidth}{(c)}
	    \fig{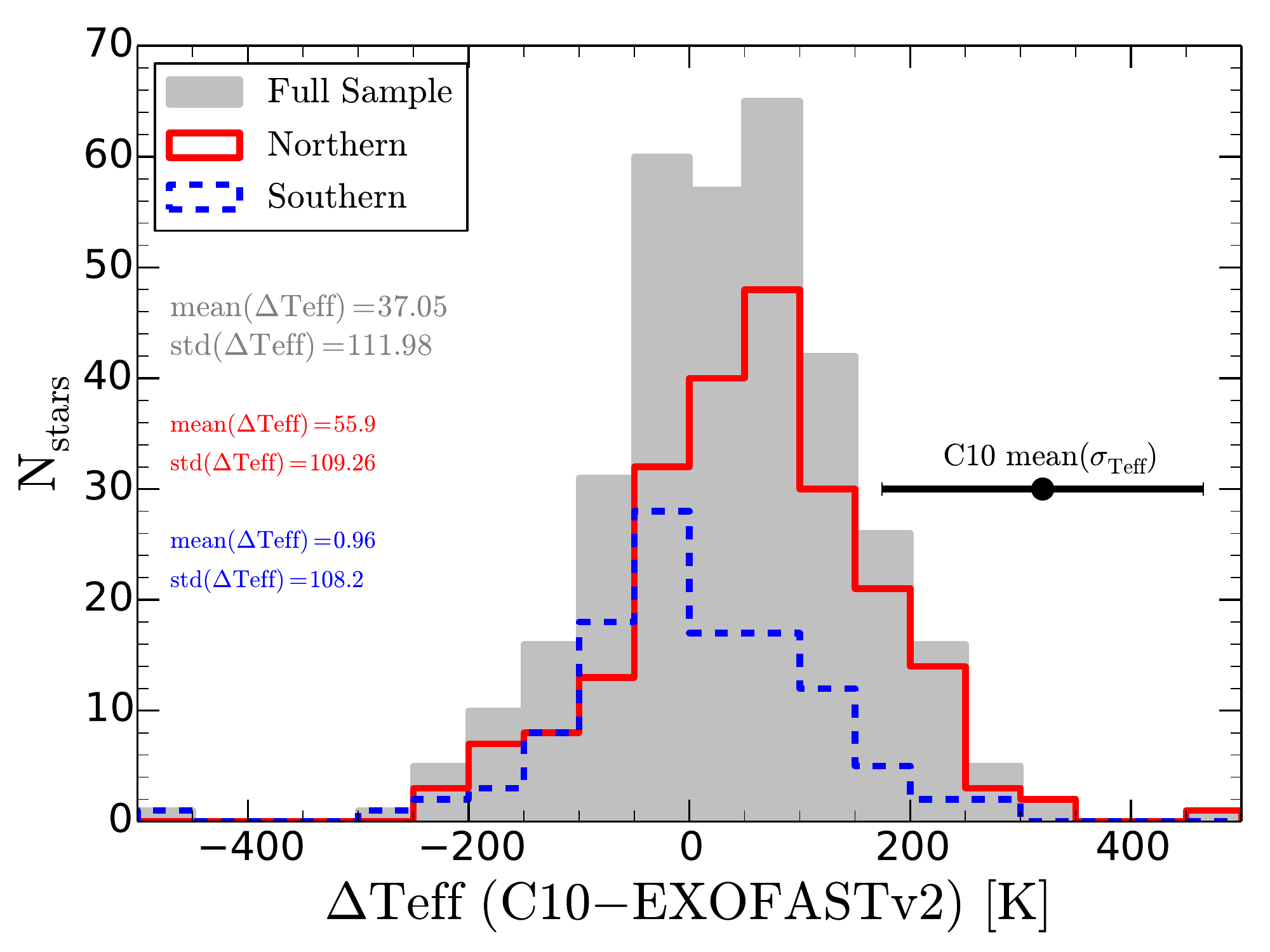}{0.45\textwidth}{(d)}
          }
\caption{Comparison of our derived temperatures with different temperature scales, analogous to Figure \ref{fig:exofast_vs_spectroscopy}. We compare our temperatures with those obtained using the dwarf ($J-K$) color-{\Teff} relation of \citet{gonzalezhernandez09} (top) and \citet{casagrande10} (bottom). These comparisons further confirm that our derived temperatures are approximately on the APOGEE scale (which in turn is by construction on the \citet{gonzalezhernandez09} scale), but also show clear {\Teff}-dependent offsets (see \S\ref{subsubsec:methodresults_errors_systematics}).}
\label{fig:applications_temperature_scale_comparison}
\end{figure*}
\subsubsection{Random Errors}
\label{subsubsec:methodresults_errors_random}

In Table \ref{tab:table_exofast_derived_parameters} (and Figures \ref{fig:exofast_results_LTlogg} and \ref{fig:exofast_results_HRccfracage}) we report the random errors from the SED fitting calculation. These values are a direct output from {\exofast}'s Markov Chain Monte Carlo (MCMC) method \citep{eastman19}, and they are dominated by the measurement errors in the photometry and astrometry. They do not consider the systematic errors we discuss next. The mean values of the random errors are: 4.5\% for luminosity, 33 K for temperature, 2.2\% for radius, 0.049 dex for surface gravity, 10\% for mass, and 40\% for age. We find the mean random errors in luminosity, temperature, and radius to be sensible representations for the full sample. For mass and age, as noted in \S\ref{subsec:methodresults_results}, the random errors are a strong function of HR diagram location (see Figure \ref{fig:exofast_results_HRccfracage}).
\subsubsection{Systematic Errors}
\label{subsubsec:methodresults_errors_systematics}

We first discuss the systematic errors in our temperature values. As explained in \S\ref{subsec:methodresults_extinction}, we have obtained our global per-hemisphere extinctions by using the APOGEE spectroscopic temperatures. This implies that, by construction, the average temperatures we have derived for the full sample are approximately on the APOGEE temperature scale. We evaluate this comparison in the top panels of Figure \ref{fig:exofast_vs_spectroscopy}, finding that in a global sense our derived temperatures are extremely close to the APOGEE values. The mean temperature difference is $\approx$ 4 K, and although the scatter increases for {\Teff}$> 5800$ K, no strong {\Teff}-dependent offsets are seen.

It is important to note that APOGEE itself is on the \citet{gonzalezhernandez09} scale \citep{holtzman18,jonsson20}, and we would therefore expect a consistent answer between our temperatures and those obtained using their relations. For context, \citet{gonzalezhernandez09} performed an implementation of the infrared flux method (IRFM) using the 2MASS filters by comparing fluxes computed from models with observed magnitudes. We show the comparison between our temperatures and those calculated using their dwarf ($J-K$)-{\Teff} relation (using the $A_V$ values from \S\ref{subsec:methodresults_extinction} and spectroscopic metallicities if available or assuming solar if otherwise) in the top panels of Figure \ref{fig:applications_temperature_scale_comparison}. Although we find a good global agreement, with a mean offset of $\approx$ -3 K considering the full sample, some clear {\Teff}-dependent offsets are seen. The $\Delta T_{\text{eff}}$ values are centered around $\sim$ 0 K at 5000 K (and mostly contained within $\pm$100 K differences), increase to $\sim +70$ K around 5500 K, and systematically decrease until reaching $\sim -200$ at 6500 K. For {\Teff}$<$ 6000 K, these differences hint to potential systematic effects in our temperature values that we discuss in more detail below. For {\Teff}$>$ 6000 K, the differences are perhaps unsurprising given the known offsets between temperature scales in hot stars \citep{johnson17,pinsonneault12}, which may be related to the Kraft break between slowly rotating cooler stars and rapidly rotating hotter ones \citep{kraft67b}.

Another point of comparison for our temperatures is the scale of \citet{casagrande10}. Somewhat similarly to \citet{gonzalezhernandez09}, \citet{casagrande10} provided an absolute temperature scale by using the IRFM on a sample of solar twins. We show the comparison between our temperatures with those calculated from their ($J-K$)-{\Teff} relation in the bottom panels of Figure \ref{fig:applications_temperature_scale_comparison}. The results from this comparison are extremely similar to those obtained by comparing with \citet{gonzalezhernandez09}, but with a mean offset of $\approx$ 37 K for the full sample. The same $\Delta T_{\text{eff}}$ trends as a function of temperature are seen, but with the differences being larger by $\sim +40$ K. All of this is in excellent agreement with the analysis by \citet{casagrande10}, who found that due to differences on the absolute calibration and zero points adopted for Vega, their scale and that of \citet{gonzalezhernandez09} differ at the 30$\textendash$40 K level (with the \citet{casagrande10} temperatures being hotter).

Ultimately, all of the temperature systematics seen above are subject to an error floor that is set by uncertainties in the accuracy of the temperature scales themselves \citep{casagrande14}. These are typically examined using stars with precise angular diameter measurements \citep{casagrande10,gonzalezhernandez09}. Comparisons of interferometric angular diameters for the same stars using different instruments show a typical scatter of $\sim$ 4\%, which translates into a $\sim$ 2\% temperature uncertainty \citep{tayar20}. For near-solar {\Teff} stars this corresponds to $\sim$ 120 K, which globally dominates over the previously described $\Delta T_{\text{eff}}$ trends, and we adopt it as a representative 2$\sigma$ systematic uncertainty (or a 1$\sigma$ uncertainty of 60 K).

For surface gravity, the comparison with the spectroscopic values is more nuanced. As detailed in \citet{jonsson20}, the APOGEE $\log(g)$ scale itself is not a homogeneous one, but rather a combination of a number of calibrations on different parts of the HR diagram (e.g., they differ for dwarfs, giants, and stars in between). Given their intermediate evolutionary stage, our subgiant stars fall across these different categories. Of the 168 stars with spectroscopic data, according to their $\log(g)$ calibration, APOGEE classifies 55 of them as main sequence, 42 as RGB, 70 in the intermediate RGB/main sequence category, and 1 as a red clump (RC) star. The bottom panels of Figure \ref{fig:exofast_vs_spectroscopy} show the comparison of our surface gravity values with those from APOGEE. In a global sense, there is good agreement between both samples, but with strong $\log(g)$-dependent offsets (see panel \textbf{(c)}). Color-coding the stars by their APOGEE $\log(g)$ calibration shows clear trends among the different categories, as well as within each one of them.

While in principle we could calibrate our gravities to force a better agreement, we opt not to do this due to potential inaccuracies that could be present in the heterogeneous APOGEE calibrations. Instead, we leave this as future work to be done when external and homogeneous comparison samples (e.g., asteroseismic gravities from {\tess} light curves) are available, and the APOGEE gravity scale can be uniformly tested. Thus, for our derived surface gravities, we assign a systematic error at the $\approx$ 0.01 dex level with a standard deviation of $\approx$ 0.13 dex. 

For luminosity and radius systematics, we follow the approach of \citet{zinn19c}. They explored the effects of different temperature scales and choices of bolometric corrections. Given the similar techniques used, we expect their 1$\sigma$ values (of order $\sim$ 2\% for both luminosity and radius) to be representative of our derived properties, and we adopt them.

\begin{figure}[h]
\epsscale{1.2}  
\plotone{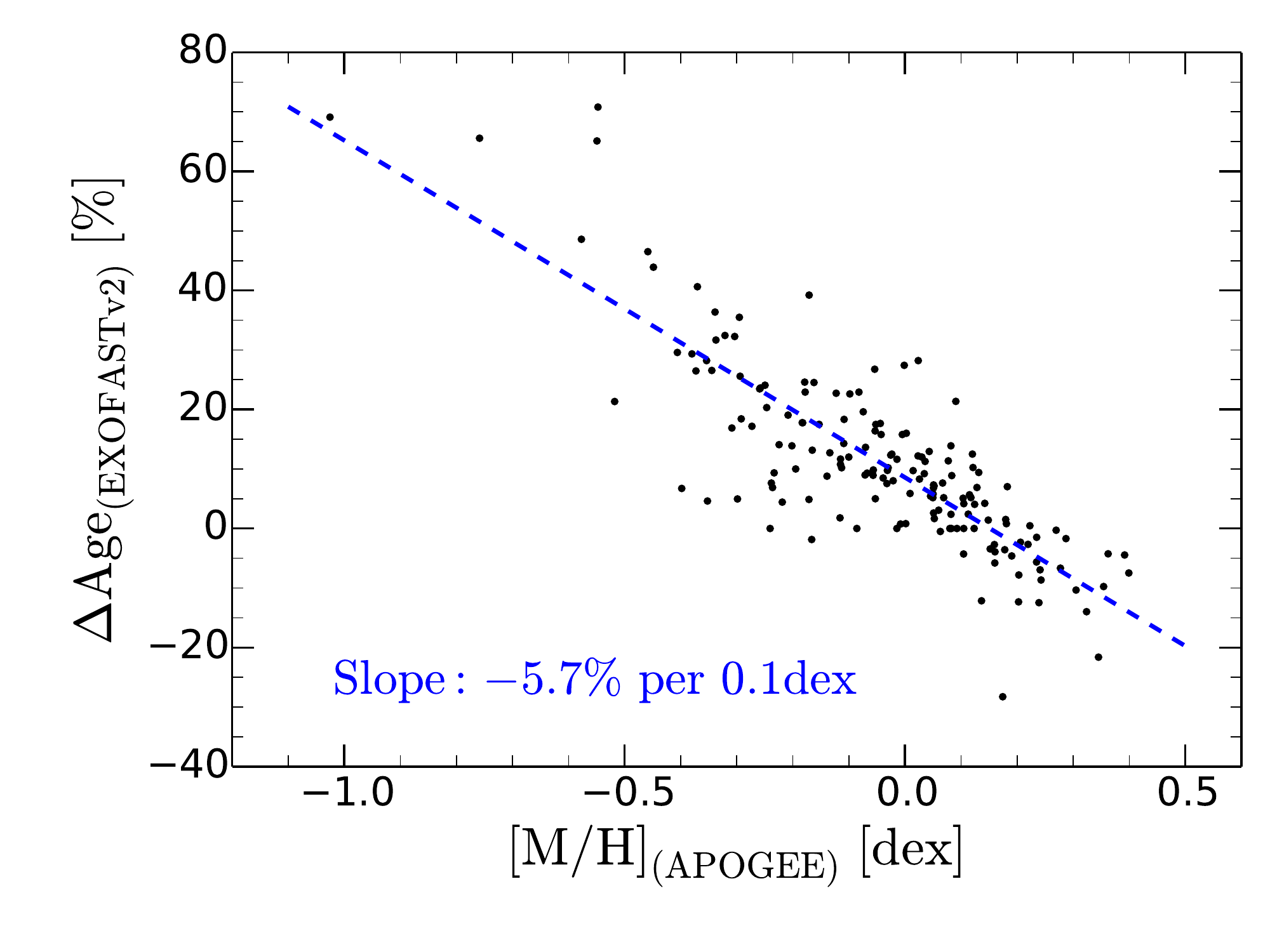}
\caption{Percentage change in the age values obtained by {\exofast} as a function of metallicity. Each data point corresponds to one of our subgiants with APOGEE data. The $y$-axis values are calculated as: $(\text{Age}_{\text{with [M/H]}} - \text{Age}_{\text{without [M/H]}})/\text{Age}_{\text{without [M/H]}}$, where $\text{Age}_{\text{with [M/H]}}$ are the values obtained using the APOGEE metallicities in the fit, and $\text{Age}_{\text{without [M/H]}}$ are the values obtained assuming a solar metallicity prior (see Appendix \ref{appendix:app_sed_fitting}). A linear fit to the data returns a slope of $-5.7\%$ change in age per $+0.1$ dex in metallicity, and the analogous figure for mass returns a slope of $+2.5\%$ per $+0.1$ dex in metallicity.}
\label{fig:exofast_results_age_vs_MoH}
\end{figure}
\subsubsection{A Cautionary Word on Masses and Ages}
\label{subsubsec:methodresults_errors_cautionmassage}

We note that, while derived together with the rest of the parameters, the masses and ages listed in Table \ref{tab:table_exofast_derived_parameters} only represent first order estimates obtained with the tools in hand. As previously discussed, these parameters rely more heavily on stellar models, and the values reported here have been calculated only using the default MIST models employed by {\exofast}. They have not been vetted or calibrated against estimates from independent methods, or from a different suite of models. 

Some of the model assumptions that could impact the derived masses and ages include the helium fraction, mixing-length parameter, and metallicity (e.g., \citealt{valcarce12}). Specifically for subgiants, \citet{li20b} studied the dependences of asteroseismic properties on these parameters, and demonstrated that metallicity is the predominant factor, with changes of $\Delta\text{[M/H]}=+0.1$ dex inducing a change of $+4.3\%$ in mass and $-6.5\%$ in age. A quick test using the APOGEE metallicities in our fitting (when available) produces mass and age trends that go in the same sense and are of similar magnitude as those obtained by \citet{li20b}, with $\Delta\text{[M/H]}=+0.1$ dex inducing a change of $+2.5\%$ in mass and $-5.7\%$ in age. We illustrate the latter of these in Figure \ref{fig:exofast_results_age_vs_MoH}. Given the metallicity distribution of our sample (centered at solar with $\sigma_{\text{[M/H]}}=0.23$ dex), an unknown metallicity with a true $1\sigma$ discrepancy would add an error of $\approx 6\%$ in mass and $\approx 13\%$ in age (in quadrature) to the error budget, which are still secondary for the vast majority of our sample given the random errors shown in Figure \ref{fig:exofast_results_HRccfracage}.

Presumably, we could compare our masses with independent estimates obtained by combining the radii derived from the SED fitting (which are mostly model independent), with the spectroscopic $\log(g)$ values. This, however, requires extremely accurate surface gravities, something that is lacking in the current APOGEE data for this comparison to be meaningful on a star-by-star basis (see \S\ref{subsubsec:methodresults_errors_systematics}; see also \citealt{torres12} for a complementary example). Therefore, although precise on some parts of the HR diagram (see Figure \ref{fig:exofast_results_HRccfracage}), the accuracy of our masses and ages has not been tested, and they should be used with caution. We leave a thorough calibration and systematic error analysis for these parameters as future work to be done with more specialized tools, and when asteroseismic results from {\tess} data become available for our sample.
\section{Applications} 
\label{sec:applications}

A large number of astrophysical investigations depend on reliable stellar temperatures, radii, and luminosities. In general, there are advantages in using stellar properties derived from large and homogeneous catalogs. However, these catalogs are constructed with less information than we have available for our targets, and few of them have been carefully validated in the subgiant regime. Furthermore, they are designed for the study of more distant objects where the precise treatment of extinction can have a major impact on the derived results. A comparison of the stellar parameters from these catalogs with our results is therefore a useful test of the leverage that can be gained for smaller samples with more specialized studies such as ours.
\subsection{Comparison with {\gaia} DR2}
\label{subsec:applications_comparison_with_catalogs_gaia}

The derivation of the stellar parameters reported in {\gaia} DR2 is described in \citet{andrae18}. In short, effective temperatures were derived from the {\gaia} color information, luminosities from the observed photometry, measured parallax, and bolometric corrections, and radii were obtained from these using the Stefan-Boltzmann law. Given their wide availability, they are an outstanding resource that is being widely used by the community (e.g., \citealt{hattori18,huang18,moschella21,ramsay20,traven20}). Nevertheless, it is important to keep in mind some considerations regarding their inference, as they assume zero extinction, do not consider the parallax zero point (e.g., \citealt{lindegren18,schonrich19,stassun18a,zinn19a,chan20,riess18}), and only use information from the 0.3\textendash1 $\mu$m wavelength range covered by the $G$, $G_{\text{BP}}$, and $G_{\text{RP}}$ magnitudes.

\begin{figure*}[ht!]
\gridline{
	    \fig{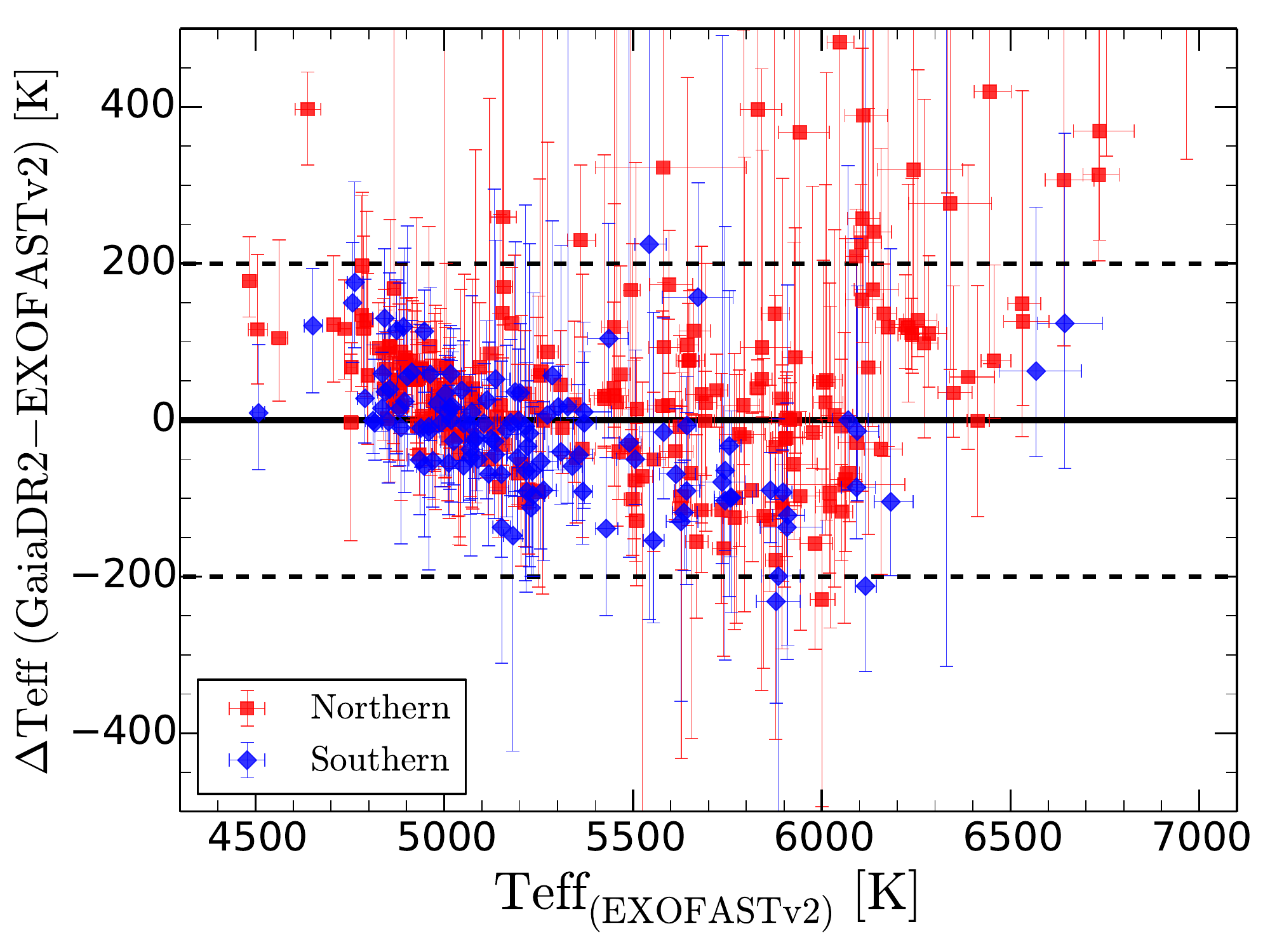}{0.40\textwidth}{(a)}
	    \fig{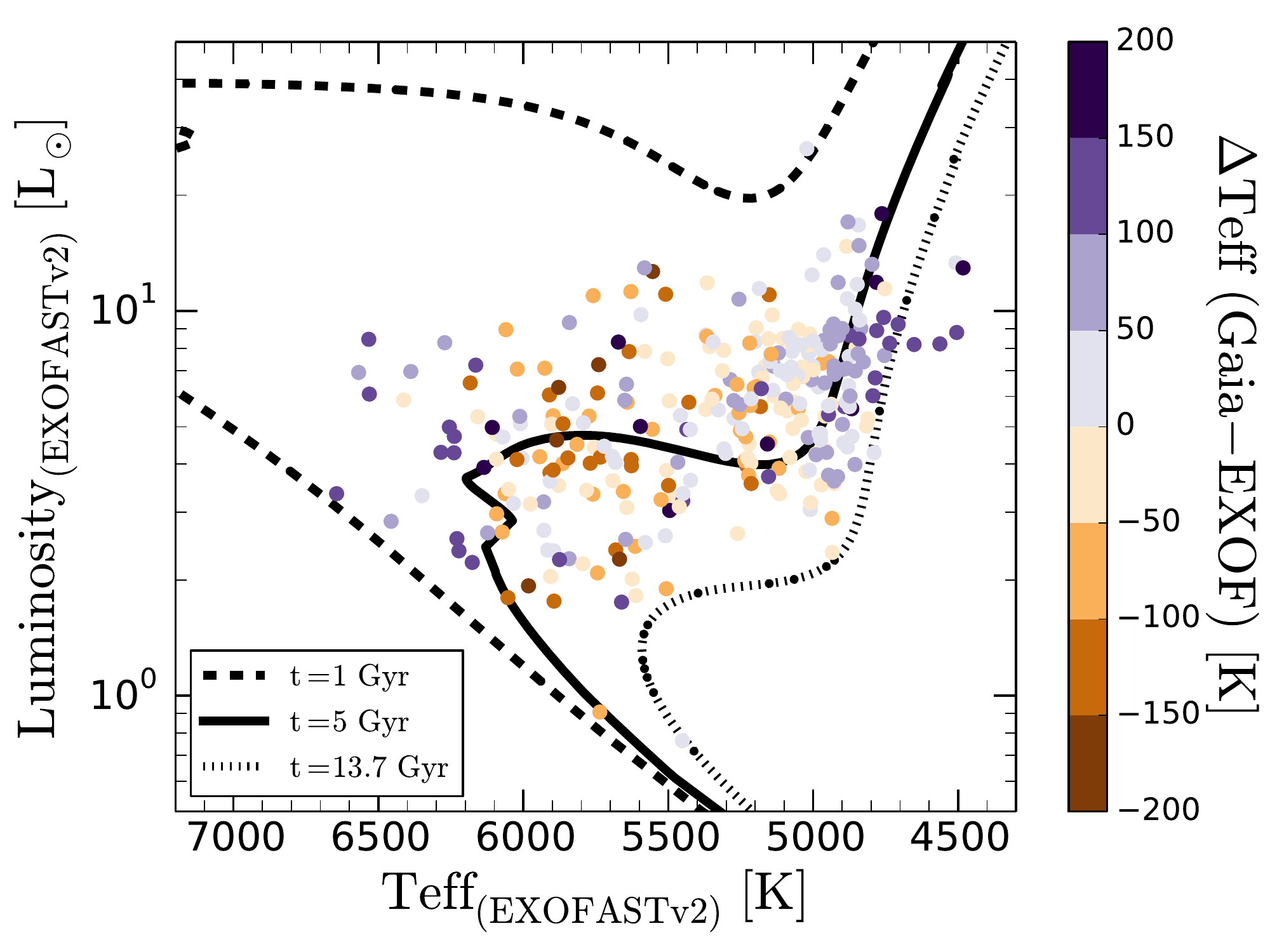}{0.40\textwidth}{(b)}
          }
\gridline{
	    \fig{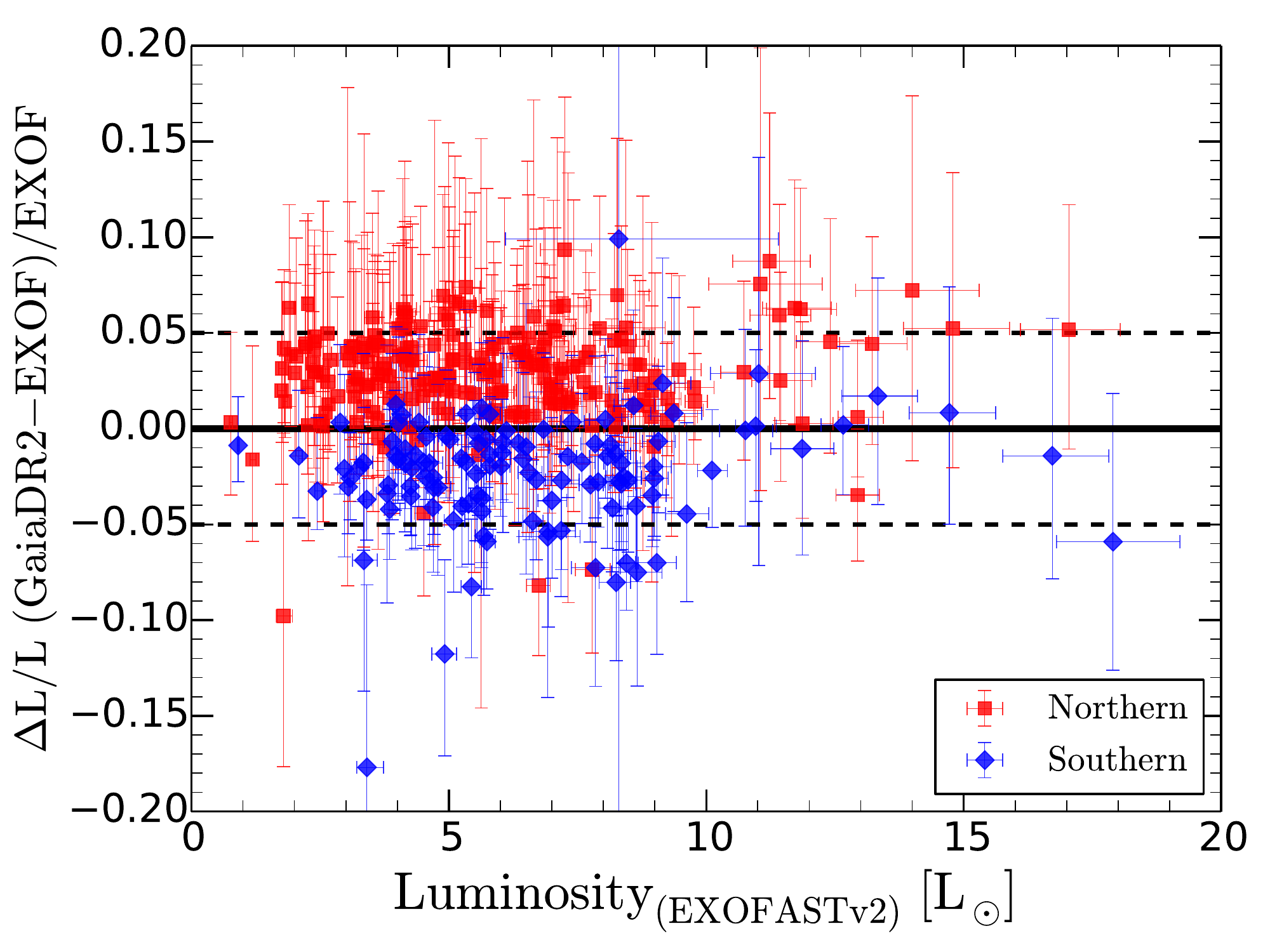}{0.40\textwidth}{(c)}
	    \fig{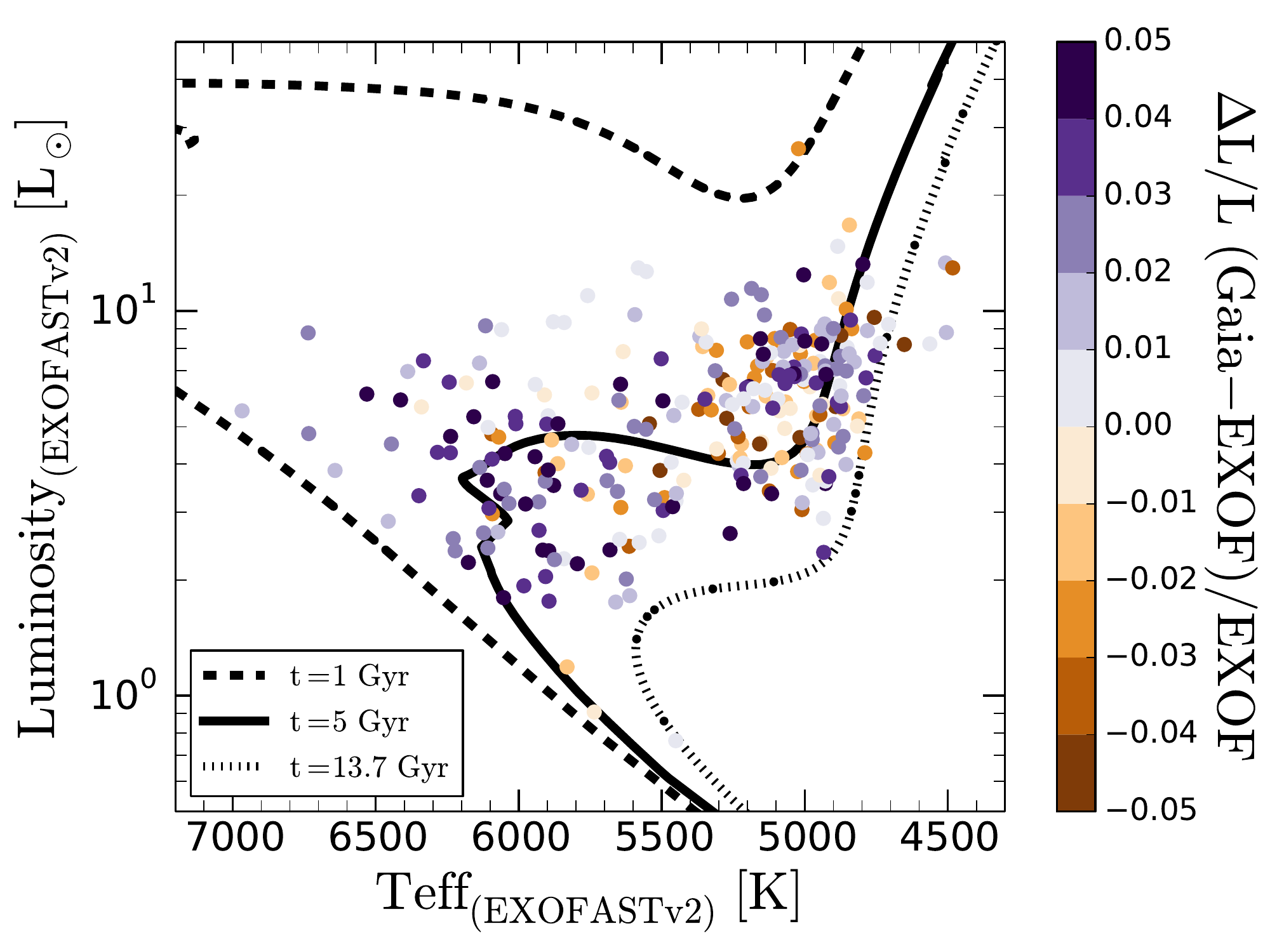}{0.40\textwidth}{(d)}
          }
\gridline{
	    \fig{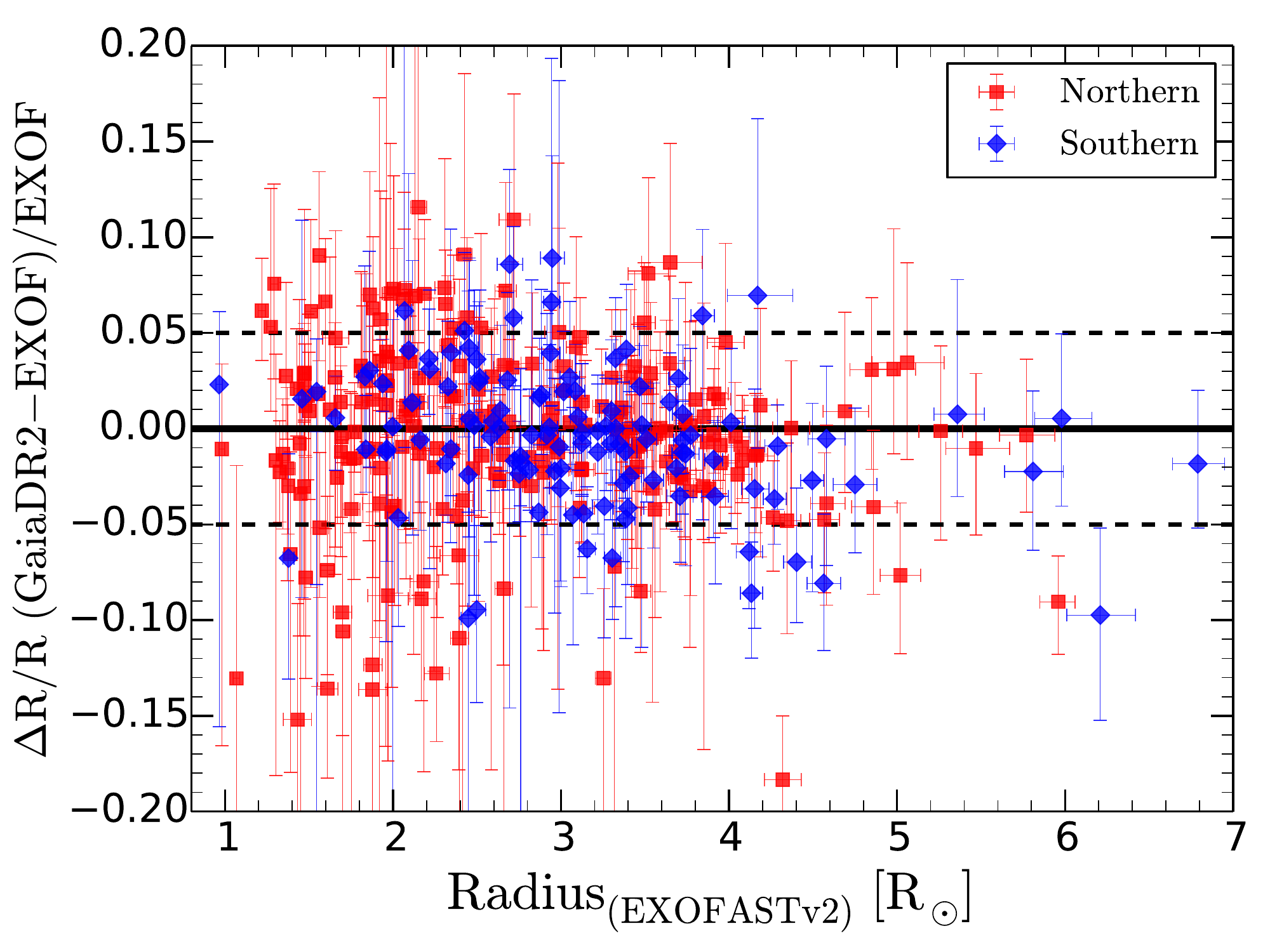}{0.40\textwidth}{(e)}
	    \fig{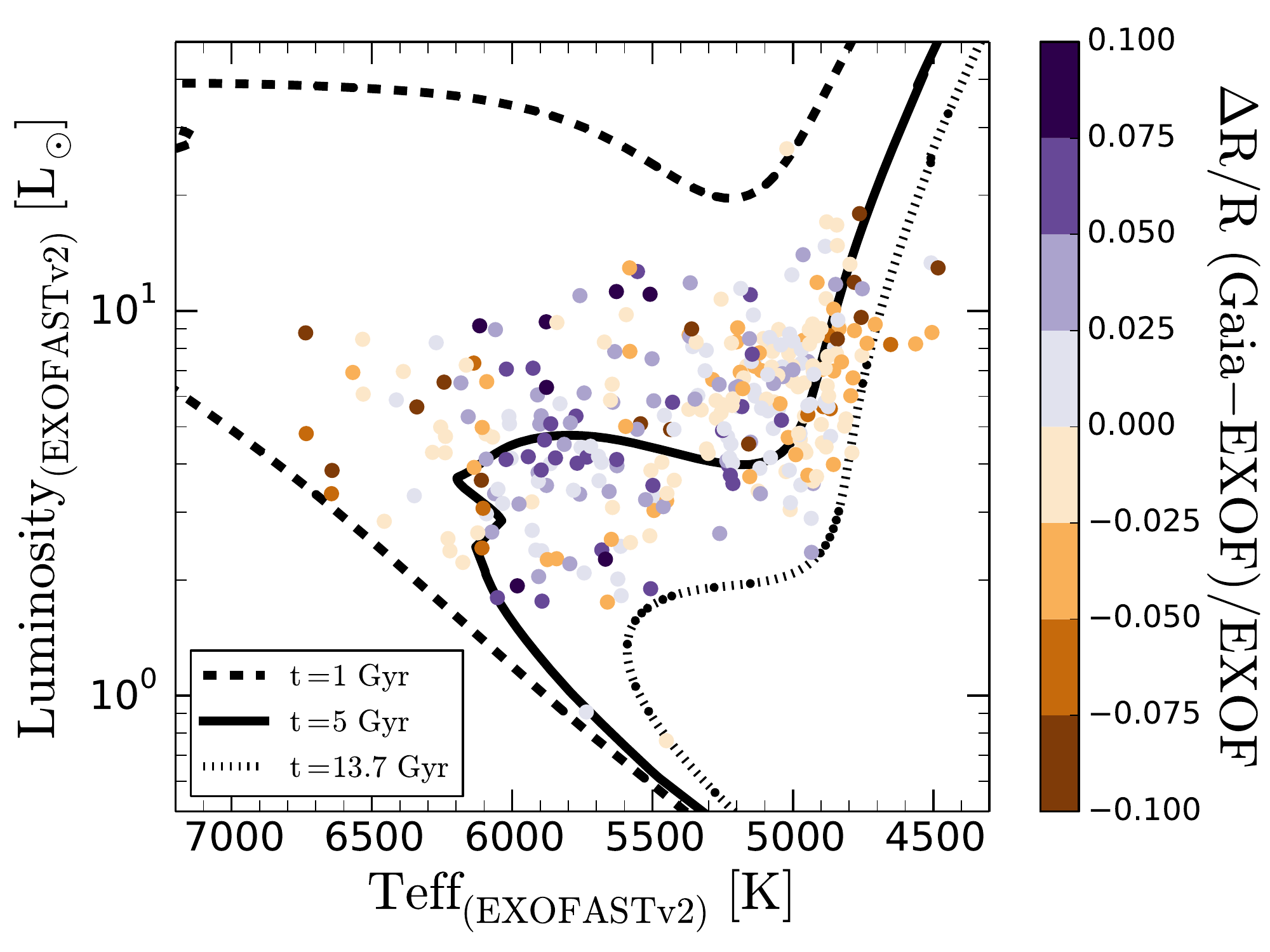}{0.40\textwidth}{(f)}
          }
\caption{Comparison of our derived stellar parameters with those from the {\gaia} DR2 catalog. We show comparisons for temperature (top panels), luminosity (middle panels), and radius (bottom panels). We display these as differences (left column), and as a function of position on the HR diagram (right column). When compared with our stellar parameters, we find the {\gaia} values to show various systematic trends depending on the stars' hemisphere and HR diagram location, highlighting the advantages of detailed studies for smaller samples (such as ours) over large and homogeneous catalogs (see \S\ref{subsec:applications_comparison_with_catalogs_gaia}).}
\label{fig:applications_GaiaDR2}
\end{figure*}

We show a comparison of the {\gaia} DR2 stellar parameters with our values in Figure \ref{fig:applications_GaiaDR2}. The temperature comparison reveals strong systematic differences. Taking our values as reference, in the 4700 K $<$ {\Teff} $<$ 5300 K range (at the base of the RGB), the {\gaia} values go from overestimating {\Teff} by $\sim$ 150 K, to underestimating it by a similar amount (in agreement with Figure 10 of \citealt{andrae18}). For hotter temperatures, 5300 K $<$ {\Teff} $<$ 6200 K (in the middle of the subgiant branch), the {\gaia} values also tend to underestimate {\Teff} and the point to point scatter increases. Closer to the MSTO ({\Teff} $>$ 6200 K), the trend reverses and the {\gaia} values consistently overestimate {\Teff} (sometimes reaching up to $\sim$ 400 K offsets). Regarding luminosity, we see clear hemisphere-dependent offsets across the entire range, with mean values of $\approx +3\%$ for the northern sample (in the sense of {\gaia} minus this work) and $-2\%$  for the southern sample. This is not surprising, as the \citet{andrae18} luminosities do not consider the parallax zero point that we include (which we would analytically expect to cause our luminosities to be intrinsically fainter by about $\approx$ 3\%), or extinction values (which, given the higher $A_V$ adopted for the southern sample, makes it more luminous than the northern sample by about $\approx$ 5\% from similar analytical expectations).  Finally, the radius comparison is mostly contained within $\pm$5\% differences, but with the {\gaia} radii showing similar systematic trends as a function of HR diagram position as those seen for {\Teff}, although in the opposite direction (as $R \propto 1/T_{\text{eff}}^2$). 

\begin{figure*}[ht!]
\gridline{
	    \fig{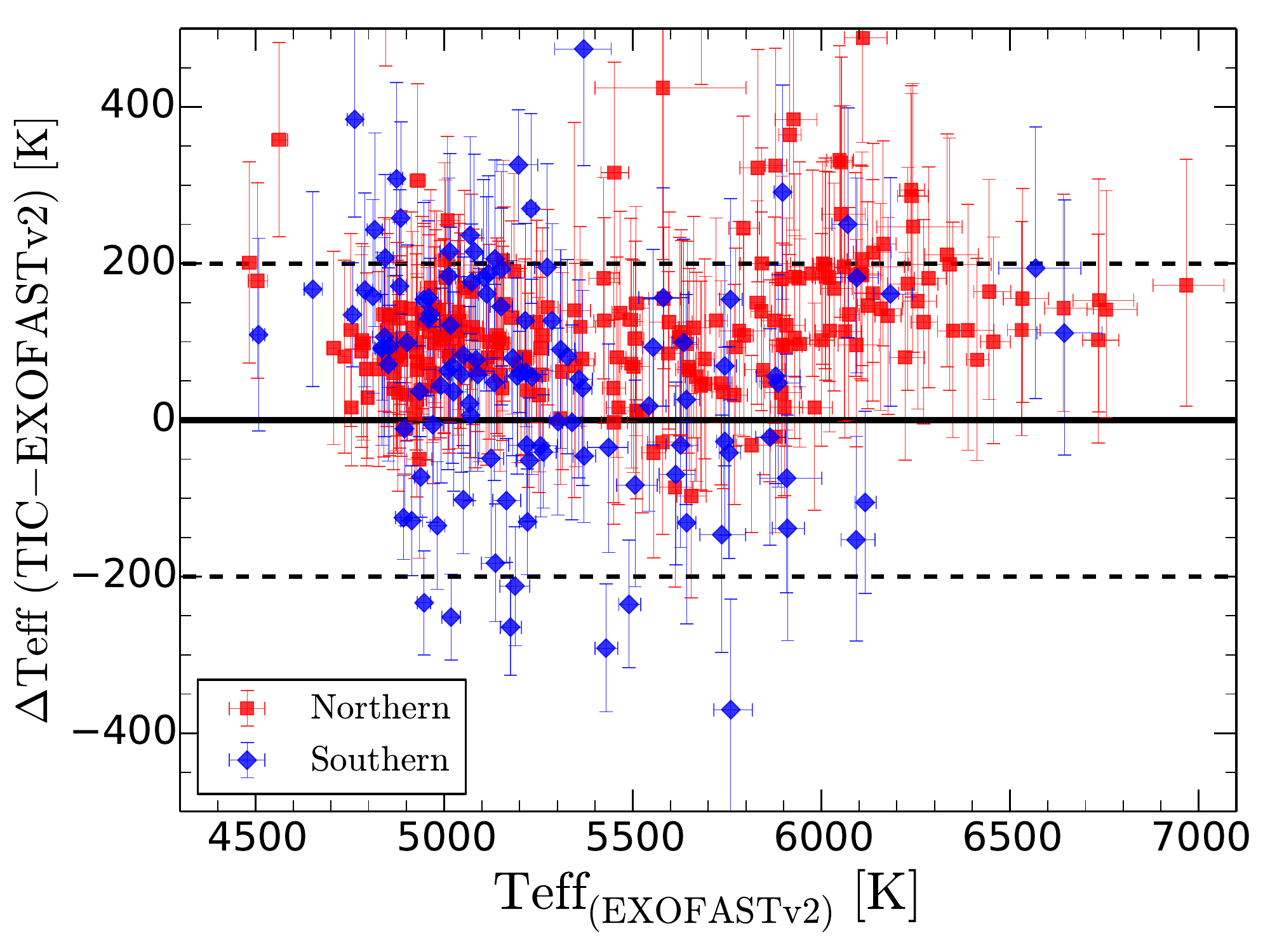}{0.40\textwidth}{(a)}
	    \fig{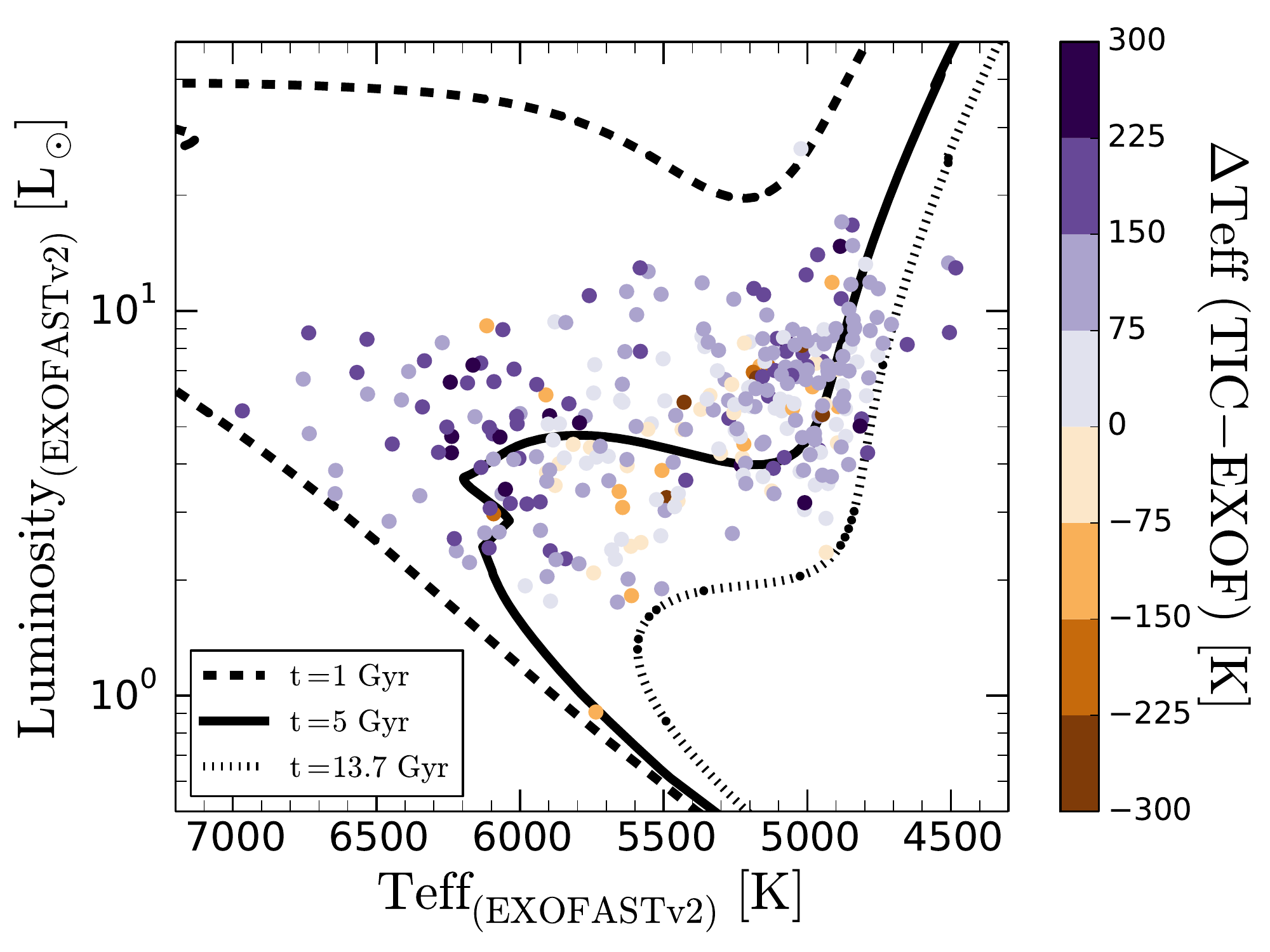}{0.40\textwidth}{(b)}
          }
\gridline{
	    \fig{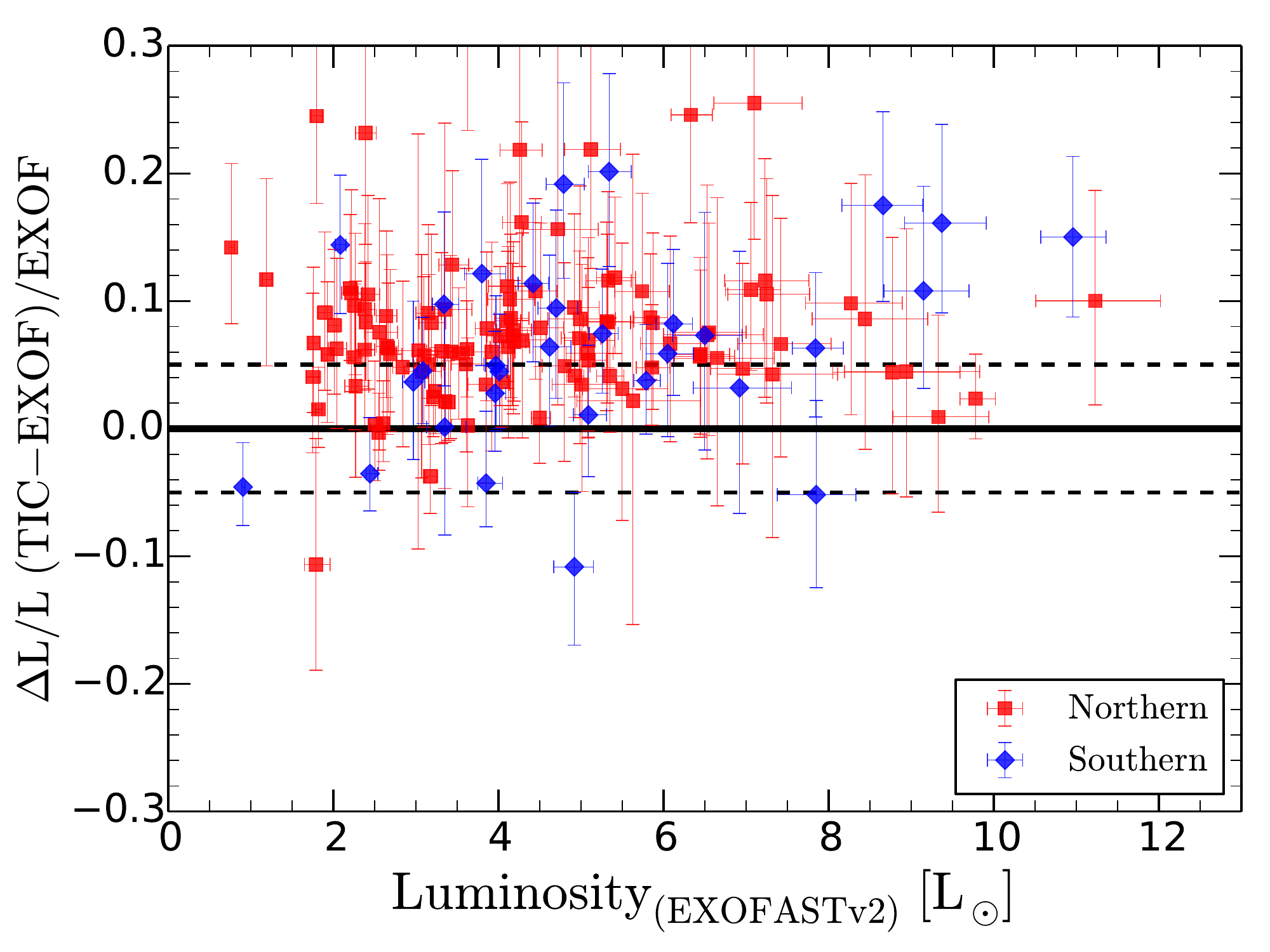}{0.40\textwidth}{(c)}
	    \fig{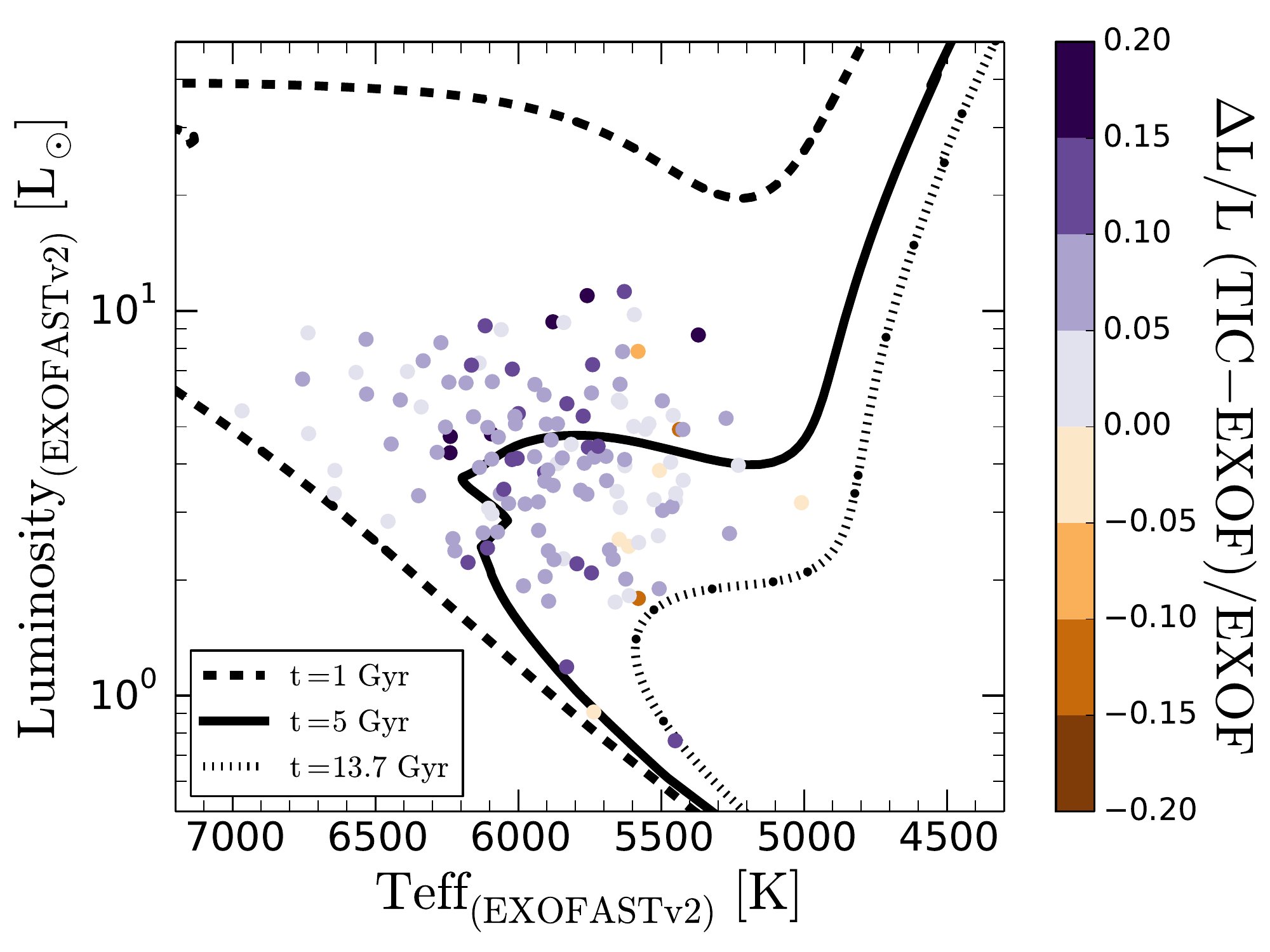}{0.40\textwidth}{(d)}
          }
\gridline{
	    \fig{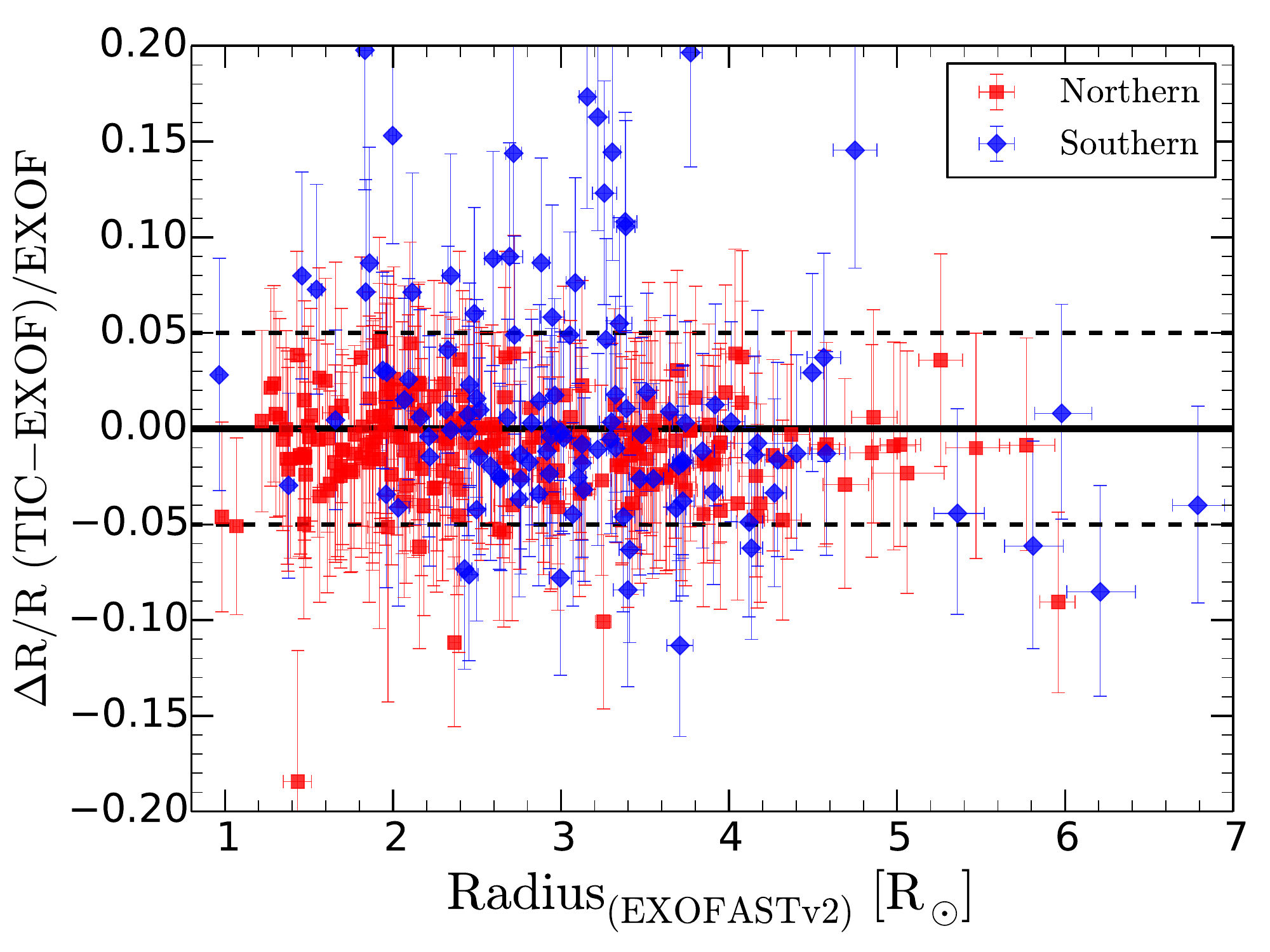}{0.40\textwidth}{(e)}
	    \fig{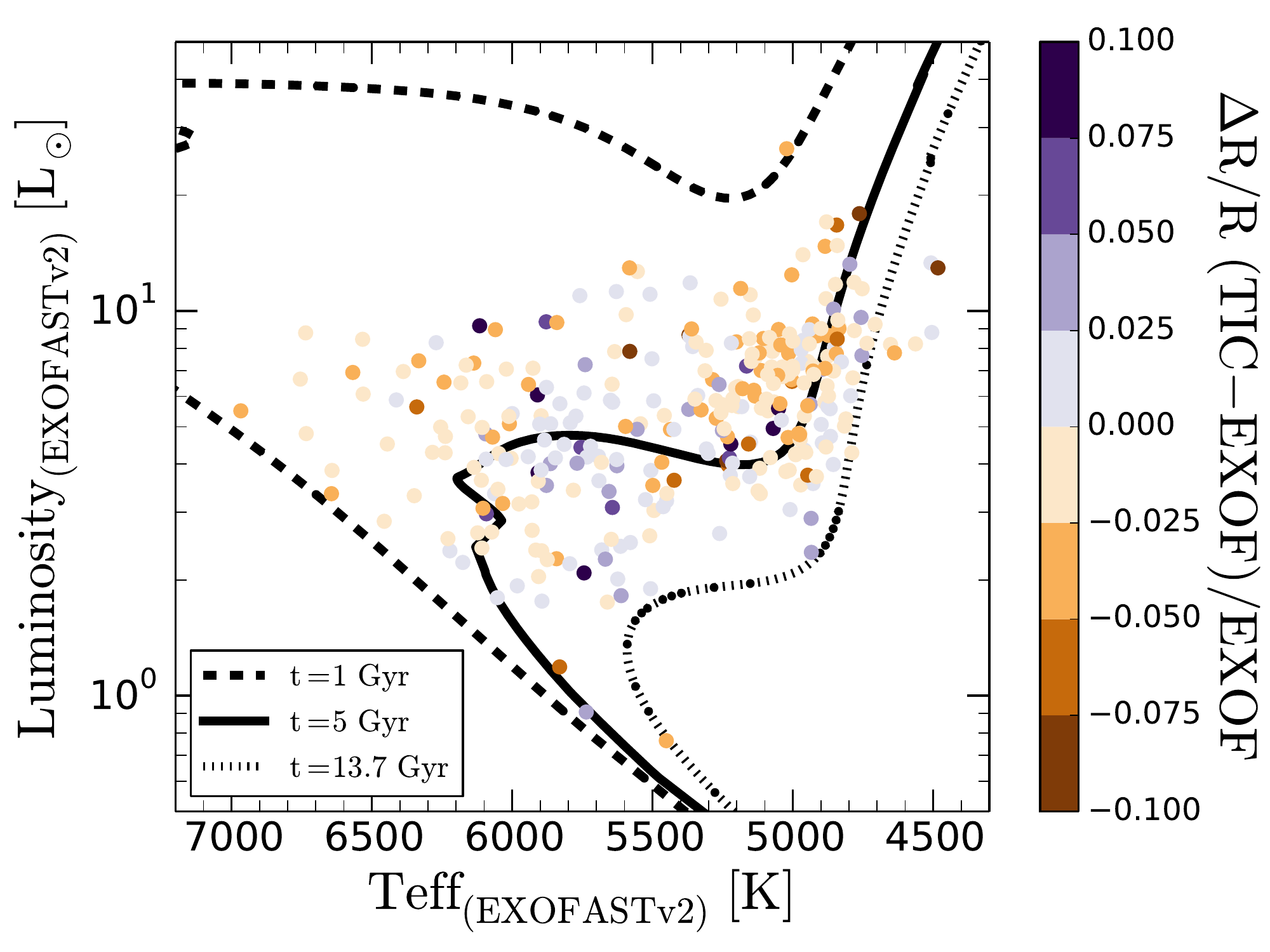}{0.40\textwidth}{(f)}
          }
\caption{Comparison of our derived stellar parameters with those from the TICv8 catalog, analogous to Figure \ref{fig:applications_GaiaDR2}. The discrepancies seen in {\Teff} are likely due to the different temperature scales adopted, and the wider wavelength coverage of the photometry we use, which then propagate onto the luminosity and radius comparison (see \S\ref{subsec:applications_comparison_with_catalogs_TIC}).}
\label{fig:applications_TICv8}
\end{figure*}

\begin{figure*}[ht!]
\gridline{
	    \fig{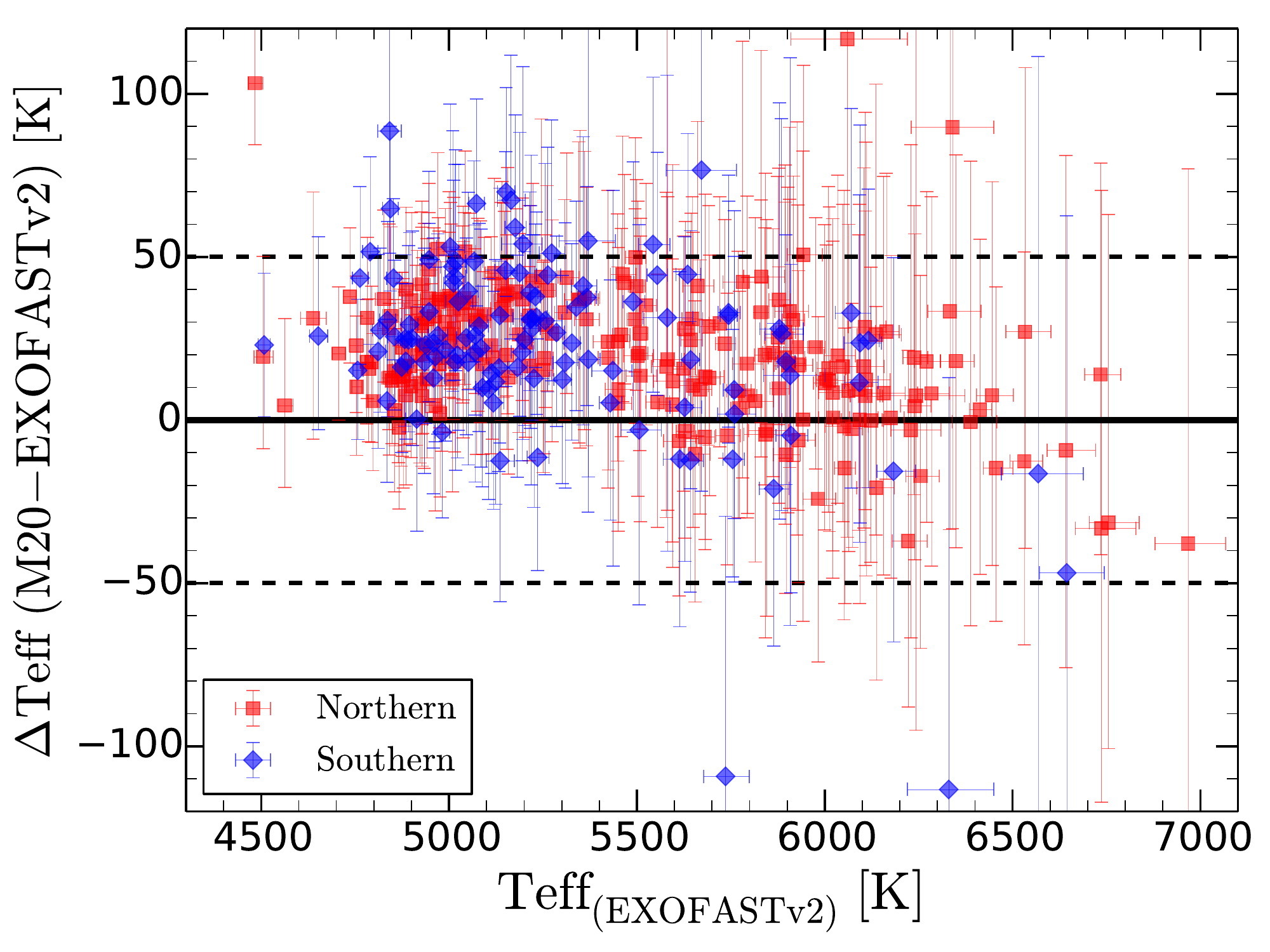}{0.45\textwidth}{(a)}
	    \fig{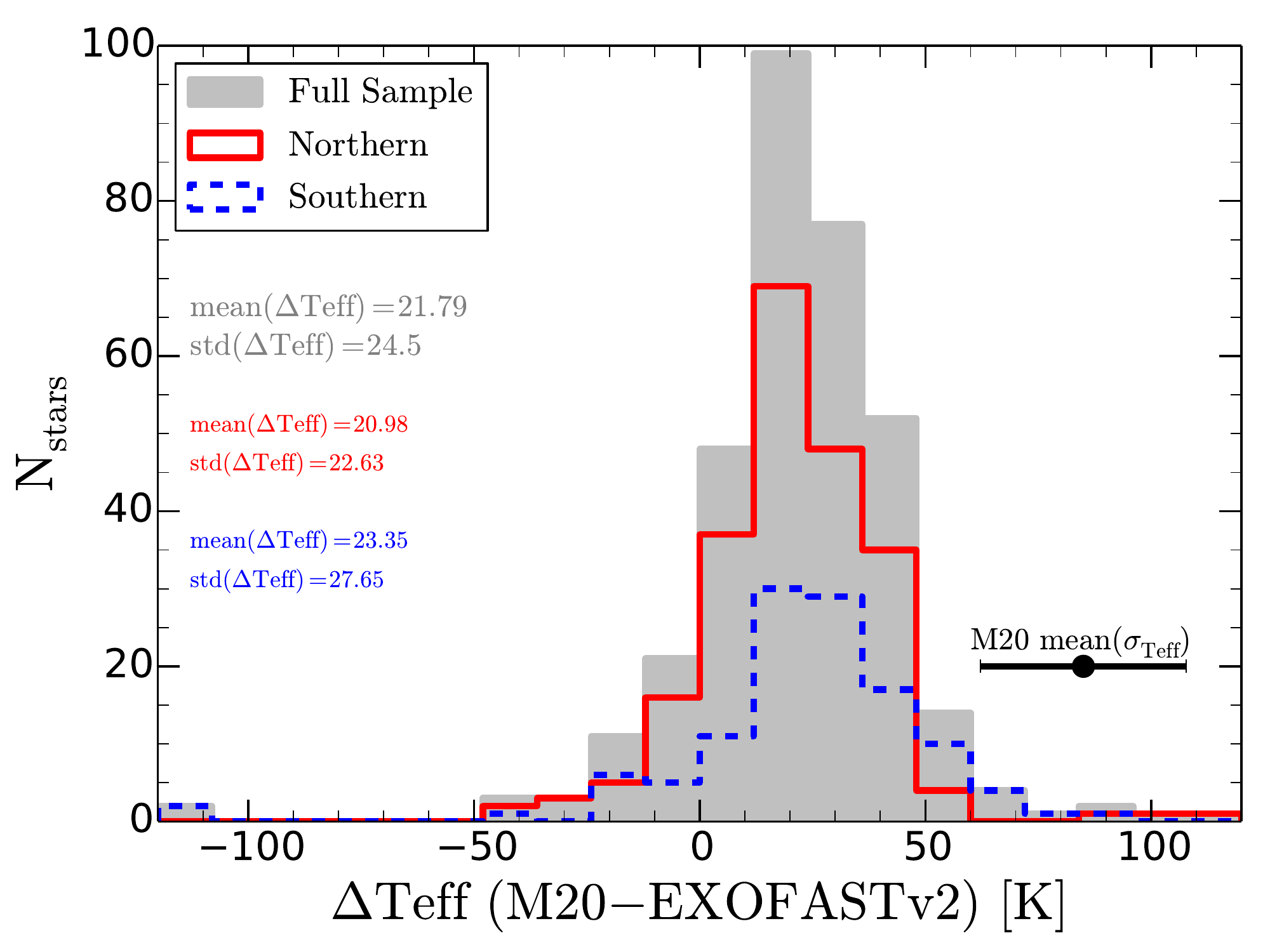}{0.45\textwidth}{(a)}
          }
\caption{Comparison of our derived temperatures with those obtained using the dwarf color-{\Teff} relations from \citet{mucciarelli20}, analogous to Figures \ref{fig:exofast_vs_spectroscopy} and \ref{fig:applications_temperature_scale_comparison}. Note the smaller $\Delta${\Teff} scale compared to previous figures. Given the small difference seen (mean offset of $\approx$ 22 K), we find these dwarf relations to be a practical and reliable tool to estimate temperatures for subgiants stars purely based on {\gaia} and 2MASS photometry.}
\label{fig:applications_Mucciarelli20}
\end{figure*}

For interested readers, under the previously stated assumptions regarding parallax zero point and extinction distributions, we suggest a global correction formula of the {\gaia} luminosities in the TESS CVZs of:
\begin{equation}
L_{\text{Corrected}} = \frac{L_{{\gaia} \text{DR2}}}{(1+x)},
\label{eq:correction_gaia_luminosity}
\end{equation}
where $x=+0.03$ for the northern hemisphere, and $x=-0.02$ for the southern hemisphere.

The above differences highlight some of the advantages of meticulous studies for smaller samples, where better precision and a higher degree of homogeneity can be reached. As expected, choices regarding temperature scales, the parallax zero point, and extinction values propagate into the derived stellar parameters. For instance, in the context of Galactic archaeology, luminosity systematics of order 10\% propagate to differences of order 10\% in age or 4\% in mass, which could introduce undesired biases for highly precise subgiant studies.
\subsection{Comparison with the TICv8}
\label{subsec:applications_comparison_with_catalogs_TIC}

The derivation of the stellar parameters reported in the TICv8 catalog are described in \citet{stassun19}. In summary, for the targets without spectroscopic parameters (the overwhelming majority), effective temperatures were calculated from an empirical {\Teff}-$(G_{\text{BP}}-G_{\text{RP}})$ relation (based on stars with spectroscopic values), with photometry dereddened according to the \citet{green18} and \citet{schlegel98} extinction maps (which, however, we found to be inaccurate for nearby stars, see \S\ref{subsec:methodresults_extinction}). Radii were calculated based on these {\Teff} estimates, {\gaia} distances (from \citealt{bailerjones18}), apparent $G$ band magnitudes, and similar bolometric corrections to those used by \citet{andrae18}. Finally, bolometric luminosities were calculated from the {\Teff} and radius estimates.

We show a comparison of the TICv8 stellar parameters with our values in Figure \ref{fig:applications_TICv8}. Regarding temperature, taking our values as reference, we find the TICv8 to systematically overestimate {\Teff} by $\approx$ 110 K, a feature that is seen across most of the subgiant branch. This is somewhat similar to the comparison with the \citet{casagrande10} values (see \S\ref{subsubsec:methodresults_errors_systematics}), which is unsurprising given that the TICv8 photometric temperatures were calculated based on a relation derived from spectroscopic values, which are often calibrated onto the \citet{casagrande10} {\Teff} scale. Regarding luminosity, the TICv8 does not report values for stars classified as giants \citep{stassun19}, and therefore the comparison is restricted to stars with {\Teff} $\gtrsim$ 5300 K (see panel \textbf{(d)} of Figure \ref{fig:applications_TICv8}). We find a median offset of $\approx$ +7\% in luminosity (in the sense of TICv8 minus this work), which can be explained due to the aforementioned {\Teff} offset, the different parallax zero points employed in the calculations (0.029 mas from \citealt{lindegren18} for the TICv8 versus 0.054 mas from \citealt{schonrich19} for us), and the potential differences arising from the extinction treatment. Finally, for the radius comparison, we find most of the offsets to be within $\pm$5\%, although some systematic trends appear as a function of HR diagram position (as expected from the dependence on temperature and luminosity). Compared to our values, the TICv8 tends to underestimate the radius closer to the base of the RGB, overestimate it more frequently on the middle of the subgiant branch, and underestimate it again closer to the MSTO.

Similarly as in Equation (\ref{eq:correction_gaia_luminosity}), under the adopted parallax and extinction choices, we suggest a global correction factor of the TICv8 luminosities in the northern TESS CVZ of:
\begin{equation}
L_{\text{Corrected}} = \frac{L_{\text{TICv8}}}{(1+x)},
\label{eq:correction_TIC_luminosity}
\end{equation}
where $x=+0.07$. Note that we do not provide a value for $x$ in the southern TESS CVZ, as only a handful of our stars located in that hemisphere have luminosities in the TICv8 (see Figure \ref{fig:applications_TICv8}).

Complementary to the comparison with {\gaia} DR2, the discrepancies of the TICv8 stellar parameters with respect to our values allow us to quantify differences that can arise between a method designed to be used all-sky with one that can take advantage of similarities between closely located stars in smaller samples. It also demonstrates the level of systematic differences still present in modern high quality techniques for inferring stellar parameters (e.g., when more photometric bands with wider wavelength coverage are used, as well as differences arising from color-based relations versus dedicated SED fitting).
\subsection{Comparison with color-{\Teff} relations by \citet{mucciarelli20}}
\label{subsec:applications_comparison_mucciarelli20}

Recently, \citet{mucciarelli20} published a set of six relations that use the {\gaia} $G$, $G_{\text{BP}}$, $G_{\text{RP}}$ and 2MASS $K$ magnitudes to estimate effective temperatures for dwarfs and giants. These relations were constructed based on the \citet{gonzalezhernandez09} sample and IRFM temperatures, and they performed a fit to map these onto color and metallicity dependences. If proven to be accurate, these relations could provide a practical way for estimating temperatures for a wide range of astrophysical studies. 

We compare the temperatures obtained using these relations with our values in Figure \ref{fig:applications_Mucciarelli20}. Depending on the available magnitudes, for each star we use the maximum possible number of all six dwarf color-{\Teff} relations ($G_{\text{BP}}-G_{\text{RP}}$, $G_{\text{BP}}-G$, $G-G_{\text{RP}}$, $G_{\text{BP}}-K$, $G_{\text{RP}}-K$, $G-K$). We then obtain a single temperature value by calculating the weighted mean (and its standard error) of the individual temperatures estimates. 

We find an excellent overall agreement, with a mean offset of $\approx$ 22 K, and only a slight {\Teff}-dependent trend. The range of $\Delta${\Teff}, however, is considerably smaller when compared to those obtained when using the ($J-K$) relations from \citet{gonzalezhernandez09} and \citet{casagrande10} (see Figure \ref{fig:applications_temperature_scale_comparison}). This is not surprising, as in this case we are comparing with an average of up to six different {\Teff} estimates instead of just one. Regardless, given the good agreement seen in Figure \ref{fig:applications_Mucciarelli20}, and considering the homogeneity and wide availability of the {\gaia} and 2MASS photometry, the dwarf \citet{mucciarelli20} relations represent a straightforward and trustworthy tool to estimate temperatures for subgiants without the need for more complex calculations.
\section{Rotation and Activity} 
\label{sec:rotation_and_activity}

Rotation in late-type stars is correlated with the strength of surface magnetic fields, which can both directly impact stellar structure and also modify color-{\Teff} relationships significantly \citep{pecaut13}. This can have a direct and significant effect on stellar parameter estimation. For subgiants, however, this effect is counterbalanced by the majority of them being relatively slow rotators and inactive from a combination of prior angular momentum loss and slowdown from radius expansion.

\begin{figure}[h]
\epsscale{1.2}  
\plotone{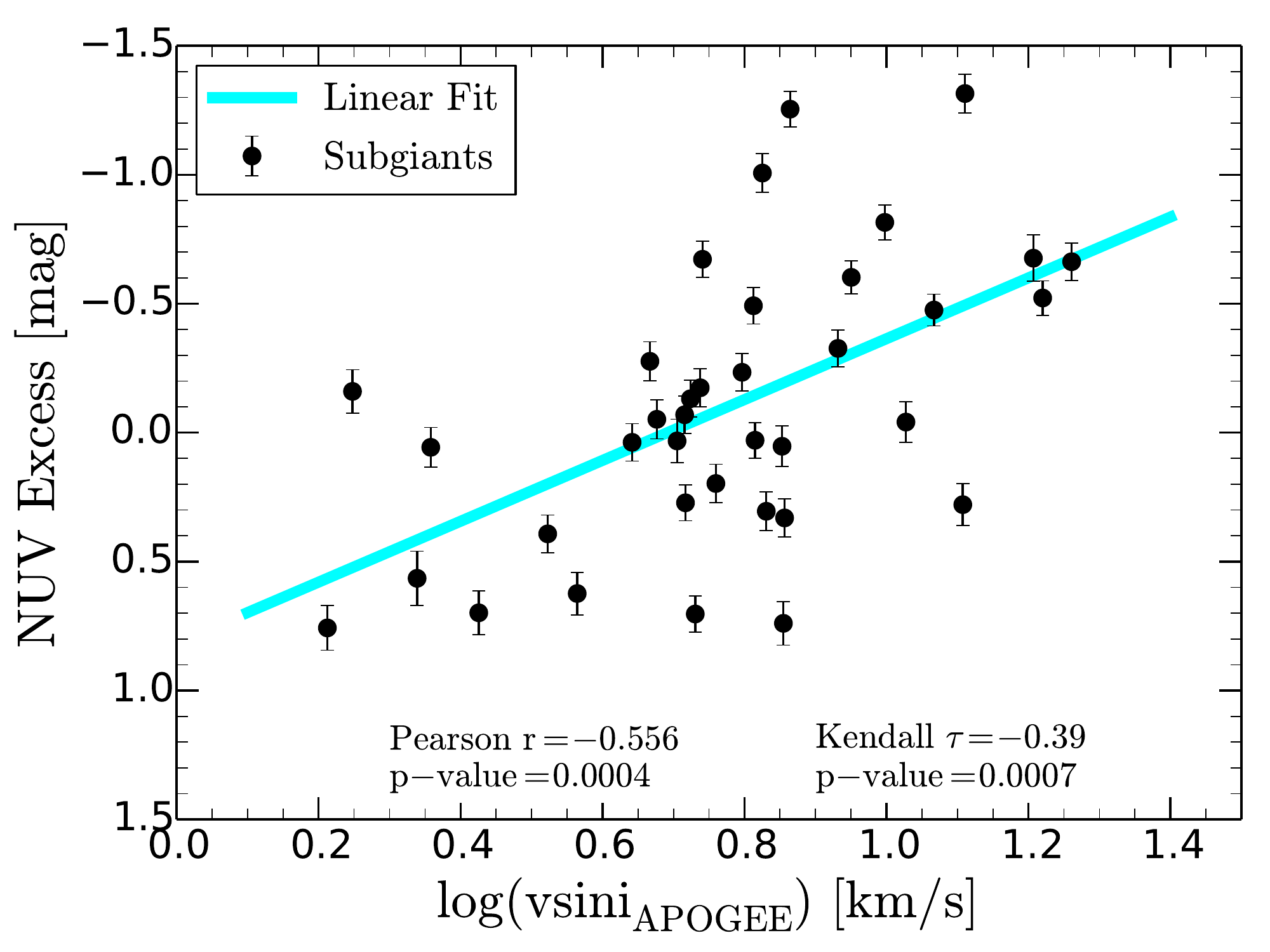}
\caption{NUV excess versus $\log(v \sin i)$ for the subset of subgiants with measured GALEX $NUV$ magnitudes and APOGEE projected rotational velocities (black circles). We find a tentative correlation, with the NUV excess increasing with increasing $v \sin i$, and a linear fit to the data is shown in the  cyan line. We find a qualitatively similar correlation to that reported by \citet{dixon20}, in spite of our sample being less active and less evolved than theirs, reinforcing the connection between stellar rotation and activity in post main sequence stars.}
\label{fig:NUV_excess_vsini}
\end{figure}

The connection between rotation and activity in stars is also a probe of their underlying dynamo mechanism, and its dependence on magnetism and convection \citep{kraft67b,middelkoop82,noyes84,skumanich72}. For stars with convective envelopes, rotation periods and a variety of activity indicators (e.g., Ca {\sc ii} H\&K and H$\alpha$ chromospheric emission, X-ray coronal emission, stellar flares, and NUV excess) have been shown to correlate with each other \citep{findeisen11,magaudda20,pizzocaro19,raetz20,rodriguezmartinez20,stelzer13,stelzer16,wright11,wright13,zhang20}. These comparisons are sometimes parametrized by the Rossby number (the ratio of the rotation period to the convective overturn timescale), or by alternative measurements of rotation such as projected rotational velocities ($v\sin i$). Recently, the rotation-activity correlations have been extended to fully convective low-mass stars \citep{wright16,wright18a}, as well as to evolved stars on the subgiant branch and RGB \citep{dixon20,lehtinen20,wang20}. All of the above hints towards a common dynamo mechanism across a range of masses and evolutionary stages. Therefore, if properly calibrated, stellar rotation and activity indicators could provide useful age diagnostics for single and multiple star systems \citep{barnes03b,barnes03a,barnes10a,chaname12,godoyrivera18,godoyrivera21,janes17,mamajek08,zhang19b}, potentially aiding the understanding of exoplanetary systems and Galactic evolution.

In this context, our sample of subgiant stars offers an opportunity to further explore the rotation-activity connection at an intermediate phase of evolution. As an initial test, we use the subset of stars with measured $v \sin i$ from APOGEE and $NUV$ magnitudes from GALEX (38/347 stars, or $\approx$ 11\% of our sample). We follow \citet{dixon20} and use the latter to calculate an NUV excess from the observed $(NUV-J)$ color, which measures the displacement of a given star with respect to the stellar locus in the NUV-NIR color-color space \citep{findeisen10}. In practice, a stronger NUV excess corresponds to a more negative $(NUV-J)$ color. We present our resulting NUV excess versus $v \sin i$ diagram in Figure \ref{fig:NUV_excess_vsini}, and report both quantities in Table \ref{tab:table_astrometry_photometry_spectroscopy}\footnote{Although reported in Table \ref{tab:table_astrometry_photometry_spectroscopy}, TIC 366487943 is not shown in Figure \ref{fig:NUV_excess_vsini} as its $v \sin i$ value of $\approx$ 79 km s$^{-1}$ most likely corresponds to a measurement error or suggests a binary companion is influencing the results. Additionally, while listed in Table \ref{tab:table_astrometry_photometry_spectroscopy}, TIC 233203155 and TIC 289571725 are not shown in Figure \ref{fig:NUV_excess_vsini}. Although these stars have $v \sin i$ measurements, APOGEE does not report {\Teff} or $\log(g)$ values for them, and we exclude them to avoid stars with potentially poor spectroscopic fits.}.

Figure \ref{fig:NUV_excess_vsini} shows a tentative correlation, in the sense of the NUV excess increasing (becoming more negative) for larger projected rotational velocities. We fit a linear relation to the data, and obtain $y=-1.180x+0.816$ (with $x$ and $y$ representing $\log(v \sin i)$ and NUV excess, respectively). We show this fit as the cyan line, and report its Pearson's $r$ and Kendall's $\tau$ coefficients (and their respective $p$-values), which confirm the correlation and its statistical significance. Although our subgiants are less active and less evolved than the sample studied by \citet{dixon20}, we nonetheless find a similar qualitative correlation. Therefore, our results provide supporting evidence of a connection between stellar rotation and activity on the subgiant branch (see also \citealt{lehtinen20}).

Finally, we note that the limiting quantity in Figure \ref{fig:NUV_excess_vsini} is the availability of $v \sin i$ measurements, as the ASPCAP pipeline is not fully optimized for this purpose ($NUV$ magnitudes, on the other hand, are available for 223/347 stars, or $\approx$ 64\% of the sample). To further explore this rotation-activity correlation with our subgiants, a more meticulous treatment to extract the broadening signals from the spectra is needed (e.g., \citealt{tayar15}), in addition to performing comparisons in other parameter spaces (e.g., using Rossby number). Moreover, this is a regime where the presence of close binaries could resemble the signature of stellar rotation \citep{simonian20}. Performing all of the above, however, is beyond the scope of this paper. Nevertheless, we suggest that future work on this subject could be fruitful.
\section{Conclusions} 
\label{sec:conclusions}

Given their distinctive location on the HR diagram, between the main sequence and the RGB, subgiant stars can place interesting constraints on numerous astrophysical puzzles at an intermediate evolutionary stage. In this paper we study a sample of 347 subgiants located in the {\tess} CVZs that were previously selected based on their probability of showing asteroseismic oscillations. Our procedures and findings are summarized as follows:

\begin{itemize}
\item We perform a thorough characterization based on parallaxes from {\gaia}, photometry from {\gaia}, 2MASS, and ALLWISE, and spectroscopy from APOGEE. While other further sources of photometry (e.g., Tycho-2, APASS) and stellar parameters (e.g., ATL temperatures) are available for our sample, they do not meet our accuracy, precision, and homogeneity requirements. 
\item We fit SEDs using the {\exofast} software and derive accurate and precise stellar parameters, finding that our sample spans a range of luminosities, temperatures, and ages (Figure \ref{fig:exofast_results_LTlogg}). The derived luminosities, effective temperatures, and radii have mean $1\sigma$ random (systematic) uncertainties of 4.5\% (2\%), 33 K (60 K), and 2.2\% (2\%).
\item Our results demonstrate that subgiants are ideal targets for precise determination of masses and ages based on HR diagram location alone (Figure \ref{fig:exofast_results_HRccfracage}). In agreement with other references, we find that metallicity uncertainties of $\sim 0.2$ dex can induce mass and age uncertainties of $\sim 5\%$ and $\sim 10\%$, respectively (Figure \ref{fig:exofast_results_age_vs_MoH}).
\item We test the 3D extinction maps that are generally used in the literature, and find them to consistently overestimate the extinction for our sample of nearby subgiants (distance $\lesssim$ 500 pc), raising concerns about their overall accuracy (Figure \ref{fig:Av_map_comparison}). Given the close spatial proximity of our targets, and based on a calibration using the APOGEE data, we adopt global extinction values of $A_V=0.048$ for the northern hemisphere and $A_V=0.107$ for the southern hemisphere.
\item We use our resulting stellar parameters to explore the advantages of detailed studies for smaller samples (such as ours), in comparison with the parameters reported by large and homogeneous catalogs such as {\gaia} DR2 and the TICv8, with special emphasis on mapping out systematic differences on the subgiant branch (Figures \ref{fig:applications_GaiaDR2} and \ref{fig:applications_TICv8}). We provide correction formulas for the {\gaia} and TICv8 CVZs luminosities in Equations (\ref{eq:correction_gaia_luminosity}) and (\ref{eq:correction_TIC_luminosity}).
\item We examine the subset of stars with measured projected rotational velocities and NUV magnitudes, finding a qualitative correlation that provides supporting evidence for the rotation-activity connection in subgiants (Figure \ref{fig:NUV_excess_vsini}).
\end{itemize}

Our subgiant sample holds immense astrophysical prospects, as it can be used as a calibration or training set for future subgiant studies, as well as to perform novel experiments in a number of fields (ranging from angular momentum transport in stars, to the evolution of extrasolar planets). This study represents a first step in a larger effort to comprehensively characterize and use subgiants. In future works we will improve the age derivation using more specialized tools, complement our data with chemical composition and rotational information from spectroscopic surveys, and fully exploit the synergies with upcoming surface rotation and asteroseismic results from {\tess}. All of this will place subgiants as potentially the best characterized stars, initiating a new era of unprecedented astrophysical studies.
\acknowledgments

We thank Jason Eastman for a large number of useful discussions regarding the functioning and capabilities of {\exofast}. We thank Dan Stevens for valuable conversations regarding the principles of SED fitting. We are deeply thankful to David Will, for always having provided us with exceptional IT support, and for making the realization of this project possible. We thank Daniel Huber for useful comments regarding temperature systematics. We thank Jack Warfield, Jennifer Johnson, and the Ohio State stars group for useful conversations. We thank Travis Berger for sharing his discrete color bar plotting script. We thank the referee for the useful comments and suggestions provided.

DGR acknowledges support from NASA grant 80NSSC19K0367, and DGR and MHP acknowledge support from NASA grant 80NSSC18K1582. JT acknowledges that support for this work was provided by NASA through the NASA Hubble Fellowship grant \#51424 awarded by the Space Telescope Science Institute, which is operated by the Association of Universities for Research in Astronomy, Inc., for NASA, under contract NAS5-26555 and by NASA Award 80NSSC20K0056. JvS acknowledges support from the NASA TESS Guest Investigator Program grant 80NSSC18K18584. Support for this work was provided by NASA through Hubble Fellowship grant \#51386.01 awarded to RLB by the Space Telescope Science Institute, which is operated by the Association of  Universities for Research in Astronomy, Inc., for NASA, under contract NAS 5-26555. DAGH acknowledges support from the State Research Agency (AEI) of the Spanish Ministry of Science, Innovation and Universities (MCIU) and the European Regional Development Fund (FEDER) under grant AYA2017-88254-P. Support for this work was provided to JKT by NASA through Hubble Fellowship grant HST-HF2-51399.001 awarded by the Space Telescope Science Institute, which is operated by the Association of Universities for Research in Astronomy, Inc., for NASA, under contract NAS5-26555.

This work has made use of data from the European Space Agency (ESA) mission {\it Gaia} (\url{https://www.cosmos.esa.int/gaia}), processed by the {\it Gaia} Data Processing and Analysis Consortium (DPAC, \url{https://www.cosmos.esa.int/web/gaia/dpac/consortium}). Funding for the DPAC has been provided by national institutions, in particular the institutions participating in the {\it Gaia} Multilateral Agreement. 

This publication makes use of data products from the Two Micron All Sky Survey, which is a joint project of the University of Massachusetts and the Infrared Processing and Analysis Center/California Institute of Technology, funded by the National Aeronautics and Space Administration and the National Science Foundation. 

This publication makes use of data products from the Wide-field Infrared Survey Explorer, which is a joint project of the University of California, Los Angeles, and the Jet Propulsion Laboratory/California Institute of Technology, funded by the National Aeronautics and Space Administration. 

Funding for the Sloan Digital Sky Survey IV has been provided by the Alfred P. Sloan Foundation, the U.S. Department of Energy Office of Science, and the Participating Institutions. SDSS-IV acknowledges
support and resources from the Center for High-Performance Computing at
the University of Utah. The SDSS web site is www.sdss.org.

SDSS-IV is managed by the Astrophysical Research Consortium for the 
Participating Institutions of the SDSS Collaboration including the 
Brazilian Participation Group, the Carnegie Institution for Science, 
Carnegie Mellon University, the Chilean Participation Group, the French Participation Group, Harvard-Smithsonian Center for Astrophysics, 
Instituto de Astrof\'isica de Canarias, The Johns Hopkins University, Kavli Institute for the Physics and Mathematics of the Universe (IPMU) / 
University of Tokyo, the Korean Participation Group, Lawrence Berkeley National Laboratory, 
Leibniz Institut f\"ur Astrophysik Potsdam (AIP),  
Max-Planck-Institut f\"ur Astronomie (MPIA Heidelberg), 
Max-Planck-Institut f\"ur Astrophysik (MPA Garching), 
Max-Planck-Institut f\"ur Extraterrestrische Physik (MPE), 
National Astronomical Observatories of China, New Mexico State University, 
New York University, University of Notre Dame, 
Observat\'ario Nacional / MCTI, The Ohio State University, 
Pennsylvania State University, Shanghai Astronomical Observatory, 
United Kingdom Participation Group,
Universidad Nacional Aut\'onoma de M\'exico, University of Arizona, 
University of Colorado Boulder, University of Oxford, University of Portsmouth, 
University of Utah, University of Virginia, University of Washington, University of Wisconsin, 
Vanderbilt University, and Yale University.

\appendix
\section{Specifics of SED Fitting with {\exofast}}
\label{appendix:app_sed_fitting}
\subsection{Input photometry}
\label{appendixsub:photometry}

As described in \S\ref{subsec:methodresults_exofast}, we only use photometry from {\gaia}, 2MASS, and ALLWISE in our calculations, effectively overriding {\exofast}'s {\tt MKSED} program, which automatically queries several other photometric catalogs that do not meet our quality requirements (e.g., Tycho-2 and APASS, see \S\ref{subsec:data_photometry_and_astrometry}). Additionally, although in Table \ref{tab:table_astrometry_photometry_spectroscopy} and in the CMDs of Figure \ref{fig:data_sample_astrometry_photometry} we show the photometric uncertainties as reported in the catalogs, for the SED fitting calculation we follow \citet{eastman19} and apply the same photometric error floors that are included by default when using {\tt MKSED}. In practice, this means the assigned errors are the maximum between 0.02 mag and the individual errors for {\gaia} and 2MASS, and the maximum between 0.03 mag and the individual errors for ALLWISE. The purpose of this is to avoid using possibly underestimated photometric uncertainties, and to avoid pushing the SED fitting to a regime where it is difficult to reach convergence.
\subsection{Priors and settings in common for all tests}
\label{appendixsub:priors}

{\exofast} allows the user to provide priors on a number of parameters. We highlight the different types of priors we used throughout this work: 1) a Gaussian prior (i.e., a central value with an uncertainty); 2) a uniform prior within a specified range that starts from an initial guess; and 3) an initial guess alone (where no range is specified). 

We now describe the priors that are common to all our fits in both \S\ref{subsec:methodresults_extinction} and \S\ref{subsec:methodresults_results}. Given that most of our sample are nearby stars (see Figure \ref{fig:data_sample_astrometry_photometry}), {\gaia} DR2 provides stringent parallax values that we use as a Gaussian priors. We follow \citet{schonrich19} and add a 0.054 mas zero point to the {\gaia} parallaxes, as well as a 0.033 mas uncertainty in quadrature to the parallax uncertainties (but note that these corrections have not been applied to the values reported in Table \ref{tab:table_astrometry_photometry_spectroscopy}). We use a reasonable EEP (Equivalent Evolutionary Phase) initial guess for subgiants of 470. For luminosity, radius, and mass, we take a similar approach and use uniform priors with initial guesses of $5 L_{\odot}$, $3 R_{\odot}$, and $1.4 M_{\odot}$ (typical values for subgiants), and wide ranges of $0.001\textendash 40 L_{\odot}$, $0.5\textendash10 R_{\odot}$, and $0.5\textendash 3 M_{\odot}$. The minimum and maximum values for mass, radius, and luminosity have been set with the purpose of excluding main sequence and evolved giant solutions.

To control the settings of the {\exofast} runs, we prepare a file that contains a number of instructions of how to perform each fit. The maximum number of steps taken in the MCMC chain is defined by the parameter {\tt maxsteps}, and we set it to 1000 for every star. Closely tied to the number of steps is {\tt nthin}, which is a factor by which the length of the chain can be thinned. This parameter means that the run will only keep the Nth link in the chain as it is calculated, and it is implemented to reduce the amount of data stored in the computer. In our work we set {\tt nthin} to 10. Lastly, we enable parallel tempering in all our fits. This is set by the {\tt ntemps} parameter, and it is the number of {\it temperatures}. The higher {\tt ntemps} is, the more parameter space the MCMC will explore. We set {\tt ntemps} to 8 for all our stars. Interested readers can find a more in-depth description of these parameters in \citet{eastman13}, \citet{eastman17}, and \citet{eastman19}.
\subsection{Priors used when testing the 3D maps with the spectroscopic sample}
\label{appendixsub:test_spectroscopic_sample}

For this experiment, described in \S\ref{subsec:methodresults_extinction}, we use the subset of subgiants with spectroscopic parameters to solver for star-by-star extinctions. In addition to the priors described in \S\ref{appendixsub:priors} (parallax, EEP, $M$, $R$, $L$), we include as priors the APOGEE {\Teff} and [M/H] values (the latter being fed to {\exofast} as {\tt feh}) as Gaussian priors with small uncertainties ($\sigma_{T_{\text{eff}}}=10$ K, $\sigma_{\text{[M/H]}}=0.01$ dex). These small uncertainties ensure that the {\Teff} and [M/H] returned by {\exofast} will be, by construction, virtually identical to our input parameters, and therefore the actual SED fitting is focused on constraining the extinction.  For extinctions we use a uniform prior in the $0\textendash0.62$ range (the upper value given by the envelope of the \citet{bovy16} $A_V$ distribution for our sample) with an initial guess of $A_V=0$, but otherwise let {\exofast} perform the fit unrestricted.
\subsection{Deriving global and unbiased per-hemisphere extinction values}
\label{appendixsub:unbiased_extinctions}

As described in \S\ref{subsec:methodresults_extinction}, after comparing our star-by-star {\exofast}-derived extinction values with the all-sky 3D maps, we decided to adopt representative per-hemisphere extinction values for the full sample SED calculation. The previous test yielded individual-star extinctions for a total of 165 stars, with 99 and 66 of them belonging to the northern and southern hemisphere, respectively. We now use these stars to calculate representative $A_V$ values for each hemisphere.

One caveat of performing SED fitting with {\exofast} is the hard-coded $A_V \geq 0$ lower limit on the extinction \citep{eastman19}. When deriving extinction values for populations (e.g., star clusters), however, it is statistically acceptable to retrieve negative extinction values for some stars, particularly for systems where the global extinction is low (as in our case). To account for this potential zero point bias on the global extinction values, we repeat the previous fiducial test but inflating the input temperatures from APOGEE by offsets of +50, +100, +150, and +200 K. By doing this we artificially make the stars' SEDs bluer, which in turn generates higher extinction estimates. We then examine the change that a given temperature offset has on the resulting extinction distribution, which allows us to remove the potential zero point bias.

\begin{figure*}
\gridline{
	    \fig{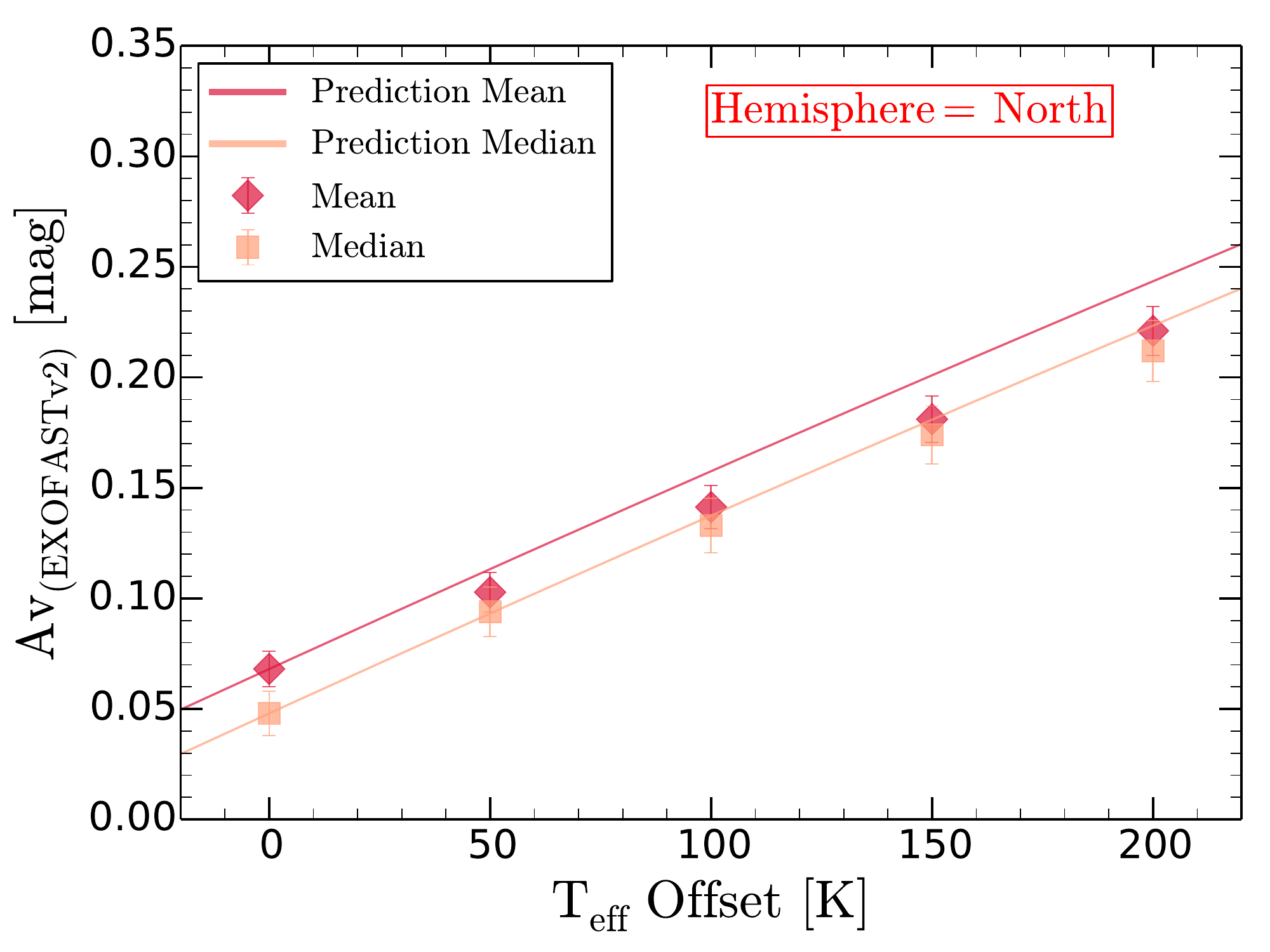}{0.50\textwidth}{(a)}
	    \fig{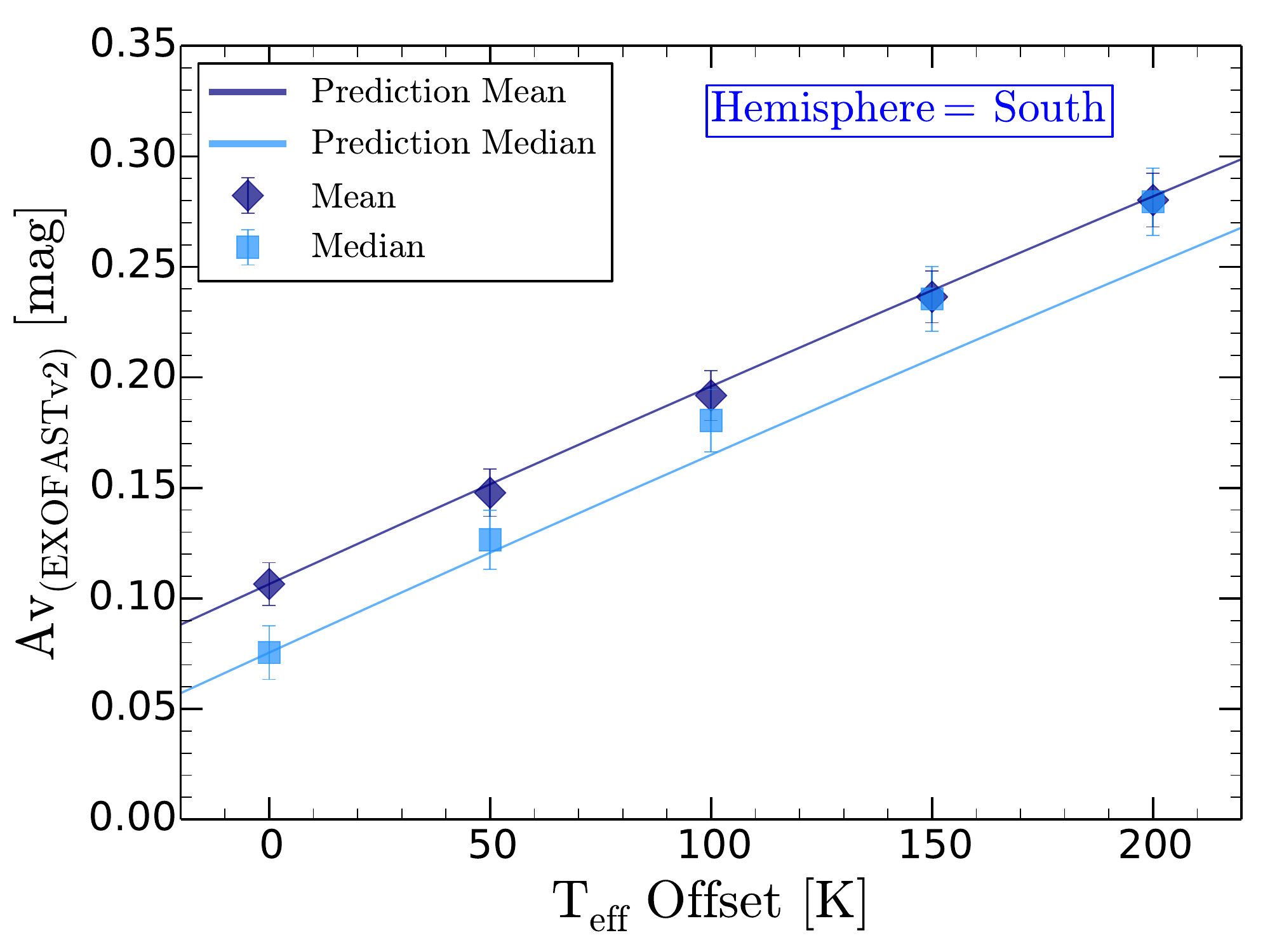}{0.50\textwidth}{(b)}
          }
\caption{Mean and median of the resulting extinction distribution obtained by perturbing the fiducial test of \S\ref{subsec:methodresults_extinction} by offsets of +50, +100, +150, and +200 K for the northern (left) and southern (right) hemisphere. The mean and median values are shown as the diamonds and squares, respectively. The solid lines are the predictions obtained by applying a continuum of extinction offsets to the \citet{mucciarelli20} $G_{\text{BP}}-G_{\text{RP}}$ to {\Teff} relation for an illustrative subgiant with $G_{\text{BP}}-G_{\text{RP}}=0.9$ and solar metallicity. In the northern hemisphere, for large values of the temperature offset (where we do not expect the potential zero point bias inherited from {\exofast} to affect our results), we see that both the mean and median values converge on the median-prediction line, indicating that for the fiducial case (no {\Teff} offset), the median is a better estimator than the mean. On the other hand, in the southern hemisphere the situation is the opposite, and the mean and median values converge on the mean-prediction line, indicating that for the fiducial case, the mean is a better estimator than the median.}
\label{fig:test_Av_zero_point}
\end{figure*}

The results of these tests for both hemispheres are shown in Figure \ref{fig:test_Av_zero_point}, where we plot the resulting mean and median extinction values (with their respective standard errors) as a function of the input temperature offset. We compare these results with analytic predictions based on the \citet{mucciarelli20} $G_{\text{BP}}-G_{\text{RP}}$ to {\Teff} relation for an illustrative example of a subgiant with a $G_{\text{BP}}-G_{\text{RP}}=0.9$ color and solar metallicity (i.e., located in the middle of the subgiant branch of Figure \ref{fig:data_sample_astrometry_photometry}), which we use to perturb the fiducial test (i.e., the test without the temperature offset). 

Figure \ref{fig:test_Av_zero_point} shows that, for the northern hemisphere, the mean and median values converge on the median-prediction line for large values of the temperature offset, indicating that for the fiducial test (no {\Teff} offset), the median is a better estimator than the mean. On the contrary, for the southern hemisphere the mean and median values converge on the mean-prediction line for large temperature offsets, indicating that for the fiducial test the mean is a better estimator than the median. Accordingly, the global extinction values we adopt are the median ($A_V = 0.048 \pm 0.010$) for the northern hemisphere and the mean ($A_V = 0.107 \pm 0.010$) for the southern hemisphere.

\subsection{Priors used when fitting the full sample}
\label{appendixsub:running_full_sample}

For this experiment, described in \S\ref{subsec:methodresults_results}, we use the full sample of subgiant stars (not just the subset with spectroscopic parameters). In order to perform a uniform and unbiased calculation that is valid for the entire sample, we only feed {\exofast} the photometric and astrometric data (which are available for $\approx$ 98\% of the sample). In addition to the priors described in \S\ref{appendixsub:priors} (parallax, EEP, $M$, $R$, $L$), we include as priors values for extinction, temperature, and metallicity.

For extinction, we adopt the global per-hemisphere values calculated in \S\ref{subsec:methodresults_extinction} and \S\ref{appendixsub:unbiased_extinctions}. We feed these to {\exofast} in the form of Gaussian priors, i.e., central values and uncertainties of 0.048 $\pm$ 0.010 for the northern hemisphere and 0.107 $\pm$ 0.010 for the southern hemisphere. For temperature, since we need a prior that depends only on photometry but not on spectroscopy, we use the \citet{casagrande10} ($J-K$) color-{\Teff} relation (assuming [Fe/H]$=0$ and $A_V$ equal to the global per-hemisphere values; see their Table 4 and Equation 3). The advantage of using this relation is that it has been calibrated for subgiants, and that we can calculate it for all stars with measured 2MASS photometry. We feed this {\Teff} value as a prior in the form of an initial guess alone, without any uncertainty or range limiting its resulting value returned by {\exofast}. Finally, for metallicity, we assume that the distribution of spectroscopic metallicities is representative for the full sample, and adopt a Gaussian prior with a central value of [M/H]$=0$ dex and an uncertainty of $\sigma_{\text{[M/H]}} =0.23$ dex. These values are obtained from the mean and standard deviation of the APOGEE [M/H] distribution. 

Our full sample SED fit produces results for 340/347 stars ($\approx$ 98\% of the sample), with the remaining 7 stars lacking a {\gaia} parallax and/or photometry in some of the {\gaia}, 2MASS, or ALLWISE (W1, W2, or W3) bands. We do not attempt to fit these stars. For every star that {\exofast} does fit, it produces a table with the following parameters: $M$, $R$, $L$, $\rho$, $\log(g)$, {\Teff}, [Fe/H], [Fe/H]$_{0}$, Age, EEP, $\log(M/M_{\odot})$, $A_V$, parallax, and distance. In \S\ref{subsec:methodresults_results} and Table \ref{tab:table_exofast_derived_parameters} we only report a subset of these, but we are glad to provide the rest of the parameters to interested readers upon request.
\bibliography{bibliography.bib}{}
\bibliographystyle{aasjournal}



\end{document}